\documentclass{llncs}
\usepackage{graphicx}
\usepackage{hyperref}
\usepackage{tikz}
\usepackage{boxedminipage}
\usepackage{amsmath}
\usepackage{amsfonts}
\usepackage{framed}

\def\llncs{1}       

\def\fullversion{1}       
\def\addtoc{1}      

\makeatletter
\renewcommand*\l@author[2]{}
\renewcommand*\l@title[2]{}
\makeatletter
\DeclareMathAlphabet{\mathsl}{OT1}{cmr}{m}{sl}

\ifnum\llncs=0

\newtheorem{thm}{Theorem}[section]
\newtheorem{lem}[thm]{Lemma}
\newtheorem{cor}[thm]{Corollary}
\newtheorem{propo}[thm]{Proposition}
\newtheorem{clm}[thm]{Claim}
\newtheorem{defn}[thm]{Definition}
\newtheorem{assumption}{Assumption}
\newtheorem{rem}[thm]{Remark}
\newtheorem{fct}[thm]{Fact}
\newtheorem{expr}{Experiment}
\newtheorem{cons}[thm]{Construction}
\newtheorem{nte}[thm]{Note}

\newenvironment{theorem}{\begin{thm}\begin{rm}}%
{\end{rm}\end{thm}}
\newenvironment{lemma}{\begin{lem}\begin{rm}}%
{\end{rm}\end{lem}}
\newenvironment{corollary}{\begin{cor}\begin{rm}}%
{\end{rm}\end{cor}}
\newenvironment{proposition}{\begin{propo}\begin{rm}}%
{\end{rm}\end{propo}}
\newenvironment{definition}{\begin{defn}\begin{rm}}%
{\end{rm}\end{defn}}
\newenvironment{assumption}{\begin{assm}\begin{rm}}%
{\end{rm}\end{assm}}
\newenvironment{claim}{\begin{clm}\begin{rm}}%
{\end{rm}\end{clm}}
\newenvironment{remark}{\begin{rem}\begin{em}}%
{\end{em}\end{rem}}
\newenvironment{fact}{\begin{fct}\begin{em}}%
{\end{em}\end{fct}}
\newenvironment{construction}{\begin{cons}\begin{rm}}%
{\end{rm}\end{cons}}
{\end{rm}\end{nte}}

\else

\newtheorem{thm}{Theorem}
\newtheorem{lem}[thm]{Lemma}
\newtheorem{cor}[thm]{Corollary}
\newtheorem{propo}[thm]{Proposition}
\newtheorem{clm}{Claim}
\newtheorem{defn}{Definition}
\newtheorem{assm}{Assumption}
\newtheorem{rem}{Remark}
\newtheorem{fct}[thm]{Fact}

\newtheorem{cons}[thm]{Construction}

\renewenvironment{theorem}{\begin{thm}\begin{rm}}%
{\end{rm}\end{thm}}
{\end{rm}\end{lem}}
\renewenvironment{corollary}{\begin{cor}\begin{rm}}%
{\end{rm}\end{cor}}
{\end{rm}\end{propo}}
\renewenvironment{definition}{\begin{defn}\begin{rm}}%
{\end{rm}\end{defn}}
{\end{em}\end{assm}}
\renewenvironment{claim}{\begin{clm}\begin{rm}}%
{\end{rm}\end{clm}}
\renewenvironment{remark}{\begin{rem}\begin{em}}%
{\end{em}\end{rem}}
{\end{em}\end{fct}}
{\end{rm}\end{cons}}

\fi

\ifnum\llncs=0

\newlength{\saveparindent}
\newlength{\saveparskip}
\newcount\proofqeded
\newcount\proofended

\def\qed{{\hspace{2pt}\rule[-1pt]{3pt}{9pt}}
\end{rm}\addtolength{\parskip}{-0pt}
\setlength{\parindent}{\saveparindent}
\global\advance\proofqeded by 1 }
\newenvironment{proof}
 {\proofstart}
 {\ifnum\proofqeded=\proofended\qed\fi \global\advance\proofended by 1
  \medskip}
 {\ifnum\proofqeded=\proofended\qed\fi \global\advance\proofended by 1
  \medskip}
\makeatletter
\def\proofstart{\@ifnextchar[{\@oprf}{\@nprf}}
\def\proofsketchstart{\@ifnextchar[{\@osprf}{\@nsprf}}
\def\@oprf[#1]{\begin{rm}\protect\vspace{6pt}\noindent{\bf Proof of #1:\ }%
\addtolength{\parskip}{5pt}\setlength{\parindent}{0pt}}
\def\@osprf[#1]{\begin{rm}\protect\vspace{6pt}\noindent{\bf Sketch of
Proof of #1:\ }
\addtolength{\parskip}{5pt}\setlength{\parindent}{0pt}}
\def\@nprf{\begin{rm}\protect\vspace{6pt}\noindent{\bf Proof:\ }%
\addtolength{\parskip}{5pt}\setlength{\parindent}{0pt}}
\def\@nsprf{\begin{rm}\protect\vspace{6pt}\noindent{\bf Proof Sketch:\ }%
\addtolength{\parskip}{5pt}\setlength{\parindent}{0pt}}

\fi

\ifnum\llncs=1

\else

\fi

\newcommand{\calA}{{\cal A}}

\newcommand{\calM}{{\cal M}}

\def\defeq{\buildrel\triangle\over=}

\newcommand{\Prob}[1]{{\Pr\left[\,{#1}\,\right]}}
\newcommand{\probb}[2]{{\Pr}_{#1}[\,{#2}\,]}

\newcommand{\etal}{{\em et al.}}

\newcommand{\remove}[1]{}

\ifnum\llncs=0

\makeatletter
\def\appearsin#1{\gdef\@appearsin{#1}}

\def\maketitle{\par
 \begingroup
 \def\thefootnote{\fnsymbol{footnote}}
 \def\@makefnmark{\hbox
 to 0pt{$^{\@thefnmark}$\hss}}
 \if@twocolumn
 \twocolumn[\@maketitle]
 \else \newpage
 \global\@topnum\z@ \@maketitle \fi\thispagestyle{plain}\@thanks
 \endgroup
 \setcounter{footnote}{0}
 \let\maketitle\relax
 \let\@maketitle\relax
 \gdef\@thanks{}\gdef\@author{}\gdef\@title{}\gdef\@appearsin{}
          \let\thanks\relax}
\def\@maketitle{\newpage
 \vskip 0.5in \begin{center}
 {\LARGE \@title \par} \vskip 1.5em {\large \lineskip .5em
\begin{tabular}[t]{c}\@author
 \end{tabular}\par}
 \vskip 1em {\normalsize \@date} \end{center}
 \par
 \vskip 1.5em}

\fi

\newcommand{\IND}{{\sf IND}}
\newcommand{\wIND}{{\sf wIND}}
\newcommand{\Priv}{{\sf Priv}}

\newcommand{\A}{\mathcal{A}}
\newcommand{\C}{\mathcal{C}}
\newcommand{\B}{\mathcal{B}}

\newcommand{\erase}[1]{}
\newcommand{\ignore}[1]{}

\def\from{\leftarrow}
\newcommand{\zu}{\{0,1\}}

\newcommand{\prob}[1]{{\rm Prob}[#1]}

\newcommand{\e}{\ensuremath{\mathbf{e}}}

\newcommand{\pk}{{\sf pk}}

\newcommand{\SK}{{\sf Sk}}

\newcommand{\CT}{{\sf Ct}}
\newcommand{\Rencf}{{\sf R}^{\sf enc, full}}
\newcommand{\Rdec}{{\sf R}^{\sf dec}}
\newcommand{\Rdecf}{{\sf R}^{\sf dec, full}}
\newcommand{\Ct}{{\sf Ct}}
\newcommand{\OK}{{\sf OK}}
\newcommand{\Abst}{{\sf Abst}}
\newcommand{\BT}{{\sf Blt}}

\newcommand{\ct}{{\sf ct}}

\newcommand{\PK}{{\sf Pk}}

\newcommand{\Verify}{{\sf Verify}}

\newcommand{\EVOTEF}{\mathsf{EVOTE}_{\mathsf {full}}}
\newcommand{\EVOTE}{\mathsf{EVOTE}}

\newcommand{\NIWI}{\mathsf{NIWI}}
\newcommand{\NIWIEncf}{\mathsf{NIWI}^{\sf enc, full}}
\newcommand{\NIWIDec}{\mathsf{NIWI}^{\sf dec}}
\newcommand{\NIWIDecf}{\mathsf{NIWI}^{\sf dec, full}}
\newcommand{\Prove}{\mathsf{Prove}}
\newcommand{\ProveEncf}{\mathsf{Prove}^{\sf enc, full}}
\newcommand{\ProveDec}{\mathsf{Prove}^{\sf dec}}
\newcommand{\ProveDecf}{\mathsf{Prove}^{\sf dec, full}}
\newcommand{\VerifyEncf}{\mathsf{Verify}^{\sf enc, full}}
\newcommand{\VerifyDec}{\mathsf{Verify}^{\sf dec}}
\newcommand{\VerifyDecf}{\mathsf{Verify}^{\sf dec, full}}

\newcommand{\PKE}{\mathcal{E}}

\newcommand{\adv}{\calA}
\newcommand{\Adv}{{\sf Adv}}

\newcommand{\M}{\calM}

\newcommand{\nota}[1]{{\bf Nota: } #1 \par}
\renewcommand{\nota}[1]{}

\newcommand{\Com}{{\sf Com}}
\newcommand{\com}{{\mathsf{com}}}

\newcommand{\Setup}{{\sf Setup}}
\newcommand{\Setupf}{{\sf Setup}_{\sf full}}
\newcommand{\NP}{{\sf NP}}
\newcommand{\CircuitSat}{{\sf CircuitSat}}
\newcommand{\Cast}{{\sf Cast}}
\newcommand{\Castf}{{\sf Cast}_{\sf full}}
\newcommand{\VerifyBallot}{{\sf VerifyBallot}}
\newcommand{\VerifyBallotf}{{\sf VerifyBallot}_{\sf full}}
\newcommand{\VerifyTally}{{\sf VerifyTally}}
\newcommand{\VerifyTallyf}{{\sf VerifyTally}_{\sf full}}
\newcommand{\EvalTally}{{\sf EvalTally}}
\newcommand{\EvalTallyf}{{\sf EvalTally}_{\sf full}}

\newcommand{\SecGame}{{\sf Priv}}
\newcommand{\SecwGame}{{\sf WeakPriv}}

\newcommand{\Dec}{{\sf Decrypt}}

\newcommand{\Enc}{{\sf Encrypt}}

\renewcommand{\nota}[1]{}

\newcommand{\negl}{{\sf negl}}

\newcounter{itemcount}

\definecolor{rosa}{RGB}{255, 102, 178}

\title{(Universal) Unconditional Verifiability in E-Voting without Trusted Parties}
\ifnum\fullversion=1
\author{Gina Gallegos-Garcia\inst{1} \and Vincenzo Iovino\inst{2} \and Alfredo Rial\inst{2} \and Peter B. R\o nne\inst{2} \and Peter Y. A. Ryan\inst{2}}

\institute{
     Instituto Politecnico Nacional, Mexico\\
      {\tt ggallegosg@ipn.mx}\\
      \and University of Luxembourg\\
      {\tt vinciovino@gmail.com, \{alfredo.rial@uni.lu, peter.roenne, peter.ryan\}@uni.lu}\\
}
\fi
\begin{document}
\maketitle
\date{}
\begin{abstract}

In e-voting protocol design, cryptographers must balance usability and strong security guarantees, such as privacy and verifiability. In traditional e-voting protocols, privacy is often provided by a trusted authority that learns the votes and computes the tally. Some protocols replace the trusted authority by a set of authorities, and privacy is guaranteed if less than a threshold number of authorities are corrupt. For verifiability, stronger security guarantees are demanded. Typically, corrupt authorities that try to fake the result of the tally must always be detected.

To provide verifiability, many e-voting protocols use Non-Interactive Zero-Knowledge proofs (NIZKs). Thanks to their non-interactive nature, NIZKs allow anybody, including third parties that do not participate in the protocol, to verify the correctness of the tally. Therefore, NIZKs can be used to obtain universal verifiability. Additionally, NIZKs also improve usability because they allow voters to cast a vote using a non-interactive protocol.

The disadvantage of NIZKs is that their security is based on setup assumptions such as the common reference string (CRS) or the random oracle (RO) model. The former requires a trusted party for the generation of a common reference string. The latter, though a popular methodology for designing secure protocols, has been shown to be unsound.

In this paper, we address the design of an e-voting protocol that provides verifiability without any trust assumptions, where verifiability here is meant without eligibility verification. We show that Non-Interactive Witness-Indistinguishable proofs (NIWI) can be used for this purpose. The e-voting scheme is private under the Decision Linear assumption, while verifiability holds unconditionally. To our knowledge, this is the first private e-voting scheme with {\em perfect} universal verifiability, i.e. one in which the probability of a fake tally not being detected is 0, and with {\em non-interactive} protocols that does not rely on trust assumptions.

\end{abstract}
\ifnum\addtoc=1
\setcounter{tocdepth}{3}
   \setcounter{page}{1}

    \tableofcontents
   \newpage
   \fi

	    \setcounter{footnote}{0}


\section{Introduction}\label{sec:intro}

\subsection{Background and Statement of the Problem}
The parties participating in a standard e-voting protocol are multiple voters and one authority. First, the authority sets up a public key retaining a corresponding secret key. A voter computes a ballot on input the public key of the authority and her intended vote and sends the ballot to a write-only public bulletin board (PBB),  which records it in an entry associated with that voter. In case of abstention, a special symbol $\bot$ is recorded on the PBB. The authority uses its secret key to compute the tally on input all the ballots on the PBB, which could possibly be $\bot$ in case of abstention. Finally, the correctness of the tally can be checked by running a verification algorithm.\footnote{In this description we skipped some details (e.g., eligibility and authentication) that are not relevant to our setting. See below for more discussion.}

E-voting protocols must provide two security properties: privacy and verifiability. Privacy should protect the secrecy of the votes. Verifiability should prevent a corrupt authority from faking the tally.  We will provide a formal definition of verifiability that is  stronger than previous ones in some respects.

Privacy protection assumes the existence of a trusted authority in many e-voting systems~\cite{CACM:Cha81,EC:CraGenSch97,PKC:DamJur01,ryan006,Helios,DBLP:journals/tifs/ChaumCCEPRRSSV09,Ryan09prettygood,JCJ10}. As for schemes that distribute the trust among several authorities, privacy protection still requires that not all of the authorities are corrupt. Nevertheless, {\em verifiability} (also called integrity) should be guaranteed even if the authorities are corrupt.


Many e-voting systems
make use of Non-Interactive Zero-Knowledge Proofs (NIZK)~\cite{STOC:BluFelMic88,C:DesMicPer87,C:RacSim91,Goldreich01,C:SCOPS01} to provide verifiability. NIZK must provide two properties: soundness and zero-knowledge. Soundness prevents a corrupt prover from proving a false statement, i.e., a statement for which no witness exists. Zero-knowledge ensures that the verifier does not learn any information about the witness.

Zero-knowledge is defined following the simulation paradigm, i.e., it requires the existence of a simulator that computes a valid proof without knowledge of the witness. However, if such a simulator existed, soundness would not hold. This apparent contradiction is solved by resorting to trust assumptions like the Common Reference String (CRS) model~\cite{STOC:BluFelMic88}. In the CRS model, a trusted party generates a CRS that is used by both provers and verifiers. The simulator is given the additional power of computing the CRS. Thanks to that, the simulator knows trapdoor information that allows it to simulate proofs for all statements.

For some applications of NIZK, the CRS model is not problematic. For instance, in IND-CCA public key encryption schemes~\cite{STOC:NaoYun90,C:SCOPS01,CraSho03}, zero-knowledge does not need to hold for the receiver of ciphertexts because the receiver must be able to decrypt anyway. Therefore, the CRS is computed by the receiver, while the NIZK proofs are computed by the sender of ciphertexts.  However, in e-voting, the authority cannot compute the CRS because it must compute proofs that show the correctness of the tally.


An alternative to the CRS model is the Random Oracle (RO) model~\cite{CCS:BelRog93}. The RO model assumes the availability of a perfect random function available to all parties. NIZKs that use the RO model are constructed following the Fiat-Shamir heuristic~\cite{C:FiaSha86}. To prove that a NIZK proof constructed following this heuristic is zero-knowledge, we need programmability of the RO, i.e., the ability of the simulator to change the input/output of the RO.

To compute a proof, in practice, the prover replaces the RO by some ``secure'' hash function. Therefore, this hash function must be chosen honestly for zero-knowledge to hold. Consequently, all the parties must trust the implementation of a concrete hash function (e.g., SHA-3~\cite{Keccak}). We note that a hash function could have been designed in a malicious way (e.g., ``programmed'' like in the simulation) to allow the computation of a proof for a false statement. Currently, this needed trust on the implementation of hash functions does not exist. In fact, different political entities have developed their own hash functions because they do not trust the hash functions designed by others. For instance, the Russian government discourages the use of SHA-3 and encourages the use of its own hash function~\cite{Streebog}.



Moreover, even when programmability is not needed, the RO methodology has been shown to be unsound~\cite{STOC:CanGolHal98}. Further problems are known regarding the programmability of the RO in the context of NIZK~\cite{FOCS:GolKal03,PHD:Kalai06,TCC:BDGJKLW13}. The current techniques to avoid the need of programmability resort to the CRS model~\cite{TCC:DamFazNic06,TCC:Lin15,PKC:ChaGro15,TCC:CPSV16}.

This motivates our main question: is it possible to design an e-voting scheme that is verifiable without assuming any trust assumption (like CRS and RO)?

In a survey~\cite{Lipmaa05}, Lipmaa asks whether Non-Interactive Witness Indistinguishable Proofs (NIWI) can be used to replace NIZKs.
NIWIs can be constructed without using any trust assumptions~\cite{C:GroOstSah06,FOCS:DwoNao00,C:BarOngVad03,TCC:BitPan15}.\footnote{Note that, in the literature, there are both NIWIs in the CRS model, like the ones of Groth and Sahai~\cite{EC:GroSah08}, and one-message NIWIs without CRS (see the citations above). Henceforth, unless specified otherwise, we denote by NIWI the (one-message) variant without CRS, and in particular we refer to the NIWIs for $\CircuitSat$ of Groth \etal\ \cite{C:GroOstSah06}.}


NIWI is a non-interactive proof/argument system that provides weaker security guarantees in comparison to NIZKs. While NIZKs ensure that a proof does not reveal any information about the witness, NIWIs only guarantee that, for any two witnesses $w_1$ and $w_2$ for the same statement, a proof computed with $w_1$ is computationally indistinguishable from a proof computed with $w_2$.
Note that this notion only makes sense for languages with multiple witnesses for each statement, which is not always the case.


To our knowledge, it was not known how to use NIWI to construct an e-voting scheme (eVote, in short) that is both private and verifiable.
Usually, it is very difficult to use NIWI because of its weaker security guarantee. Nonetheless, inspired by a recent result on functional encryption~\cite{TCC:BonSahWat11,FOCS:GGHRSW13} of Badrinarayanan, Goyal, Jain and Sahai~\cite{EPRINT:BGJS16}, we are surprisingly able to profitably use NIWI to answer our main question affirmatively.






\subsection{Our Results}

First, we define correctness, privacy and verifiability properties for an eVote. We define two flavors of privacy and verifiability: weak and full. We propose an eVote that is (fully) private and (fully) verifiable.   Its privacy can be reduced to the Decision Linear assumption~\cite{C:BonBoySha04}. Its verifiability is {\em perfect} (see below) and thus is not based on any assumption. Moreover, its verifiability is {\em universal}, i.e., even a third party who did not participate in the election process should be able to verify the correctness of the tally. As a warm-up, we also describe an eVote that fulfills the weak privacy and weak verifiability properties.

Our eVote uses as building blocks a NIWI proof system, a public key encryption scheme with perfect correctness and unique secret key, and a perfectly binding commitment scheme. It can be instantiated by using just bilinear groups~\cite{BonFra03,JC:Joux04}. For instance, we can instantiate our construction with the NIWI of Groth, Ostrovsky and Sahai \cite{C:GroOstSah06} and the Decision Linear encryption scheme of Boneh \etal\ \cite{C:BonBoySha04}. When instantiated with those building blocks, our construction is the first eVote with {\em non-interactive} algorithms for casting and verifying ballots and for computing and verifying the tally that provides {\em perfect} (weak and full) verifiability (as defined in Def. \ref{def:evote}) and that fulfills the (weak and full) privacy property under the Decision Linear assumption~\cite{C:BonBoySha04}. The Decision Linear assumption is a well-studied assumption over bilinear groups. Our construction attains {\em universal} verifiability, i.e., even third parties who did not participate in the election process are able to verify the tally.

We prove that our weakly verifiable eVote fulfills the weak verifiability and weak privacy properties in Corollary~\ref{cor:weak}, and we prove that our (fully) verifiable eVote fulfills the (full) verifiability and (full) privacy properties in Corollary~\ref{cor:full}. We remark that the computational assumption is only needed to prove that our eVotes fulfill the (weak or full) privacy properties. In contrast, no assumption at all is necessary to prove that they fulfill the (weak or full) verifiability properties.

Therefore, our eVote with non-interactive algorithms is the first eVote whose perfect verifiability is not based on any trust and that is provably secure under a well-studied and falsifiable assumption~\cite{C:Naor03}. The latter is a key point of our results because otherwise one could just claim that an eVote in the RO model is secure when instantiated with any hash function. However, even when using such ``unfalsifiable'' assumptions, {\em perfect} verifiability cannot be achieved against unbounded adversaries because, in practice, for any hash function whose domain is larger than the range, the probability of finding a collision is not 0.

In Section \ref{sec:threshold}, we outline how to adapt our (fully) verifiable construction to a model with multiple authorities. In this model, the tally evaluation algorithm is run by a set of authorities and the privacy property must hold if at least one authority is honest. (As this is not the main focus of our work, we do not present formal definitions and details for its construction.) An important advantage of our construction is that no interaction among the authorities is required. In this respect, our techniques completely diverge from previous approaches to the problem and may be of independent interest. We stress that the multi-string model of Groth and Ostrovsky~\cite{JC:GroOst14}, though conceptually appealing in this scenario, fails to provide a solution.

In this work, we use cryptographic primitives to demonstrate the achievability of perfect verifiable systems. However, we are not concerned about usability and ``human-friendly'' verifiability, as dealt with in~\cite{threeballot,Randell06votingtechnologies,selene}. Furthermore, we only consider traditional e-voting systems and hence we neglect other approaches~\cite{PKC:KiaYun02,ACISP:DamJur03,FC:Gro04,HRZ10,EV:KSRH12,GiuIovRon16}.

Our privacy definition is inspired by the one of Benaloh~\cite{thesis:Benaloh}, also called ``PRIV'' in~\cite{BCGPW15}, which we reformulate by using modern terminology and we modify conveniently to withstand the attacks shown in~\cite{BCGPW15}. We believe that our (fully) verifiable eVote can be proven secure according to other definitions of security, like for instance the one of Chase \etal\ \cite{PKC:CKLM13}, but we did not investigate the details because it is out of the scope of this initial work.



\ifnum\fullversion=1
\subsection{Organization}
\else
\paragraph{\bf Organization.}
\fi

\ifnum\fullversion=1
We describe the concept of eVote and its verifiability and privacy properties in Section~\ref{sec:overviewdefs}. In Section~\ref{sec:def}, we present detailed definitions of an eVote and of its verifiability and privacy properties. In Section \ref{sec:buildingblocks} we present the building blocks we will use in our constructions.

In Section~\ref{sec:sketchweak} (resp. Section~\ref{sec:sketchfully}) we include all major details needed to understand our construction for a weakly verifiable eVote (resp. (fully) verifiable eVote) and its security properties. In Section \ref{sec:scheme} (resp. Section \ref{sec:fullyscheme}) we present the full details of our construction for a weakly verifiable eVote (resp. fully verifiable eVote) and of its security properties.

In Section~\ref{sec:threshold}, we outline how to adapt our (fully) verifiable eVote to a model with multiple authorities and threshold privacy. In this model, the tally evaluation algorithm is run by a set of authorities and privacy must hold if at least one of the authorities is honest.

In Section \ref{sec:reusability} we make some additional remarks about our definitions and in particular about the possibility of re-using the parameters through different elections. In Section \ref{sec:related} we discuss relevant related works. Finally, in Section \ref{sec:future} we discuss some future directions in cryptography and e-voting that our work opens up.

%
%
%
%
\else
We describe an eVote and its verifiability and privacy properties in Section~\ref{sec:overviewdefs}. In Appendix~\ref{sec:def}, we present detailed formal definitions of an eVote and its properties. In Appendix~\ref{sec:buildingblocks}, we present the building blocks we use in our constructions.

In Section~\ref{sec:sketchweak} (resp. Section~\ref{sec:sketchfully}) we include all major details needed to understand our construction for a weakly verifiable eVote (resp. (fully) verifiable eVote) and its security properties. For the interested reader, in Appendix~\ref{sec:scheme} (resp. Appendix~\ref{sec:fullyscheme}) we present the full details of our construction for a weakly verifiable eVote (resp. (fully) verifiable eVote) and of its security properties.

In Section~\ref{sec:threshold}, we outline how to adapt our (fully) verifiable eVote to a model with multiple authorities and threshold privacy. In this model, the tally evaluation algorithm is run by a set of authorities and privacy must hold if at least one of the authorities is honest.

In Section~\ref{sec:reusability} we make some additional remarks about our definitions and in particular about the possibility of reusing the public parameters through different elections.
In Section~\ref{sec:related}, we discuss relevant related works.

Finally, in Section \ref{sec:future} we discuss some future directions in cryptography and e-voting that our work opens up.
\fi



\ifnum\fullversion=1
\subsection{Our Model and Definitions}\label{sec:overviewdefs}
\else
\section{Our Model and Definitions}\label{sec:overviewdefs}
\fi

In this section, we introduce our definitions of privacy and verifiability. We use a simple e-voting model with a single authority.  We remark that, even for this model, it was not known how to avoid the use of CRSs or ROs. In Section~\ref{sec:threshold}, we outline how to adapt our constructions to a model with multiple authorities. Formal definitions of an eVote and of its privacy and verifiability are given in Section \ref{sec:defevote}.


We use a general tally function $F:(\M\cup \{\bot\})^N\rightarrow \zu^\star\cup\{\bot\}$, where $\M$ is the message space. The special symbol $\bot$ denotes either an invalid vote or a blank ballot, when it is input to the function, or an error, when it is output by the function. A voter casts $\bot$ to denote a blank ballot, i.e., a valid ballot where no candidate is chosen. A voter abstains from voting by not casting any ballot or by casting an invalid ballot, which is replaced by $\bot$ in the evaluation phase.

Our general tally function $F$ must satisfy a very natural property given in Def.~\ref{def:tally}. The messages belong to a message space $\M$ that is not specified. As byproduct, our constructions can be instantiated, for instance, to the case of a YES/NO election with the sum as tally function. As shown in \cite{BCGPW15}, care has to be taken when considering general tally functions.

\ifnum\fullversion=1
\subsubsection{Privacy}
\else
\subsection{Privacy}
\fi

Our {\em privacy} definition is indistinguishability-based and states that no PPT adversary can win the following game with non-negligible advantage. The adversary receives the public key generated by a challenger and chooses two tuples of strings that encode either valid votes in the message space $\M\cup\{\bot\}$ or arbitrary ballots, which are cast by possibly corrupt voters. We require that the tally function outputs the same result on input any of the tuples of strings.

The challenger chooses at random one of the two tuples. The challenger runs the ballot verification algorithm on input each of the arbitrary ballots and replaces the arbitrary ballot in the tuple by $\bot$ if verification is unsuccessful.  The challenger runs the cast algorithm on input each of the valid votes in the message space to compute a ballot and replaces the valid vote in the tuple by the ballot. Then the challenger computes the tally and a proof of correctness of the tally.

The new tuple, which replaces valid votes by ballots and invalid arbitrary votes by $\bot$, is given to the adversary along with a proof of the correctness of the tally. The adversary guesses which of the two tuples was chosen by the challenger.


More formally, the adversary sends two tuples $V_0 = (m_{0,1},\ldots,m_{0,N})$ and $V_1=(m_{1,1},\ldots,m_{1,N})$ and a set $S\subset[N]$. The set $S$ contains the indices of the strings of arbitrary ballots. For each $j\in S$, $m_{0,j}=m_{1,j}$ must hold. For each $j\notin S$, $m_{0,j},m_{1,j}\in\M\cup \{\bot\}$ must hold. Moreover, we require that for all $d_1,\ldots,d_N\in\M\cup\{\bot\}$,    $F(m'_{0,1},\ldots, m'_{0,N})=F(m'_{1,1}, \ldots, m'_{1,N})$ must hold, where, for each $j\in S, m'_{0,i}= m'_{1,i}= d_i$ must hold, and for each $j\notin S$, $b\in\zu, m'_{b,j}= m_{b,j}$ must hold.


Our definition can be viewed as a variant of Benaloh's ballot privacy definition~\cite{thesis:Benaloh} (also called ``PRIV'' in~\cite{BCGPW15}) reformulated by using modern terminology and corrected to rule out some known attacks~\cite{BCGPW15}.


We also define {\em weak privacy}. The difference between the definitions of {\em weak privacy} and privacy is that, in {\em weak privacy}, the set $S$ must be empty, i.e., the adversary cannot submit arbitrary ballots.



The privacy definitions that we use here are simple and do not capture vote replay attacks, see e.g. \cite{EPRINT:CorSmy10}. Such attacks are easily prevented by enforcing ballot independence. This can e.g. be done by appending a proof of knowledge of the plaintext in the ballots of the voters. Presently, this has not been done in the NIWI setting, so we will disregard this point for clarity. However, we stress that it is easy to change the schemes to satisfy full privacy definitions within the framework of having trust for privacy, but not for verifiability.

\ifnum\fullversion=1
\subsubsection{Verifiability}
\else
\subsection{Verifiability}
\fi

We define a ballot verification and a tally verification algorithm. In our definition of verifiability,  we require two conditions to hold. The first condition states that, if each ballot and the proof of correctness of the tally issued by the authority are verified successfully by the respective algorithms, then each ballot $C_i$ (possibly computed on input a maliciously generated public key) must be associated with a unique message $m_i \in \M\cup\{\bot\}$, and the result $y$ claimed by the authority equals $F(m_1,\ldots,m_n)$.

The second one requires that, even when the adversary generates the public key, if honest voters cast a ballot that is accepted by the ballot verification algorithm, then the ballot has to be ``counted''.
More concretely, consider that some ballots are computed by honest voters and are accepted by the ballot verification ballot algorithm. (These ballots could be ill-formed if they are computed on input a public key generated by the adversary.) Consider also that the remaining ballots are computed by corrupt voters. In this situation, the tally evaluation algorithm outputs a tally $y$ and a proof of correctness that, along with the public key and the ballots, is accepted by the tally verification algorithm. Then, it must be the case that the ballots sent by honest voters were counted to obtain $y$. For example, if the tally function is a sum function that sums binary votes and three honest voters cast three $1$'s, then the authority should not be able to claim that $y < 3$.


Remarkably, our construction provides {\em perfect} verifiability. {\em Perfect} verifiability means that the probability that a malicious authority computes an incorrect tally and a proof that are accepted by the tally verification algorithm is {\em null}.

We also define {\em weak verifiability}. In {\em weak verifiability}, the authority can incorrectly claim that $y = \bot$. The second condition described above is still guaranteed for all the tallies $y \neq \bot$. For a weakly verifiable eVote, we only require weak privacy.


Although our weakly verifiable eVote  satisfies weaker properties, it represents a worthwhile warm-up.  Our (fully) verifiable eVote is based on it, though with some relevant modifications.
Our weakly verifiable construction does not need a ballot verification algorithm, but for simplicity we use the same syntax for both the weakly verifiable and (fully) verifiable schemes.
In the next subsection, we describe a definition of correctness that is stronger than previous ones. This definition is needed to exclude the case that an eVote that is intuitively not verifiable fulfills formally the definition of verifiability.



In Section \ref{sec:threshold}, we outline how to adapt our (fully) verifiable construction to a model with multiple authorities. In this model, the tally is computed by a set of authorities. Privacy must hold if at least one authority is honest. Because this model is not the main focus of our work, we do not present formal definitions or a detailed description of its construction. We note that such a construction would satisfy a different, but still without trust assumptions, definition of verifiability that essentially states that if there is at least one honest voter, the verifiability holds with overwhelming probability over the random coins of such voter, a very minimal assumption.


In this paper, for simplicity, we do not directly address issues of {\em eligibility}. We assume that a ballot is associated with a voter uniquely and that the adversary cannot submit a ballot on behalf of some voters. Unconditional eligibility verifiability seems hard or impossible to achieve, since we normally use some commitment, e.g. a PKI, and digital signatures on the ballots to prove eligibility, but such an approach is not secure against a computationally unbounded adversary.

Our construction can however easily be extended to take into account such attacks by using digital signatures in a standard, but non-perfect, way (see e.g. \cite{ESORICS:CGGI14}). The resulting construction would nonetheless satisfy a meaningful notion of verifiability secure against computationally bounded adversaries not based on any trust assumption, which advances the state of the art. In fact, to our knowledge, it is not even known how to construct an eVote protocol with computational verifiability without trusted parties.

\ifnum\fullversion=1
\subsubsection{On the Need of a Stronger Correctness Property}\label{sec:strongcorrectness}
\else
\subsection{On the Need of a Stronger Correctness Property}\label{sec:strongcorrectness}
\fi

We justify here why a stronger correctness property is needed. Traditionally, the correctness property guarantees both (1) that the ballot verification algorithm accepts the ballots computed by the cast algorithm, and (2) that the tally verification algorithm accepts the tally and the proof computed by the tally evaluation algorithm. In the latter, the ballots taken as input by the tally evaluation algorithm are computed by the cast algorithm. Therefore, it is not guaranteed that the tally verification algorithm accepts the output of the tally evaluation algorithm when the ballots are not computed by the cast algorithm.

We explain now that this is an issue. In our weakly verifiable scheme, the ballot verification algorithm accepts any ballot. Therefore, it would be possible to say that such a scheme is (fully) verifiable by just changing the tally verification algorithm so that it accept $y\defeq\bot$ only when all the ballots equal $\bot$. As can be seen, condition (1) in the definition of (full) verifiability (cf. Def. \ref{def:evote}) is fulfilled because the ``if part'' of the condition never holds.

However, intuitively, such a scheme is incorrect. Namely, if an honest authority that runs tally evaluation algorithm and gets $y=\bot$ (because some ballots where ill-formed), the tally verification algorithm should accept that result.

To address this issue, we add condition (2) to the definition of correctness (cf. Def. \ref{def:evote}). This condition states that the tally verification algorithm must accept the output of the tally evaluation algorithm when run on input ballots that are accepted by the ballot verification algorithm (as opposed to ballots computed by the cast algorithm). We point out that in some works on definitional foundations (e.g., Bernhard \etal\ \cite{BCGPW15}) this issue has been overlooked.

\ifnum\fullversion=1
\subsection{Warm-up: Our Weakly Verifiable eVote (Sketch)}\label{sec:sketchweak}
\else
\section{Warm-up: Our Weakly Verifiable eVote (Sketch)}\label{sec:sketchweak}
\fi

In this section, we sketch our construction for a weakly verifiable eVote, i.e.\ an eVote that fulfills the weak verifiability and weak privacy properties. A (fully) verifiable eVote, which satisfies (full) verifiability and (full) privacy, is presented in Section~\ref{sec:sketchfully}. We stress that in practice such weakly verifiable eVote lacks fundamental security guarantees, but nonetheless it serves as a worthwhile warm-up to our (fully) verifiable eVote.

\ifnum\fullversion=1
\subsubsection{Intuition}
\else
\subsection{Intuition}
\fi

Our weakly verifiable eVote uses $3$ instances of a public key encryption (PKE) scheme in parallel. We require that the PKE scheme fulfills two properties: perfect correctness and unique secret key (see Def. \ref{def:pke}). PKE schemes with those properties are known in the literature~\cite{DifHel76,C:BonBoySha04} and can be constructed, e.g., from the Decision Linear assumption~\cite{C:BonBoySha04}.
The voter encrypts her vote $3$ times using the PKE scheme {\em without} adding any proof of ciphertext well-formedness. Therefore, a ballot consists of three ciphertexts.


To compute the tally, the authority proceeds as follows. The authority decrypts the first ciphertext and the second ciphertext in a ballot. The authority replaces decrypted messages that do not belong to the message space by $\bot$.


The authority evaluates the tally function twice. First, the authority uses as input the messages encrypted in the first ciphertext of each ballot. Second, it uses the messages encrypted in the second ciphertext of each ballot. If both tallies are equal, the authority outputs the tally along with a proof of correctness, else the authority returns $\bot$ to indicate an error.

The property of unique secret key guarantees that the decrypted message will be unique for each ciphertext. Without this property, it could be possible that a voter cast an invalid ciphertext $\Ct$ not belonging to the ciphertext space such that the decrypted message is different when using two well-formed secret keys $\SK_1$ and $\SK_2$ for the same public key. Note that this is not prevented by the correctness property, which only holds when the ciphertext is an output of the encryption algorithm.

\ifnum\fullversion=1
\subsubsection{Sketch of the construction}
\else
\subsection{Sketch of the construction}
\fi

Let $N$ be the number of voters and let $F$ be a tally function with message space $\M$.
The public key $\PK$ of our eVote consists of the $3$ PKs $(\PK_1,\ldots,\PK_3)$ of the underlying PKE. The secret key consists of the $3$ corresponding SKs $(\SK_1,\ldots,\SK_3)$ of the PKE.

Our cast algorithm takes as input the public key $(\PK_1,\ldots,\PK_3)$, the index $j$ of the voter\footnote{The index is needed to associate a ballot with a unique voter. For instance, an eVote could require that each voter encrypts her ballot with a different PKE public key, adding a proof of well-formedness. The public key of the eVote would contain $N$ PKE's public keys, one for each voter, and so the statement of the proof would have to contain the index of the voter in the set $N$.} (for $j\in[N]$), and a vote $v$. The cast algorithm outputs a ballot for the $j$-th voter.
Our cast algorithm just encrypts the vote $v$ with the $3$ instances of the PKE to produce the ciphertexts $\Ct_1,\ldots,\Ct_3$.
The ballot given as output is  $\BT\defeq(\Ct_1,\ldots,\Ct_3)$.

The tally evaluation algorithm works as follows. For all $j\in[N]$,  if the corresponding voter cast her vote, for all $l\in[2]$, decrypt $\CT_{j,l}$ with $\SK_l$ to get $m_{j,l}$.
Then, for all $l\in[2]$ compute $y_l=F(m_{1,l},\ldots,m_{N,l}),$ where for indices $j$ such that either $m_{j,l}\notin\M$ or the $j$-th voter did not cast her vote, we set $m_{j,l}=\bot$.
If the two $y_l$'s are equal to the {\em same} string $y$ then return this as the tally, otherwise return an error $y=\bot$.
Finally, compute a NIWI proof $\gamma$ of the fact that $x= (\BT_1,\ldots,\BT_N,\PK_1,\ldots,\PK_3)$ satisfies the relation $\Rdec$ in Fig.~\ref{relationdec} using as witness $(\SK_1,\SK_2,s_1,s_2)$. Another part of the witness is the two indices $i_1,i_2\in[3]$, $i_1< i_2$, which determine the two columns of ciphertexts that are used to compute the tally. In the real mode described above, we have $i_1=1, i_2=2$, but we can also have trapdoor modes with other index choices which will be essential for privacy.

\begin{figure}
\begin{framed}

\noindent Relation $\Rdec(x,w)$: \\

\noindent Instance: $x=(\BT_1,\ldots,\BT_N,\PK_1,\ldots,\PK_3, y).$ (Recall that a ballot is set to $\bot$ if the corresponding voter did not cast her vote.) \\

\noindent Witness: $w = (\SK_1',\SK_2',s_1,s_2, i_1, i_2)$, where the $s_i$'s are the randomness used to generate the secret key/public key pairs, which is known by the authority that set up the system. \\

\noindent $\Rdec(x, w) = 1$ if and only if the following condition holds: $2$ of the secret keys corresponding to indices $\PK_{i_1}, \PK_{i_2}$ are constructed using honestly generated public and secret key
		pairs and are equal to $\SK_1',\SK_2'$; and either $y=\bot$ or for all $l\in[2]$, $y=F(m_1^l,\ldots,m_N^l)$ and for all $j\in[N]$, if $\BT_j\neq\bot$ then for $l\in[2]$, $\SK_{l}'$ decrypts ciphertext $\Ct_{j,i_l}$ in $\BT_j$ to $m_j^{i_l}\in\M$; and for all $l\in[2]$, $m_j^l=\bot$ if either $\BT_j=\bot$ or $\SK_{l}'$ decrypts $\Ct_{j,i_l}$ to a string $\notin \M$.

\end{framed}

\caption{Relation $\Rdec$.}
\label{relationdec}

\end{figure}

Note that the proof $\gamma$ can be computed using as witness the randomness used to compute the public and secret key pairs.
Finally, the algorithm outputs the pair $(y, \allowbreak \gamma)$. The tally verification algorithm verifies $(y, \allowbreak \gamma)$ by using the verification algorithm of the NIWI system.


\ifnum\fullversion=1
\subsubsection{Weak Verifiability of the Construction}\label{sec:sketchproofsweakscheme}
\else
\subsection{Weak Verifiability of the Construction}\label{sec:sketchproofsweakscheme}
\fi

Weak verifiability (cf. Def. \ref{def:evote}) requires that, given a public key and a set of messages decrypted from the ballots, the authority cannot output a pair $(y,\gamma)$, $y\neq\bot$, such that $y$ is an incorrect tally, but $\gamma$ is accepted by the tally verification algorithm. However, the authority is able to claim that $y = \bot$ even if that is not the correct tally. The construction described above suffers from this problem.


We give a detailed proof that our construction fulfills the weak verifiability property in Theorem~\ref{thm:weakver}. In the following, we explain why our construction above fulfills the two conditions required by the weak verifiability property. First, we show that it fulfills the first condition. The first condition states that, if each ballot and the proof of correctness of the tally issued by the authority are verified successfully by the respective algorithms, then each ballot $C_i$ (possibly computed on input a maliciously generated public key) must be associated with a unique message $m_i \in \M\cup\{\bot\}$, and the result $y$ claimed by the authority equals $F(m_1,\ldots,m_n)$.


We use a contradiction to show that our construction fulfills the first condition. Let us assume that there exist two results $y_0,y_1\neq \bot$  such that $y_0\neq y_1$, and two proofs $\gamma_0,\gamma_1$ that are accepted by the tally verification algorithm. By the unique secret key property, the decryption of the ciphertexts in the ballots produces a unique result. By the pigeon principle,  there exists one index $i^\star\in[3]$ used by both proofs. Therefore, it must be the case that either $y_0 = y_1 = \bot$ or $y_0$ and $y_1$ are equal to the evaluation of the tally function $F$ on input the messages obtained by decrypting the ciphertexts. Consequently, $y_0,y_1\neq \bot$ such that $y_0\neq y_1$ is a contradiction.


The second condition requires that, even when the adversary generates the public key, if honest voters cast a ballot that is accepted by the ballot verification algorithm, then the ballot has to be ``counted''. We recall that an honest ballot for the $j$-th voter consists of three ciphertexts that encrypt the same message $m$. The perfect soundness of the NIWI ensures that the public key for the PKE scheme is honestly generated. The perfect correctness of the PKE scheme ensures that a ballot that encrypts $m$ will be decrypted to $m$. Therefore, if the claimed tally $y$ does not equal $\bot$, $y$ has to be in the range of the function $F$ restricted to $m$ at index $j$.


\ifnum\fullversion=1
\subsubsection{Weak Privacy of the Construction}\label{sec:sketchproofsweakschemeprivacy}		
\else
\subsection{Weak Privacy of the Construction}\label{sec:sketchproofsweakschemeprivacy}		
\fi

	We explain how we prove that our construction fulfills the weak privacy property. The proof consists of a sequence of hybrid experiments~\cite{GolMic84}, which are summarized in Table~\ref{table}.


	\setlength{\fboxsep}{1.5pt}
{

	\begin{table}
\begin{center}
	\begin{small}
\caption{Sequence of hybrid games to prove fulfillment of the weak privacy property.}
	  \begin{tabular}{ | c|c|c|c|c| }
		      \hline
		      Exp & $(\Ct_{j,1},\Ct_{j,2},\Ct_{j,3})$ & $\SK$ index  & $\gamma$ & Security \\ \hline
		      $H_1$ & ($m_{0,j},m_{0,j},m_{0,j}$) & (\textcolor{blue}{1,2},3)  & R & - \\ \hline
	      $H_2$ & $(m_{0,j},m_{0,j},\textcolor{red}{m_{1,j}})$   & (\textcolor{blue}{1,2},3)  & R & IND-CPA \\ \hline
	      $H_3$ & $(m_{0,j},m_{0,j},m_{1,j})$  & (\textcolor{blue}{1},\fcolorbox{red}{white}{2,\textcolor{blue}{3}})  & \textcolor{red}{T}  & WI \\ \hline
		      $H_4$ & $(m_{0,j},\textcolor{red}{m_{1,j}},m_{1,j})$ & (\textcolor{blue}{1},2,\textcolor{blue}{3})  & T & IND-CPA \\ \hline
	      $H_5$ & $(m_{0,j},m_{1,j},m_{1,j})$   & (\fcolorbox{red}{white}{1,\textcolor{blue}{2}},\textcolor{blue}{3})  & T & WI \\ \hline
		      $H_{6}$ & $(\textcolor{red}{m_{1,j}},m_{1,j},m_{1,j})$   & (1,\textcolor{blue}{2,3})  & T & IND-CPA \\ \hline
		      $H_7$ & $(m_{1,j},m_{1,j},m_{1,j})$   & (\fcolorbox{red}{white}{\textcolor{blue}{1,2}},3)  & \textcolor{red}{R} & WI \\ \hline
				        \end{tabular}
				\end{small}
\end{center}
		\label{table}
			\end{table}
		
		}

For simplicity, in this sketch we assume that the adversary submits a challenge that consists of two tuples $(m_{0,1},\ldots,m_{0,N})$ and $(m_{1,1},\ldots,m_{1,N})$, where each of the messages belongs to the message space $\M$. In the table, the first column shows the name of the hybrid experiment. The second column shows the three messages that are encrypted in the $3$ ciphertexts $\Ct_{j,1},\ldots,\Ct_{j,3}$ contained in the challenge ballot $\BT_j=(\CT_{j,1},\ldots,\Ct_{j,3})$ associated with voter $j$. The text in blue in the ``$\SK$ index''  column denotes the indices used as witness in the proof $\gamma$. As mentioned above, if such blue indices correspond to the set $\{1,2\}$ (resp. to a set different from \{$1,2\}$)  we say that the statement or proof is in real mode (resp. trapdoor mode), which we denote by $R$ (resp. $T$) in the column $\gamma$.
The text in red indicates the difference from the previous hybrid experiment.


The proofs of indistinguishability between the hybrid experiments $H_1$ and $H_2$, $H_2$ and $H_3$, and $H_3$ and $H_4$ are symmetrical to the proofs of indistinguishability between   $H_4$ and $H_5$, $H_5$ and $H_6$, and $H_6$ and $H_7$. Therefore, it suffices to explain how we prove indistinguishability between the first four hybrid experiments.




Hybrid $H_1$ corresponds to the real experiment, except that the challenger sets the bit $b=0$.

In hybrid $H_2$, we switch the third message (in red) in any ballot to encrypt $m_{1,j}$. This is possible because the witness used to compute the proof $\gamma$ does not contain the randomness used to compute the third secret key. Thanks to that, we can show indistinguishability between  $H_1$ and $H_2$ by using the IND-CPA property of the PKE scheme.

In hybrid $H_3$, the witness used to compute the proof $\gamma$ contains the indices $\{1,3\}$ instead of $\{1,2\}$. Therefore, $\gamma$ is in trapdoor mode. The witness-indistinguishability property of the NIWI allows as to show that $H_3$ cannot be distinguished from $H_2$. Note that the result of the decryption does not change thanks to the constraint in the weak privacy definition that $F(m_{0,1},\ldots,m_{0,N})=F(m_{1,1},\ldots,m_{1,N})$ must hold.

In hybrid $H_4$, we switch the second message (in red) in any ballot to encrypt $m_{1,j}$. This is possible because the witness used to compute the proof $\gamma$ does not contain the randomness used to compute the second secret key.  Thanks to that, we can show indistinguishability between  $H_4$ and $H_3$ by using the IND-CPA property of the PKE scheme.

We remark that, in order to switch the encrypted messages to $m_{1,j}$'s in every ballot, we use a simple property: the witness of the proof $\gamma$ contains the randomness used to compute {\em two} of the secret keys. Thanks to that, we can show indistinguishability between  $H_1$ and $H_2$ and between $H_3$ and $H_4$ by using the IND-CPA property of the PKE scheme whose randomness is not needed to compute $\gamma$. We point out that, to prove indistinguishability between those hybrid experiments, we need to use $N$ ``sub-hybrids''. In each ``sub-hybrid'', we switch the message encrypted in just one ballot.

\ifnum\fullversion=1
In Section \ref{sec:scheme}, we present our weakly verifiable eVote in a more detailed manner.
\else
In Appendix \ref{sec:scheme}, we present our weakly verifiable eVote in a more detailed manner.
\fi



\ifnum\fullversion=1
\subsection{Our Fully Verifiable eVote (Sketch)}\label{sec:sketchfully}
\else
\section{Our Fully Verifiable eVote (Sketch)}\label{sec:sketchfully}
\fi

The scheme sketched in Section~\ref{sec:sketchweak} suffers from a severe problem: the authority can claim that the tally is $\bot$ when it is not.  That is, there can be two tallies $y_0 \neq \bot$ and  $y_1 = \bot$ and  two proofs $\gamma_0$ and $\gamma_1$ such that both proofs are accepted by the tally verification algorithm.


For instance, consider the following case. The ballots submitted by the voters are such that the tally $y_1$ obtained by evaluating the tally function on input the messages decrypted from the first ciphertext of each ballot equals the tally $y_2$ obtained when using the second ciphertext of each ballot, but differs from the tally $y_3$ obtained when using the third ciphertext. Then, by using the indices $(1,2)$, the authority can prove successfully that the result of the election is $y_1 = y_2$, and  by using indices $(1,3)$, the authority can claim that the result of the election was $\bot$. The voters do not learn the indices that the authority used in the NIWI proof.

This also allows severe DoS attacks. For example, if just one voter submits a wrong ballot that makes the two tallies $y_1$ and $y_2$ be different from each other, then an honest authority has to output $\bot$.   Furthermore, this scheme only fulfills the weak privacy property, which does not take into account corrupt voters.

Therefore, we propose a scheme that fulfills the (full) verifiability and the (full) privacy properties. This scheme solves the above-mentioned problems in an elegant way. Here we show a sketch of the scheme.
\ifnum\fullversion=1
In Section \ref{sec:fullyscheme}, we present our (fully) verifiable eVote in a more detailed manner.
\else
In Appendix \ref{sec:fullyscheme}, we present our (fully) verifiable eVote in a more detailed manner.
\fi


\ifnum\fullversion=1
\subsubsection{Sketch of the Construction}
\else
\subsection{Sketch of the Construction}
\fi 

In addition to the three public keys of the PKE scheme, the public key of the authority contains a perfectly binding commitment $Z$ to the bit $1$, i.e., the public key is $\PK=(\PK_1,\PK_2,\PK_3,Z)$, where  $Z=\Com(1)$.

A ballot consists of three ciphertexts, which are computed as in the weakly verifiable scheme, and of a proof that either the three ciphertexts encrypt the same message in the message space $\M\cup\{\bot\}$ or $Z$ is a commitment to $0$. Formally, the ballot contains a NIWI proof for the relation in Fig.~\ref{relationenc}.


\medskip
\begin{figure}

\begin{framed}
\noindent Relation $\Rencf(x,w)$: \\

\noindent Instance: $x\defeq(j,\Ct_1,\ldots,\Ct_3,\PKE.\PK_1,\ldots,\PKE.\PK_3, Z)$. \\

\noindent Witness : $w \defeq (m, r_1,\ldots,r_3, u)$, where the values $r_l$'s are the random values used to encrypt the ciphertexts $\CT_l$'s and $u$ is the random value used to compute the commitment $Z$.\\

\noindent $\Rencf(x, w) = 1$ if and only if either of the following two conditions hold:

\begin{enumerate}
	\item{\bf Real mode.} All $3$ ciphertexts $(\Ct_1,\ldots,\Ct_3)$ encrypt the same message in $\M\cup\{\bot\}$.
		
		$$\text{OR}$$
	\item{\bf Trapdoor mode.} $Z$ is a commitment to $0$.

\end{enumerate}
\end{framed}
\caption{Relation $\Rencf$.}
\label{relationenc}

\end{figure}

\medskip

The ballot verification algorithm runs the verification algorithm for the NIWI proof system for the relation $\Rencf$. (We recall that the ballot verification algorithm of our weakly verifiable eVote accepts any ballot.) The tally evaluation algorithm is the same as in our weakly verifiable eVote.

The tally verification algorithm also follows the one of the weakly verifiable eVote with the following modification. If either (1) not all inputs are $\bot$ and $y=\bot$, or (2) all inputs are $\bot$ and $y\neq\bot$, the tally verification algorithm outputs $\bot$.

We explain the reason for this modification. First, note that, in the (fully) verifiable scheme, the ballots that are rejected by the ballot verification algorithm are replaced by $\bot$ as input to the tally evaluation algorithm. We recall that, in the weakly verifiable scheme, the tally evaluation algorithm is run on input $\bot$ only when voters do not send any ballot.

Our tally functions must fulfill a very natural property: $F(m_1,\ldots,m_N)=\bot$ iff $m_1=\bot,\ldots,m_N=\bot$  (cf. Def. \ref{def:tally}). That is, if at least one message is valid, then it has to be ``counted''.

As we show below, if the public key is honestly generated, the tally evaluation algorithm never returns $\bot$ on input a tuple of possibly dishonest ballots. Therefore, except for the case that all the ballots are invalid, a tally $y = \bot$ may only occur if the authority acted dishonestly and, consequently, the tally verification algorithm should not accept $y = \bot$.

\ifnum\fullversion=1
\subsubsection{(Full) Verifiability of the Construction}\label{sec:sketchproofsfullyscheme}
\else
\subsection{(Full) Verifiability of the Construction}\label{sec:sketchproofsfullyscheme}
\fi

We show that our scheme fulfills the (full) verifiability property. This property consists of two conditions described in Section~\ref{sec:overviewdefs} and defined in Def.~\ref{def:evote}.

First, we show that our scheme fulfills the first condition. (A detailed proof is given in Theorem \ref{thm:fullver}.) This condition requires that the authority cannot output two tallies $y_1,y_2$ such that $y_1\neq y_2$ and two proofs $\gamma_1$ and $\gamma_2$ that are accepted by the tally verification algorithm.

	\ignore{
For ease of exposition, we divide the analysis in two cases (though it could be unified in just one case).
\begin{itemize}
\item{Authority sets the commitment honestly.} In this case, the tally verification algorithm is ever run on inputs that are either $\bot$ (in the case that the corresponding ballot is ill-formed) or a well-formed ballot. So, it never happens the case that the authority needs to claim a tally $y\neq\bot$ and the properties of our previous weakly verifiable eVote still guarantee that the authority cannot claim two values $y_1,y_2$ with $y_1\neq y_2, y_1,y_2\neq\bot$.
\item{Authority sets the commitment dishonestly.}
	By construction, the value $\bot$ is not accepted by our tally verification algorithm except when the inputs are all equal to $\bot$, so the authority cannot claim a tally $\bot$.
	However, the guarantees of our previous weakly verifiable eVote still hold and thus the authority cannot claim two values $y_1,y_2$ with $y_1\neq y_2, y_1,y_2\neq\bot$.
\end{itemize}
}

Our tally verification algorithm only accepts a tally $y = \bot$ when all the ballots are invalid. Therefore, (1) the authority is not able to wrongly claim that a tally is $\bot$. Furthermore, as in our weakly verifiable eVote, (2) the authority cannot output two tallies $y_1,y_2$ such that $y_1\neq y_2, y_1,y_2\neq\bot$ along with proofs $\gamma_1$ and $\gamma_2$ that are accepted by the ballot verification algorithm. Therefore, (1) and (2) imply that the authority cannot output two tallies $y_1,y_2$ such that $y_1\neq y_2$.

We show now that the second condition also holds. First, we note that the authority can only create a dishonest public key by setting the commitment dishonestly. The reason is that the authority has to prove that the public key of the PKE scheme is honestly generated. Therefore, the perfect correctness of the NIWI and of the PKE scheme guarantee that an honestly computed ballot\footnote{Here, ``honestly computed ballot'' just means that it is computed by the voter using the Cast algorithm on input the public key of the authority, which could be honestly or dishonestly created. By design of our construction, an honestly generated ballot computed on input an honestly created public key has the same distribution of an honestly created ballot computed on input any possibly dishonest public key whenever the authority is able to compute proofs of tally correctness that are accepted by the tally verification algorithm.}  for the $j$-th voter that encrypts message $m$ will always be ``counted'', i.e., for any $(y,\gamma)$ pair that is accepted by the tally verification algorithm, $y$ will be compatible with $m$ at index $j$ according to Def. \ref{def:restriction}.

\ifnum\fullversion=1
\subsubsection{(Full) Privacy of the Construction}\label{sec:sketchproofsfullyschemeprivacy}
\else
\subsection{(Full) Privacy of the Construction}\label{sec:sketchproofsfullyschemeprivacy}
\fi

We show now that our scheme fulfills the (full) privacy property. Here we summarize the proof. In Section~\ref{sec:privacyfull}, we describe the proof in detail. We stress that, for privacy to hold, the authority must be honest and thus the public key is honestly generated.

In the security proof, we consider a sequence of hybrid experiments. First, we define an experiment $H^Z$ in which the commitment in the public key is a commitment to $0$. We show that $H^Z$ is indistinguishable  from the real experiment under the computationally hiding property of the commitment.

Second, we define an event $E^1$ in experiment $H^Z$. In $E^1$, the adversary submits a ballot that is accepted by the ballot verification algorithm but, when decrypting the three ciphertexts in the ballot, the three decrypted messages in $\M\cup\{\bot\}$ are not equal. We show that the probability of $E^1$ is negligible under the computationally hiding property of the commitment. More concretely, we show that, if $E^1$ occurs with non-negligible probability, then the adversary can be used to distinguish a commitment to $0$ from a commitment to $1$. We note that,  if $Z$ is a commitment to $1$, the perfect soundness of the NIWI guarantees that the adversary can never submit an ill-formed ballot that is accepted by the ballot verification algorithm. 

The next hybrid experiments are similar to the ones used in the security proof of the weakly verifiable scheme. Thanks to the hybrid experiment $H^Z$, we can still show indistinguishability between those hybrid experiments by using the IND-CPA property of the PKE scheme. The reason is that, thanks to $H^Z$, the NIWI proof in the ballots can be a proof that the commitment in the public key is a commitment to 0. Therefore, we avoid the computation of a proof that shows that the three ciphertexts encrypt the same message, which allows us to switch the message encrypted in one of the ciphertexts and prove indistinguishability by using the IND-CPA assumption.

There is one difference between the hybrid experiments in the weakly verifiable scheme and in the (fully) verifiable scheme. Namely, in the (fully) verifiable scheme, we have to handle possibly dishonest ballots. In particular, we have to guarantee that, when we switch the indices used as witness for the NIWI proof of tally correctness, the tally does not change. To illustrate this issue, suppose that, in an adversarial ballot, the first two ciphertexts encrypt the same message $x$ but the third one encrypts a different message $z$. Then the tally computed by the secret keys for indices $\{1,2\}$ could differ from the one computed with secret keys for indices $\{2,3\}$. In that case, we cannot prove indistinguishability between a hybrid experiment where the NIWI witness comprises $\SK_1,\SK_2$ and a hybrid experiment where the NIWI witness comprises $\SK_2,\SK_3$.


To solve this issue, we show that event $E^1$ occurs with negligible probability. Therefore, it is sufficient to analyze the advantage of the adversary in the hybrid experiments conditioned on the occurrence of $\bar E^1$ (i.e., the complement of $E^1$). 

More concretely, the sequence of hybrid experiments after $H^Z$ is as follows. We recall that the adversary sends two tuples $V_0 = (m_{0,1},\ldots,m_{0,N})$ and $V_1=(m_{1,1},\ldots,m_{1,N})$, and a set $S\subset[N]$ that contains the indices of the strings of arbitrary ballots.
\begin{itemize}

\item Hybrid experiment $H_1$ is equal to the experiment $H^Z$, except that the challenger sets the bit $b = 0$.

\item Hybrid experiment $H_2$ switches the message encrypted in the third ciphertext in any ballot to encrypt $m_{1,j}$ instead of $m_{0,j}$. More in detail, for $k=0$ to $N$, we define a sequence of hybrid experiments $H_2^k$. $H_2^k$ is identical to $H_1$, except that, for all $j=1,\ldots,k$ such that $j\notin S$, the challenger computes the third ciphertext of the ballot on input $m_{1,k}$. Therefore, $H_2^0$ is identical to $H_1$, while $H_2^N$ is identical to $H_2$. We show that $H_2^k$ and $H_2^{k+1}$ are indistinguishable thanks to the IND-CPA property of the PKE scheme.

\item Hybrid experiment $H_3$ is identical to experiment $H_2$, except that the challenger computes the NIWI proof $\gamma$ on input a witness that contains indices $(1, \allowbreak 3)$ and secret keys $\SK_1, \allowbreak \SK_3$, instead of indices $(1,\allowbreak 2)$ and secret keys $\SK_1, \allowbreak \SK_2$. We show that $H_3$ and $H_2$ are indistinguishable thanks to the witness-indistinguishability property of the NIWI proof. Because $\bar E^1$ occurs with overwhelming probability, any ballot in $S$ is either replaced by $\bot$, if the ballot verification algorithm does not accept it, or decrypted to the same value in $H_2$ and $H_3$. Therefore, the tally evaluation algorithm outputs the same tally in $H_2$ and $H_3$.

\item Hybrid experiment $H_4$ is identical to $H_3$, except that the second ciphertext in any ballot encrypts $m_{1,j}$ instead of $m_{0,j}$. More in detail, for $k=0$ to $N$, we define a sequence of hybrid experiments $H_4^k$. $H_4^k$ is identical to $H_3$, except that, for all $j=1,\ldots,k$ such that $j\notin S$, the challenger computes the second ciphertext of the ballot on input $m_{1,k}$. Therefore, $H_4^0$ is identical to $H_3$, while $H_4^N$ is identical to $H_4$. We show that $H_4^k$ and $H_4^{k+1}$ are indistinguishable thanks to the IND-CPA property of the PKE scheme.

\end{itemize}

The remaining hybrid experiments are symmetrical to the ones described above. In $H_5$, the witness used to compute the NIWI proof contains the indices $(2, \allowbreak 3)$ and secret keys $\SK_2, \allowbreak \SK_3$, and indistinguishability between $H_5$ and $H_4$ follows from the witness-indistinguishability property of the NIWI proof. In $H_6$, the first ciphertext of each ballot encrypts  $m_{1,j}$ instead of $m_{0,j}$ and indistinguishability between  $H_6$ and $H_5$ follows from the IND-CPA property of the PKE scheme. Finally, in $H_7$ the witness used to compute the NIWI proof contains the indices $(1, \allowbreak 2)$ and secret keys $\SK_1, \allowbreak \SK_2$, and indistinguishability between $H_7$ and $H_6$ follows from the witness-indistinguishability property of the NIWI proof.

We would like to remark the subtle difference between ill-formed and invalid ballots. An ill-formed ballot is a ballot that is not in the range of the cast algorithm.
However, an ill-formed ballot could be valid in the sense that, along with other (possibly ill- or well- formed) $N-1$ ballots, the authority obtains a tally, i.e., the tally obtained when decrypting the first and the second ciphertext in the ballots is the same. An ill-formed ballot can be computed when the commitment in the public is computed dishonestly.

The event $\bar E^1$  may occur even if the adversary submits an ill-formed ballot that is accepted by the ballot verification algorithm. In fact, if a (non-honestly computed) ballot is formed by strings that are not in the ciphertext space of the encryption algorithm of the PKE, but decryption of those strings outputs the same message, such a ballot is not considered invalid.


Note also that the proof of well-formedness of the ballots states that the encrypted messages may be equal to $\bot$. Ballots that encrypt $\bot$ are blank ballots. We consider tally functions in which the symbol $\bot$ indicates a blank vote.  For example, in case of an eVote for the sum function in which $\bot$ is counted as $0$, an adversary should not be able to distinguish three ballots that encrypt $(1,1,\bot)$ from three ballots that encrypt $(1,\bot,1)$.



\ifnum\fullversion=1
\subsection{eVote with Multiple Authorities and Threshold Privacy}\label{sec:threshold}
\else
\section{eVote with Multiple Authorities and Threshold Privacy}\label{sec:threshold}
\fi

In this section, we sketch how to generalize our (fully) verifiable construction to fit a model with multiple authorities. In this model, the tally evaluation algorithm is run by a set of authorities. The privacy property must hold if not all the authorities are corrupt. Our generalized scheme guarantees a statistical verifiability property (see below), which assumes that there is at least one honest voter.

First, we note that the multi-string model of Groth and Ostrovsky~\cite{JC:GroOst14} does not provide a solution to this problem. The multi-string model assumes that the majority of the parties that set up the CRSs are honest. It does not guarantee soundness, which would provide verifiability in our application, when {\em all} those parties, which would be the authorities in our application, are corrupt.
In the multi-string model, there is a trade-off between soundness and zero-knowledge. Namely, soundness could hold when all the authorities are corrupt, but then zero-knowledge does not hold.
Zero-knowledge is guaranteed only when there is a majority of honest authorities. In contrast, our generalized scheme fulfills the privacy property when at least one authority is honest.

\ifnum\fullversion=1
\subsubsection{Sketch of the Construction}
\else
\subsection{Sketch of the Construction}
\fi

Our generalized construction works for tally functions that can be represented as polynomials. Such tally functions comprise many functions of interest for e-voting. For simplicity, henceforth we only consider the case of the sum function with a binary message space.  The general case follows by using Lagrange's polynomial interpolation.

Consider the sum function over a set of integers $S^k$, which we specify later. Consider $m$ authorities. Each authority $k\in[m]$ publishes a public key that consists of the public key of our (fully) verifiable eVote and, in addition, a commitment $\com_k$ to a tuple of $N$ $0$'s. 

A ballot $\BT_j$ for the $j$-th voter consists of the ballots $(\BT_{j,1},\ldots,\BT_{j,m})$. For a vote $v_j$, each voter computes $m$ shares $v_{j,1},\ldots,v_{j,m}$ whose sum is $v_j$. (Later we describe how the shares are computed in order to preserve privacy.) The ballot $\BT_{j,k}$ for the $k$-th authority is computed following the cast algorithm of the (fully) verifiable scheme on input the share $v_{j,k}$. In addition, the voter adds a NIWI proof that either (the real statement) for all $k\in[m]$, $\BT_{j,k}$ encrypts a number in $S^k$ such that the sum of the encrypted numbers is in $\zu$ (for simplicity, here we do not consider messages equal to $\bot$) OR (the trapdoor statement) for all $k\in[m]$, $\com_k$ is a commitment to a tuple $(z_1,\ldots,z_N)$ such that $z_j$ is equal to the tuple $(\BT_{j,1},\ldots,\BT_{j,m})$.

For each $k\in[m]$, the $k$-th authority computes the tally $y_k$ as in the (fully) verifiable scheme. The proof of correctness of the tally is a proof for the following modified relation: either (the real statement) the witness satisfies the relation $\Rdecf$ of the (fully) verifiable scheme and $\com_k$ is a commitment to $0$ OR (the trapdoor statement) $\com_k$ is a commitment to a tuple $(z_1,\ldots,z_N)$ such that $z_j=\BT_j$ for all $j\in[N]$, where $\BT_1, \allowbreak \ldots, \allowbreak \BT_N$ are the $N$ ballots published on the public bulletin board.

Finally, the tally is computed by summing the tallies $y_k$'s output by each of the authorities to obtain $y$.
We give more details below.

To support functions represented as polynomials, the following modifications should be applied. To compute the shares $v_{j,1},\ldots,v_{j,m}$, the voter $j$ chooses a polynomial $p_j$ of degree $m-1$ such that $p_j(0)$ equals her vote $v_j$. The shares are the evaluation of $p_j$ on input $1,\ldots,m$. The tally is computed by using Lagrange interpolation.


\ifnum\fullversion=1
\subsubsection{Verifiability of the Construction}
\else
\subsection{Verifiability of the Construction}
\fi

We analyze now the verifiability of our generalized construction. If the commitments in the public key are computed honestly, we can show that the generalized construction fulfills the verifiability property by using the same arguments given for our construction with one authority.

Consider that w.l.o.g the $k$-th authority outputs a commitment $\com_k$ that does not commit to a tuple of $0$'s. If at least one voter $j$ is honest, the probability that this voter outputs a ballot $\BT_j$ such that $\com_k$ is a commitment to a tuple $(z_1,\ldots,z_N)$ and $z_j=\BT_j$ is negligible over the random coins of the $j$-th voter. Therefore, assuming that there is at least one honest voter, the authorities can compute proofs of tally correctness by using the witness for the ``trapdoor statement'' in the relation only with negligible probability. Similarly,  assuming that there is at least one honest voter, the voters can compute proofs of ballot correctness by using the witness for the ``trapdoor statement'' only with negligible probability over the random coins of the honest voters. In conclusion, the generalized construction fulfills (a statistical variant of) the verifiability property thanks to the verifiability of the (fully) verifiable eVote in Section~\ref{sec:fullyscheme} and to the fact that, in real mode, the sum of the messages encrypted in a ballot is equal to a number in $\zu$.

\ifnum\fullversion=1
\subsubsection{Privacy of the Construction}
\else
\subsection{Privacy of the Construction}
\fi

We use a selectively-secure model~\cite{EC:CanHalKat04} for our definition of privacy.  In the game between the challenger and the adversary, the adversary has to declare its challenge at the outset of the game before receiving the public keys of the authorities. The adversary is allowed to receive the secret keys of all except one authority.

We show that our generalized construction fulfills this definition of privacy. First, we define the sets $S^k$ and a method for computing the shares $v_{j,1},\ldots,v_{j,m}$ for a vote $v_j$. This method must guarantee that any subset of $m-1$ authorities does not get any information about $v_j$. For simplicity, consider that $m=2$. Then, the sets $S^1 = S^2= S$ are equal by definition to $\{-p,\ldots,p\}$, where $p$ is a number of size super-polynomial in the security parameter. The message space of the PKE scheme must comprise numbers between $-Np$ and $Np$. To encrypt $0$ (resp. $1$), the voter chooses a random number $v_1$ in $S$ and sets $v_2$ to $-v_1$ (resp. $-v_1+1$). It is easy to see that, except when either $v_1$ or $v_2$ equal $-p$, any value of $v_2$ (resp. $v_1$) can correspond to $v_1=-v_2$ (resp. $v_2=-v_1$) if the voter cast a vote for $0$ or to $v_1=-v_2+1$ (resp. $v_2=-v_1+1$) if the voter cast a vote for $1$. The case in which either $v_1$ or $v_2$ equal $-p$ occurs with negligible probability, which is guaranteed by choosing $p$ to be  super-polynomial in the security parameter. Consequently, each authority does not get any information on the vote $v_j$.
This method can be generalized to the case $m>2$. We skip the details.

Because the adversary receives the public keys after sending the challenge, in the security proof we can define a hybrid experiment where the commitments in the public key commit to ballots $(\BT_1\ldots,\BT_N)$ computed on input the challenge messages. Like in the reduction of Section~\ref{sec:privacyfull}, we prove that the probability that the adversary submits an ill-formed ballot that is accepted by the ballot verification algorithm is negligible by using the computationally hiding property of the commitment scheme.

In the next hybrid experiments, the NIWI proofs of ballot correctness and of tally correctness can be computed by using the witness for the trapdoor statement, i.e., the  randomness used to compute the commitments. Thanks to that, we are able to compute ballots where not all the ciphertexts encrypt the same message. This allows us to switch the message encrypted in one of the ciphertexts of the ballots and prove indistinguishability between the experiments by using the IND-CPA property of the PKE scheme.

To prove that our scheme fulfills a definition for privacy in a non-selective (i.e., full) security model, one can use complexity leveraging arguments. Such arguments can profit from the fact that, in our formulation, we required the number of voters $N$ and the size of the message space to be independent of the security parameter. This allows the challenger to just guess the challenge messages in advance with constant probability. This requirement can be weakened to the case of $N$ and size of message space logarithmic in the security parameter. We leave open how to achieve full security without complexity leveraging.

Note that we do not require any interaction between the authorities. The public keys of the authorities are completely {\em independent} from each other. Moreover, the authorities do not need any {\em coordination} (e.g., to run sequentially), i.e., the tally can be computed and publicly verified from the output of each authority individually. Thus, our techniques completely diverge from previous approaches to the problem.




\ifnum\fullversion=1
\subsection{On The Reusability of the Public Parameters}\label{sec:reusability}
\else
\section{On The Reusability of the Public Parameters}\label{sec:reusability}
\fi


Our definition of verifiability does not prevent the following undesirable case. Consider an ill-formed ballot $\BT_1$. Consider other valid ballots $\BT_2,\ldots,\BT_N$ that encrypt respectively $v_2,\ldots,v_N$. The authority is able to compute a tally $y=F(v_1,\ldots,v_N)$ and a valid proof of tally correctness. Consider now other valid ballots $\BT_2',\ldots,\BT_N'$ that encrypt $v_2',\ldots,v_N'$. The authority can possibly compute another tally $y'=F(v_1',v_2',\ldots,v_N')$ and another proof of tally correctness. The problem is that the ill-formed ballot $\BT_1$ can be decrypted to more than one message.

This does not contradict our definition because,  for $\BT_1,\ldots,\BT_N$, there still exist messages $v_1,\ldots,v_N$ that satisfy the statement of the definition, i.e., given $\PK$ and $\BT_1,\ldots,\BT_N$, the authority cannot output two different results and two valid proofs of tally correctness for each of them. However, it can occur that for $\PK,\BT_1,\BT_2',\ldots,\BT'_N$, there are different messages that satisfy the definition. We remark that the public key $\PK$ does not change.

Let us present a concrete example. Consider two $0/1$ elections with only $2$ voters. A ballot could possibly be {\em reused} in the second election, i.e.,  if the public parameters of the system are reused, the same ballot can be cast again. Given an ill-formed ballot $\BT_1$, there could exist two ballots $\BT_2$ and $\BT_2'$ such that, in an election with ballots $\BT_1$ and $\BT_2$, the result is $2$, and, in a election with ballots $\BT_1$ and $\BT_2'$, the result is $0$. This can only happen if the first ballot is ``associated'' with vote $1$ in the first election and with vote $0$ in the second election. Therefore, the first and the second elections are incoherent. More undesirable issues would emerge if different tally functions could be computed in different elections carried out with the same parameters and ballots.

A stronger definition could state that, for all $\PK$ and all $\BT_1$, there exists $m_1$ such that, for all $\BT_2,\ldots,\BT_N$, there exist $m_2,\ldots,m_n$ such that the authority is only able to output a tally $y = F(m_1,\ldots,m_N)$ along with a valid proof of tally correctness. We note that this is a simplification because a general definition should take into account multiple dishonest voters.

Fortunately, in our e-voting model, as well as in other traditional models, the parameters cannot be reused through different elections. Therefore, the above-mentioned problem does not occur.

In a stronger model in which the parameters can be reused, our constructions would not be secure. Nevertheless, in our (fully) verifiable construction, the inconsistency of results through different elections would occur only in the case that a malicious authority sets the commitment in the public key dishonestly to $0$, which allows the computation of ill-formed ballots.


This state of affairs could be paralleled to the case of garbled circuits, where the original one-time version~\cite{FOCS:Yao86,JC:LinPin09} can be based on the minimal assumption of existence of one-way functions, whereas the reusable variant~\cite{STOC:GKPVZ13} is known to be implementable only under stronger assumptions. Similarly, in functional encryption, the schemes with bounded security~\cite{CCS:SahSey10,C:GorVaiWee12} can be based just on public key encryption, whereas the unbounded secure variants are only known to be implementable under very strong assumptions~\cite{TCC:GGHZ16}. For instance, the scheme of Sahai and Seyalioglu~\cite{CCS:SahSey10} becomes completely insecure when the adversary can decrypt a ciphertext with two different secret keys, exactly as it occurs for our schemes.



\ifnum\fullversion=1
\subsection{Related Work}\label{sec:related}
\else
\section{Related Work}\label{sec:related}
\fi

Our work is inspired by the work of Badrinarayanan \etal\ \cite{EPRINT:BGJS16}, which puts forward the concept of verifiable functional encryption.  (We note that the committing IBE of~\cite{AC:GreHoh07} can be seen as a weaker variant of verifiable identity-based encryption.) Our work shares with BGJS the idea of ``engineering'' multiple witnesses, which are needed when using NIWI proofs, to enforce privacy in conjunction with verifiability.

Notwithstanding, the constructions are quite different, especially due to the different requirements of functional encryption and e-voting. For instance, in the security definition of functional encryption, the keys are handed to the adversary, so one needs a proof that each secret key and ciphertext is computed correctly. Instead, in our case, the adversary does not see the secret key. We can profit from this fact to just prove that the claimed tally equals the evaluation of the tally function over all ballots.

Such complications in functional encryption introduce a severe limitation: in the security reduction of BGJS, it is fundamental that the public key contain a commitment that in some hybrid experiment is set to the challenge ciphertext. Therefore, it is assumed that the adversary commits to the challenge before receiving the public key, i.e., security is proven in the {\em selective} model~\cite{EC:CanHalKat04}.
On the contrary, our constructions are secure in the {\em full} (i.e., non-selective) model.

In other respects, in e-voting we face new challenges. In BGJS, the challenger computes the NIWI on input a witness that comprises {\em all} the secret keys and proves the well-formedness of all the secret keys except one, but, in addition, proves that all the secret keys decrypt some challenge ciphertext correctly. This is sufficient to use the IND-CPA property of functional encryption to prove indistinguishability between two hybrid experiments where the message encrypted in one of the ciphertexts is switched from  $m_0$ to  $m_1$. The reason is that the secret keys are supposed to be for the same function $f$ such that $f(m_0)=f(m_1)$. (More concretely, in the IND-CPA property of functional encryption, the adversary is allowed to receive secret keys for a function $f$ that evaluates both challenge messages to the same value.) Therefore, the secret keys do not allow to distinguish between the two ciphertexts. In our setting, we can {\em only} input to the NIWI all the secret keys except one.  Otherwise we could not use the IND-CPA property to prove indistinguishability between two hybrid experiments where the encrypted message is switched from $m_0$ to $m_1$.

Furthermore, in our (full) privacy definition, we have to handle challenge tuples that contain ill-formed ballots, whereas in verifiable multi-input functional encryption the challenge contains only honestly computed ciphertexts. Therefore, the differences between the two settings make the respective techniques utterly incomparable.

It is tempting to think that the construction of BGJS of multi-input verifiable functional encryption (which extends multi-input functional encryption of Goldwasser \etal\ \cite{EC:GGGJKL14}) can be directly used to construct a verifiable eVote. Though it seems plausible, we did not verify that. However, this would eventually result in a verifiable eVote based on indistinguishability obfuscation~\cite{FOCS:GGHRSW13}, a very strong assumption, and would only be secure in the selective model.

Needless to say, our techniques, as well as the ones of BGJS, owe a lot to the celebrated FLS' OR trick \cite{FOCS:FeiLapSha90}. They can be viewed as a generalization of it.

Kiayias \etal\ \cite{EC:KiaZacZha15} (see also~\cite{CZZDMPDKR15} for a distributed implementation) put forth a verifiable eVote without trust assumptions that represents a breakthrough along this direction, but diverges from ours in several fundamental aspects:
\begin{itemize}
	\item It requires interaction between the voters and the board, whereas all our algorithms are {\em non-interactive}. 
\item Receipt-freeness, accountability and degree of dependence on secure channels to distribute vote codes are undetermined issues. In our scheme, voters can verify the election if they just know the ballot they cast. In particular, voters do not need to store the randomness used to compute it, and the authority cannot cheat in the tally process.
	\item It does not achieve {\em universal verifiability}, whereas ours does.
	\item Its information-theoretical verifiability is parameterized and depends on the number of honest voters, whereas ours is {\em perfect}, i.e., the probability of a wrong tally being accepted by the verification algorithm is equal to {\em zero}.
	\item Its privacy can be reduced to group-based assumptions at the cost of using complexity leveraging and assuming sub-exponential security, whereas ours only requires the standard version of Decision Linear assumption with polynomial security.\footnote{At some point in the security reduction for our (fully) verifiable eVote, we make use of the fact that the number of voters $N$ is a constant independent of the security parameter that could be viewed as a complexity leveraging trick or as problematic in the case that $N$ be large. But we stress that this is done only for simplicity of exposition and we sketch how the reduction and our results can be generalized even to the case of $N(\cdot)$ function of the security parameter.}
	\item Its definition of verifiability requires an extraction property, whereas ours does not.
\end{itemize}

Moran and Naor \cite{C:MorNao06} construct an universally verifiable e-voting protocol with very strong provable-security properties. However, it assumes either the availability of a ``random beacon'' that has to be sampled honestly or the soundness of the Fiat-Shamir's heuristic. Therefore, verifiability does not hold unconditionally, i.e., without any assumption (both physical or computational). 

We are not aware of other traditional e-voting schemes that achieve perfect verifiability without interaction and without trust assumptions. We refer to ~\cite{EPRINT:CGKMT16} and~\cite{BCGPW15} for a survey.

We point out that our definition of verifiability is motivated by the guidelines of~\cite{EPRINT:CGKMT16}. In its formalization, our definition is similar to the ones of~\cite{GiuIovRon16}, the verifiability for multi-input functional encryption of BGJS and the uniqueness of tally of Bernhard \etal\ \cite{BCGPW15}. Anyhow, the latter is formulated to hold only against computationally bounded adversaries and both BGJS16 and Bernhard \etal\ do not take into account our condition (2) for verifiability.\footnote{Needless to say, for many applications of multi-input functional encryption, the lack of condition (2) could not pose a threat.} See also~\cite{KreRyaSmy10} for symbolic approaches to verifiability.

Our privacy notion is inspired by the one of Benaloh~\cite{thesis:Benaloh}, which is called ``PRIV'' in \cite{BCGPW15}. We reformulate it by using modern terminology and we conveniently modify it to withstand the attacks shown in~\cite{BCGPW15}. We refer to~\cite{BCGPW15} for a survey on definitions of privacy for e-voting.

Perfect verifiability and perfect correctness seem incompatible with receipt-freeness~\cite{STOC:BenTui94,EC:SakKil95,AC:MicHor96a,C:MorNao06,DelKreRya09,BeleniosRF}. Notwithstanding, we think that it should be possible to define a statistical variant of verifiability achievable without any trust assumptions that could coexist with some form of receipt-freeness. Another possibility could be to resort to some voting server trusted for receipt-freeness but not for privacy, such as the server that re-randomizes the ballots in BeleniosRF of Chaidos, Cortier, Fuchsbauer and Galindo~\cite{BeleniosRF}. (We note that they also address the problem of authenticity that we neglect.) As it is out of the scope of this work, we deliberately omit receipt-freeness in our treatment.

Recently, Bellare, Fuchsbauer and Scafuro~\cite{EPRINT:BelFucSca16} started the study of security of NIZKs in face of public parameter subversion. They showed the impossibility of attaining subversion soundness while retaining zero-knowledge, thus justifying our need of sidestepping NIZKs.



\ifnum\fullversion=1
\else
\section{Future Directions}\label{sec:future}
Our work opens up new directions in e-voting and generally in cryptography. We discuss some of them.
\begin{itemize}
\item{\bf Efficiency.} In our work, in order to compute the NIWI proofs of Groth \etal\ \cite{C:GroOstSah06} for $\CircuitSat$, we need to represent the computation as a Boolean circuit and, though this can be done in polynomial time, it can be inefficient in practice. An important objective is to sidestep the reduction to circuits by employing a more direct approach. A possibility would be to explore the achievability of our results from variants of Groth-Sahai NIWIs~\cite{EC:GroSah08}. The NIWI of Groth-Sahai, as it stands, is formulated in the CRS model but it is worthy to study in which settings it can be instantiated without CRS.

Another important direction is to improve the efficiency of verification. It would be desirable that the cost for  verifiers be sub-linear in the number of voters. The verifiability guarantees attained would then be computational but hopefully it could be possible to avoid trust assumptions. A possibility would be to employ variants of succinct arguments (see~\cite{PHD:Bitansky14} for a survey).
\item{\bf Receipt-freeness.} Perfect verifiability and perfect correctness seem incompatible with receipt-freeness~\cite{STOC:BenTui94,EC:SakKil95,AC:MicHor96a,C:MorNao06,DelKreRya09,BeleniosRF}, but we think that it should be possible to define a statistical variant of verifiability that could coexist with some form of receipt-freeness. Another possibility could be to resort to some voting server trusted for receipt-freeness but not for privacy that re-randomizes the ballots, as done in BeleniosRF of Chaidos, Cortier, Fuchsbauer and Galindo \cite{BeleniosRF}.
\item{\bf Other applications of our techniques.} We think that our techniques could be of wide applicability to other settings. For instance, Camenisch and Shoup~\cite{C:CamSho03} put forth the concept of verifiable encryption (that in some sense could be also viewed as a special case of verifiable functional encryption~\cite{EPRINT:BGJS16}) and present numerous applications of it, such as key escrow, optimistic fair exchange, publicly verifiable secret and signature sharing, universally composable commitments, group signatures, and confirmer signatures. We believe that our techniques can be employed profitably to improve their results with the aim of removing the need of trust assumptions.
\end{itemize}


\fi
\ifnum\fullversion=1
\section{Definitions}\label{sec:def}
\paragraph{\bf Notation.}
A {\em negligible} function $\negl(k)$ is a function that is smaller than the inverse of any polynomial in $k$ (from a certain point and on).
We denote by $[n]$ the set of numbers $\{1,\ldots,n\}$.
If $S$ is a finite set, we denote by $a\from S$ the process of setting $a$ equal to a uniformly chosen element of $S$.
With a slight abuse of notation, we assume the existence of a special symbol $\bot$ that does not belong to $\zu^\star$.

If $A$ is an algorithm, then $A(x_1,x_2,\ldots)$ denotes the probability distribution of the output of $A$ when $A$ is run on input $(x_1,x_2,\ldots)$ and randomly chosen coin tosses. Instead, $A(x_1,x_2,\ldots; r)$ denotes 	the output of $A$ when run on input $(x_1,x_2,\ldots)$ and (sufficiently long) coin tosses $r$. All algorithms, unless explicitly noted, are probabilistic polynomial time (PPT) and all adversaries are modeled by non-uniform PPT algorithms.

If $A$ is a PPT algorithm, we say that $y\in A(x)$ iff there exists a random value $r$ such that $y=A(x;r)$; in that case, we say that $y$ is in the range of $A(x)$.
If $E$ is an event in a probability space, $\bar E$ denotes its complement.

The following definition is used in the definition of verifiability. Essentially, it states that a tally $y$ is compatible with votes $z_1,\ldots,z_k$ if the latter values are in its pre-image.
\begin{definition}\label{def:restriction}
	Given a function $F(x_1,\ldots,x_n):A^n\rightarrow B$, we say that a value $y\in B$ is compatible with $z_1,\ldots,z_k\in A$ at indices $i_1,\ldots,i_k\in[N]$ if $y$ is in the range of the restriction $F_{|C_{z_1,\ldots,z_k,i_1,\ldots,i_k}}$ of $F$ to $C_{z_1,\ldots,z_n,i_1,\ldots,i_n}\defeq\{(x_1,\ldots,x_n)| \forall j\in[k], x_{i_j}=z_j\}$.
\end{definition}

\subsection{E-Voting Schemes}\label{sec:defevote}

An e-voting scheme (eVote, in short) is parameterized by the tuple $(N,\M,\Sigma,F)$. The natural number $N>0$ is the {\em number of voters}. The set $\M$ is the {\em domain of valid votes}. The set $\Sigma$ is the {\em range of possible results}. The function  $F:(\M\cup\{\bot\})^N\rightarrow\Sigma\cup\{\bot\}$ is the {\em tally function}.
We allow the tally function to take as input the special symbol $\bot$, which denotes either an abstention, an invalid ballot or a blank vote\footnote{We note that our tally function can be made more general by assigning different symbols to an abstention, to an invalid ballot and to a blank vote.}, and to output $\bot$ to indicate an error. We require that the tally function outputs an error on input a sequence of strings iff all the strings are equal to $\bot$. Formally, the tally function is defined as follows.
\begin{definition}[Tally function]\label{def:tally}
	A function $F$ is a tally function if there exists a natural number $N>1$, and sets $\M,\Sigma\subset\zu^\star$ such that the domain of $F$ is $\M\cup\{\bot\}$, the range is $\Sigma\cup\{\bot\}$ and for all strings $m_1,\ldots,m_N\in\M\cup\{\bot\},$ it holds that $F(m_1,\ldots,m_N)=\bot$ iff $m_1=\bot,\ldots,m_N=\bot$.
\end{definition}

Before defining formally an eVote, we explain how its algorithms are used to conduct an election.
\paragraph{The voting ceremony.}
The voting ceremony occurs as follows.
\begin{itemize}
	\item{Setup phase.} An authority (also called voting authority or election authority) uses  algorithm $\Setup$ to compute a public key $\PK$ and a secret key $\SK$.
	\item{Voting phase.} Each of the $N$ voters runs an algorithm $\Cast$ on input the voter identifier $j\in[N]$, the public key $\PK$ and a vote $v\in \M$ to compute a ballot $\BT$. The voter sends $\BT$ to an append-only public bulletin board (PBB).
	\item{Tallying phase.} The well-formedness of each ballot $\BT$ published in the PBB can be publicly verified by means of an algorithm $\VerifyBallot$. If the ballot is invalid,  a new row in which the ballot is replaced by $\bot$ is appended to the PBB. Later, only the new row is used. If a voter did not cast a vote, $\bot$  is appended to the PBB.

		The authority runs {\em evaluation tally} algorithm $\EvalTally$ on input the public key, the secret key, and $N$ strings that represent either ballots or $\bot$ symbols appended to the PBB. $\EvalTally$ outputs the tally, i.e., the result of the election, and a  proof of tally correctness. The tally equals the special symbol $\bot$ to indicate an error.
\item{Verification phase.}
Algorithm $\VerifyTally$ takes as input the public key, a tuple of $N$ strings that represent either ballots or the special symbol $\bot$, the tally and the proof of tally correctness.  $\VerifyTally$ outputs a value in $\{\OK,\bot\}$.

Each participant, not necessarily a voter, can verify the correctness of the result of the election as follows.
First, verify whether the ballots cast by the voters are valid using the $\VerifyBallot$ algorithm. Check whether the authority replaced with $\bot$ only the invalid ballots. Assign $\bot$ to any voter who did not cast her vote. After that, run the $\VerifyTally$ algorithm on input the public key, the $N$ strings that represent either ballots or the special symbol $\bot$, the tally and the proof of tally correctness.
\end{itemize}

\begin{definition}[E-voting Scheme]\label{def:evote}
	A $(N,\M,\Sigma,F)$-{\em e-voting scheme} $\EVOTE$ for number of voters $N>1$, domain of valid votes $\M$, range of possible results $\Sigma$ and tally function $F:(\M\cup\{\bot\})^N\rightarrow\Sigma\cup\{\bot\}$ is a tuple
\begin{equation*}
\EVOTE\defeq(\Setup,\Cast,\VerifyBallot, \EvalTally,\VerifyTally)
\end{equation*}
 of $5$ PPT algorithms, where $\VerifyBallot$ and $\VerifyTally$ are deterministic, that fulfill the following syntax:
\begin{enumerate}
\item $\Setup(1^\lambda)$: on input the security parameter in unary, it
	outputs the {{\em public key}} $\PK$ and the {\em secret key} $\SK$.

\item $\Cast(\PK,j,v)$: on input the public key $\PK$, the voter identifier $j\in[N]$, and a vote $v \in \M$, it outputs a {\em ballot} $\BT$.
\item $\VerifyBallot(\PK,j,\BT)$: on input the public key $\PK$, the voter identifier $j\in[N]$ and a ballot $\BT$, it outputs a value in $\{\OK, \bot\}$.

\item $\EvalTally(\PK,\SK,\BT_1,\ldots,\BT_N)$: on input the public key $\PK$, the secret key $\SK$, and
	$N$ strings that are either ballots or the special symbol $\bot$, it  outputs the tally $y \in \Sigma \cup \{\bot\}$ and a proof $\gamma$ of tally correctness.

\item $\VerifyTally(\PK,\BT_1,\ldots,\BT_N,y,\gamma)$: on input the public key $\PK$, $N$ strings that are either ballots or the special symbol $\bot$, a tally $y\in\zu^\star\cup\{\bot\}$ and a proof $\gamma$ of tally correctness, it outputs a value in $\{\OK,\bot\}$.
\end{enumerate}
\end{definition}

An eVote must satisfy the following correctness, verifiability, and privacy properties. We also define the weak verifiability and weak privacy properties. A weakly verifiable eVote must satisfy correctness, weak verifiability and weak privacy.
\subsubsection{Correctness and verifiability}

\begin{itemize}
	\item{\em (Perfect) Correctness.}
		We require the following conditions (1) and (2) to hold.
		\begin{enumerate}
			\item
				Let $\Abst$ be a special symbol not in $\M\cup\{\bot\}$ that denotes that a voter did not cast her vote.\footnote{In the following definition, we need $\Abst$ to differentiate the case of a voter who did not cast a vote at all ($\Abst$) from the case of a voter who casts $\bot$ as her own vote but wishes to preserve the anonymity of her choice. However, in both cases, correctness guarantees that the result of the election equals the output of the tally function, and the input to the tally function is $\bot$ both when a voter casts $\bot$ and when a voter does not cast any vote.}
For all $\PK\in\Setup(1^\lambda)$,
all $m_1,\ldots,m_N\in \M\cup\{\bot,\Abst \}$,
all $(\BT_j)_{j=1}^N$ such that for all $j\in[N]$, $\BT_j=\bot$ if $m_j=\Abst$, $\BT_j\in\Cast(\PK,j,m_j)$ if $m_j\in\M$ and $\BT_j\in\Cast(\PK,j,\bot)$ otherwise, the following two conditions (a) and (b) hold:
\begin{enumerate}
	\item For all $j\in[N],$ if $m_j\neq\Abst$ then $\VerifyBallot(\PK,j,\BT_j)=\OK$.
	\item if $(y,\gamma)\defeq\EvalTally(\PK,\BT_1,\ldots,\BT_N)$ then it holds that:\\ $y=F(m_1,\ldots,m_N)$ and $\VerifyTally(\PK,\BT_1,\ldots,\BT_N,y,\gamma)=\OK$.

\end{enumerate}
	\item For all $\PK\in\Setup(1^\lambda)$, $\BT_1,\ldots,\BT_N\in\zu^\star\cup \{\bot\}$,
		if $S\defeq\{j|\ \BT_j\neq\bot \wedge \VerifyBallot(\PK,j,\BT_j)=\bot\}$ and $\BT_1',\ldots,\BT_N'$ are such that for all $j\in[N],$ $\BT_j'=\BT_j$ if $j\notin S$ and $\BT'_j=\bot$ otherwise, it holds that:\\
	If $(y,\gamma)\defeq\EvalTally(\PK,\BT'_1,\ldots,\BT'_N)$ then $\VerifyTally(\PK, \allowbreak\BT'_1, \allowbreak \ldots, \allowbreak \BT'_N, \allowbreak y, \allowbreak \gamma) \allowbreak = \allowbreak \OK$.
\end{enumerate}
\item{\em Weak verifiability.}
		We require the following conditions (1) and (2) to hold.
	\begin{enumerate}
		\item For all $\PK\in\zu^\star,\BT_1,\ldots,\BT_N\in\zu^\star\cup \{\bot\}$, there exist $m_1,\ldots,m_N\in\M\cup\{\bot\}$ such that for all $y\neq\bot$ and $\gamma$ in $\zu^\star$, if $S\defeq\{j|\ \BT_j\neq\bot \wedge \VerifyBallot(\PK,j,\BT_j)=\bot\}$ and $\BT_1',\ldots,\BT_N'$ are such that for all $j\in[N],$ $\BT_j'=\BT_j$ if $j\notin S$ and $\BT'_j=\bot$ otherwise, it holds that:\\
	if $\VerifyTally(\PK,\BT_1',\ldots,\BT_N',y,\gamma)=\OK$ then  $y=F(m_1,\ldots,m_N)$.
\item For all $\PK\in\zu^\lambda$, all $k\in[N]$, $i_1,\ldots,i_k\in[N]$, all $m_{i_1},\ldots,m_{i_k}\in\M\cup\{\bot\}$, all $\BT_1,\ldots,\BT_N\in\zu^\star\cup \{\bot\}$ such that for all $j\in[k]$, $\BT_j\in\Cast(\PK,m_{i_j})$ and $\VerifyBallot(\PK,\BT_j)=\OK$,
		if $S\defeq\{j|\ \BT_j\neq\bot \wedge \VerifyBallot(\PK,j,\BT_j)=\bot\}$ and $\BT_1',\ldots,\BT_N'$ are such that for all $j\in[N],$ $\BT_j'=\BT_j$ if $j\notin S$ and $\BT'_j=\bot$ otherwise, it holds that:\\
		if there exist $y\in\zu^\star$ and $\gamma\in\zu^\star$ such that $\VerifyTally(\PK,\allowbreak\BT_1',\allowbreak\ldots,\allowbreak\BT_N',\allowbreak y,\allowbreak\gamma)\allowbreak = \allowbreak \OK$, then $y$ is compatible with $m_{i_1},\allowbreak \ldots,\allowbreak m_{i_k}$ at indices $i_1,\allowbreak \ldots,\allowbreak i_k$.
\end{enumerate}
\item{\em Verifiability.}
		We require the following conditions (1) and (2) to hold.
	\begin{enumerate}
		\item For all $\PK\in\zu^\star,\BT_1,\ldots,\BT_N\in\zu^\star\cup \{\bot\}$, there exist $m_1,\allowbreak \ldots,\allowbreak m_N \allowbreak \in \allowbreak \M\cup\{\bot\}$ such that for all $y\in\zu^\star\cup\{\bot\}$ and $\gamma$ in $\zu^\star$, if $S\defeq\{j|\ \BT_j\neq\bot \wedge \VerifyBallot(\PK,j,\BT_j)=\bot\}$ and $\BT_1',\ldots,\BT_N'$ are such that for all $j\in[N],$ $\BT_j'=\BT_j$ if $j\notin S$ and $\BT'_j=\bot$ otherwise, it holds that:\\
	if $\VerifyTally(\PK,\BT_1',\ldots,\BT_N',y,\gamma)=\OK$ then  $y=F(m_1,\ldots,m_N)$.
\item For all $\PK\in\zu^\lambda$, all $k\in[N]$, $i_1,\ldots,i_k\in[N]$, all $m_{i_1},\ldots,m_{i_k}\in\M\cup\{\bot\}$, all $\BT_1,\ldots,\BT_N\in\zu^\star\cup \{\bot\}$ such that for all $j\in[k]$, $\BT_j\in\Cast(\PK,m_{i_j})$ and $\VerifyBallot(\PK,\BT_j)=\OK$,
		if $S\defeq\{j|\ \BT_j\neq\bot \wedge \VerifyBallot(\PK,j,\BT_j)=\bot\}$ and $\BT_1',\ldots,\BT_N'$ are such that for all $j\in[N],$ $\BT_j'=\BT_j$ if $j\notin S$ and $\BT'_j=\bot$ otherwise, it holds that:\\
		if there exist $y\in\zu^\star\cup\{\bot\}$ and $\gamma\in\zu^\star$ such that $\VerifyTally(\PK,\allowbreak\BT_1',\allowbreak\ldots,\allowbreak\BT_N',\allowbreak y,\allowbreak\gamma)\allowbreak=\allowbreak\OK$, then $y$ is compatible with $m_{i_1},\allowbreak\ldots,\allowbreak m_{i_k}$ at indices $i_1,\allowbreak \ldots,\allowbreak i_k$.
\end{enumerate}

\medskip

Note that the difference between condition (2) of verifiability and condition (2) of weak verifiability lies in the fact that, in the latter, $y$ cannot equal $\bot$, whereas, in the former, the condition has to hold even for $y=\bot$. In this work, we use the terms verifiability and full verifiability interchangeably to differentiate them from weak verifiability.
\end{itemize}
\subsubsection{Privacy}

We define {\em privacy} in the style of indistinguishability-based security.
Privacy for a $(N,\M,\Sigma,F)$-eVote
\begin{equation*}
\EVOTE\allowbreak\defeq\allowbreak(\Setup,\allowbreak\Cast,\allowbreak\VerifyBallot,\allowbreak\EvalTally,\allowbreak\VerifyTally)
\end{equation*}
 is formalized by means of the game
$\SecGame^{N,\M,\Sigma,F,\EVOTE}_\adv$
between a stateful adversary
$\adv$ and a {\em challenger} $\C$. We describe the game in Fig.~\ref{definitionprivacy}.

\begin{figure}
\small
\begin{center}
\begin{framed}
$\SecGame^{N,\M,\Sigma,F,\EVOTE}_\adv(1^\lambda)$
\begin{itemize}
\item{Setup phase.} $\C$ generates
	$(\PK,\SK)\from\Setup(1^\lambda)$, chooses a random bit $b\from\zu$ and runs
$\adv$ on input $\PK$.

\item{Query phase.} $\adv$ outputs two tuples
	$M_0\defeq(m_{0,1},\ldots,m_{0,N})$ and
	$M_1\defeq(m_{1,1},\ldots,m_{1,N})$, and a set $S\subset[N]$. (The set $S$ contains the indices of the strings in the tuples that are possibly dishonest ballots. The strings in the tuples whose indices are not in $S$ are supposed to be votes to be given as input to the $\Cast$ algorithm.)
\item{Challenge phase.} The challenger does the following.
	For all $j\in[N]$, if $j\in S$, then set $\BT_j\defeq m_{b,j}$, else set $\BT_j\from\Cast(\PK,j,m_{b,j})$.
	For all $j\in S,$ if $\VerifyBallot(\PK,j,\BT_j)=\bot$, set $\BT_j\defeq\bot$.	Compute $(y,\gamma) \from \EvalTally(\PK,\allowbreak\SK,\allowbreak\BT_1,\allowbreak\ldots, \allowbreak\BT_N)$ and return $(\BT_1,\ldots,\BT_N,y,\gamma)$ to the adversary.

\item{Output.} At some point the adversary outputs its guess $b'$.

\item{Winning condition.}
	The adversary wins the game if all the following conditions hold:
\begin{enumerate}
	\item $b'=b$.
	\item For all $j\in S, m_{0,j}=m_{1,j}$. (That is, if the adversary submits a dishonest ballot, it has to be the same in both tuples.)
\item For all $d_1,\ldots,d_N\in\M\cup\{\bot\}$, for all $j\in [N]$, let $m'_{0,j}\defeq m'_{1,j}\defeq d_j$ if $j\in S$, and for all $b\in\zu$ let $m'_{b,j}\defeq m_{b,j}$ if $m_{b,j}\in\M$ and $m_{b,j}'\defeq \bot$ if $m_{b,j}\notin\M$. Then, $F(m_{0,1}',\ldots,m_{0,N}')=F(m_{1,1}',\ldots,m_{1,N}')$.
		
		(That is, the tally function outputs the same result on input both tuples, even if the ballots corresponding to indices in $S$ are replaced by arbitrary messages in $\M\cup\{\bot\}$.)
\end{enumerate}
\end{itemize}
\end{framed}
\end{center}
\caption{Definition of privacy}
\label{definitionprivacy}
\end{figure}

The advantage of adversary $\A$ in the above game
is defined as
$$
\Adv_{\adv}^{\EVOTE,\Priv}(1^\lambda)\defeq
|\prob{\SecGame^{N,\M,\Sigma,F,\EVOTE}_\adv(1^\lambda)=1}-1/2|
$$

\begin{definition}\label{def:evotepriv}
An $\EVOTE$ for parameters $(N,\M,\Sigma,F)$ is {\em private} or \IND-Secure
if the advantage of all PPT adversaries
$\adv$
is at most negligible in $\lambda$ in
the above game.
\end{definition}
\begin{definition}\label{def:evoteweakpriv}
An $\EVOTE$ for parameters $(N,\M,\Sigma,F)$ is {\em weakly private} or \wIND-Secure
if the advantage of all PPT adversaries
$\adv$
is at most negligible in $\lambda$ in
a game $\SecwGame^{N,\M,\Sigma,F,\EVOTE}_\adv(1^\lambda)$ identical to the one above except that $\adv$ is required to output an empty set $S$, i.e., $\adv$ cannot submit dishonest ballots.
\end{definition}

\begin{remark}
We make some remarks on the previous definitions.
\begin{itemize}
	\item
	Our definitions suppose that algorithm $\VerifyBallot$ is run on input each ballot before running algorithm $\VerifyTally$. The ballots that are input to $\VerifyTally$ are replaced by $\bot$ if they were not accepted by $\VerifyBallot$. Another possibility would be to let $\VerifyTally$ do this task itself.

	\item We require that $\VerifyBallot$ and $\VerifyTally$ be deterministic algorithms. Alternatively, they can be defined as PPT, but then definitions of weak verifiability and verifiability would have to be changed accordingly to hold with probability $1$ over the random coins of the algorithms.
	\item Our definition is parameterized by the number of voters $N$. It is possible to define a more restricted eVote that may possibly be ``unbounded''. Note that our definition is more general and, for instance, takes into account e-voting schemes in which the public key is of size proportional to the number of voters.
	\item Both condition (2) of verifiability and condition (2) of weak verifiability lie in some sense between correctness and verifiability as they state a requirement about honest voters.
	\item In our weakly verifiable construction in Section~\ref{sec:scheme}, algorithm $\VerifyBallot$ could be completely discarded because it accepts any ballot. Both for the sake of generality (there could exist some weakly verifiable eVote that makes a non-trivial use of $\VerifyBallot$) and to avoid overburdening the presentation, we use the same syntax for weakly verifiable eVotes and for verifiable eVotes.
	\item For the necessity of condition (2) of correctness, we refer the reader to the discussion in Section~\ref{sec:strongcorrectness}.
	\item As shown in~\cite{BCGPW15}, the definition of ``Benaloh'' (recall that we restate it using modern terminology) is subject to attacks when instantiated with specific tally functions like the majority. Nonetheless, ours is strengthened to withstand such attacks. This is done by adding the $3$-rd winning condition.
\end{itemize}
	\end{remark}


\section{Building Blocks}\label{sec:buildingblocks} \label{app:dlin}\label{sec:dlin}Our constructions use perfectly binding commitment schemes,  (one-message) non-interactive witness-indistinguishable proof systems with perfect soundness for $\NP$ \cite{C:GroOstSah06} (see also \cite{FOCS:FeiLapSha90,FOCS:DwoNao00,FOCS:DwoNao00,C:BarOngVad03,TCC:BitPan15}) and IND-CPA public key encryption with perfect correctness and unique secret key. In this section, we recall the definitions of those primitives.

\begin{definition}[IND-CPA secure PKE with perfect correctness and unique secret key]\label{def:pke}
An IND-CPA (or semantically) secure Public Key Encryption (PKE) scheme consists of three PPT
algorithms $(\Setup, \Enc, \Dec)$ defined as follows.
\begin{itemize}
	\item{$\Setup(1^\lambda)$}: On input $1^\lambda$, it outputs public key $\PK$ and decryption key $\SK$.
	\item{$\Enc(m, \PK)$}: On input message $m$ and the public key, it outputs ciphertext $\CT$.
	\item{$\Dec(\CT, \SK)$}: On input ciphertext $\CT$ and the decryption key, it outputs $m$.
	\end{itemize}
	The PKE scheme is said to be IND-CPA (or semantically) secure if for any PPT adversary $\A$, there exists
	a negligible function $\nu(\cdot)$ such that the following is satisfied for any two messages $m_0, m_1$ and for
$b \in \{0, 1\}$:
\begin{equation*}
|\Prob{\A(1^\lambda, \Enc(m_0, \PK)) = b} - \Prob{\A(1^\lambda, \Enc(m_1, \PK)) = b}| \le \nu(\lambda).
\end{equation*}

{\em Perfect correctness} requires that, for all pairs $(\PK,\SK)\in\Setup$, for all messages $m$ in the message space and all ciphertexts $\Ct$ output by $\Enc(\PK,m)$, $\Dec(\CT,\SK)=m$ must hold. {\em Unique secret key} requires that, for all $\PK$, there exists at most {\em one} $\SK$ such that $(\PK,\SK)\in\Setup(1^\lambda)$.

The Decision Linear Encryption scheme~\cite{C:BonBoySha04} fulfills those properties. It is secure under the Decision Linear Assumption~\cite{C:BonBoySha04}. We recall them next.
\end{definition}



\newcommand{\BilinearSetup}{\ensuremath{\mathcal{G}}}

\newcommand{\myvar}[1]{\ensuremath{\mathit{#1}}}
\newcommand{\securityparameter}{\myvar{k}}
\newcommand{\Gp}{\mathbb{G}_p}
\newcommand{\Ga}{\mathbb{G}}
\newcommand{\Gb}{\mathbb{\tilde{G}}}
\newcommand{\Gt}{{\mathbb{G}_{t}}}
\newcommand{\p}{\myvar{p}}
\newcommand{\q}{\myvar{q}}
\newcommand{\Zp}{\mathbb{Z}_p}
\newcommand{\Zq}{\mathbb{Z}_q}
\newcommand{\g}{\myvar{g}}
\newcommand{\ga}{\myvar{g}}
\newcommand{\gb}{\myvar{\tilde{g}}}
\newcommand{\gt}{\myvar{g_t}}
\newcommand{\h}{\myvar{h}}
\newcommand{\hb}{\myvar{\tilde{h}}}
\newcommand{\hbig}{\myvar{H}}
\newcommand{\grp}{\myvar{grp}}

\paragraph{Bilinear Groups.} We assume the existence of a PPT algorithm $\BilinearSetup(1^\lambda)$, the {\em bilinear group generator}, that outputs a {\em pairing group setup} $(\p,\Ga,\allowbreak \Gt,\e,\ga)$, where $\Ga$ and $\Gt$ are multiplicative groups of prime order $\p$  and $\e: \Ga \times \Ga \rightarrow \Gt$ is a {\em bilinear map} satisfying the following three properties: (1) bilinearity, i.e., $\e(\ga^x,\ga^y)=e(\ga,\ga)^{xy}$; (2) non-degeneracy, i.e., for all generators $\ga \in \Ga$, $\e(\ga,\ga)$ generates $\Gt$; (3) efficiency, i.e., $\e$ can be computed in polynomial time.

\begin{assm}[Decision Linear Assumption for $\BilinearSetup$.\cite{C:BonBoySha04}] Let the tuple $(p,\allowbreak\Ga,\allowbreak\Gt,\allowbreak\e,\allowbreak\ga)$ be a pairing group setup output by  $\BilinearSetup$ as defined above, and let $g_1$, $g_2$ and $g_3$ be generators of $\Ga$. Given $(g_1, g_2, g_3, g_1^a, g_2^b, g_3^c)$, where $a$ and $b$ are picked randomly from $\mathbb{Z}_p$, the Decision Linear (DLIN) assumption is to decide whether $c=a+b\ \mathrm{mod}\ p$. Precisely, the advantage of an adversary $\mathcal{A}$ in solving the Decision Linear assumption is given by:
\begin{align*}
&\bigl|\mathrm{Pr}\ [\mathcal{A}(\mathbb{G},p,g_1,g_2,g_3,g_1^a,g_2^b,g_3^{a+b})=1\mid (\p,\Ga,\allowbreak \Gt,\e,\ga)\from\BilinearSetup(1^\lambda); \\& (g_1,g_2,g_3) \leftarrow\Ga; (a,b) \leftarrow \mathbb{Z}_p] - \\& \mathrm{Pr}\ [\mathcal{A}(\mathbb{G},p,g_1,g_2,g_3,g_1^a,g_2^b,g_3^{c})=1\mid (\p,\Ga,\allowbreak \Gt,\e,\ga)\from\BilinearSetup(1^\lambda); \\& (g_1,g_2,g_3) \leftarrow\Ga; (a,b,c) \leftarrow \mathbb{Z}_p]\bigr|
\end{align*}
The Decision Linear assumption states that the advantage of $\mathcal{A}$ is negligible in $\lambda$. Boneh \etal\ \cite{C:BonBoySha04} provide a bilinear group generator $\BilinearSetup$ for which such assumption is conjectured to hold.
\end{assm}
\paragraph{Decision Linear Encryption Scheme.} Consider the following PKE scheme described by a setup algorithm $\Setup$, an encryption algorithm $\Enc$ and a decryption algorithm $\Dec$.
\begin{description}
	\item[$\Setup(1^\lambda)$:] pick
	$(\p,\Ga,\allowbreak \Gt,\e,\ga)\from\BilinearSetup(1^\lambda)$, pick randomly $g_3 \leftarrow\Ga$ and $(x,y) \leftarrow \mathbb{Z}_p$. Compute $g_1 = g_3^{1/x}$ and $g_2 = g_3^{1/y}$. Output the public key $\PK=(\mathbb{G},\allowbreak p,\allowbreak g_1,\allowbreak g_2,\allowbreak g_3)$ and the secret key $\SK\allowbreak=\allowbreak(\PK,\allowbreak x,\allowbreak y)$.
\item[$\Enc(\PK,m)$:] on input a public key $\PK$ and a message $m \in\Ga$, pick random $(a,b) \leftarrow \mathbb{Z}_p$. Output a ciphertext $\CT=(g_1^a,g_2^b,m \cdot g_3^{a+b})$.
\item[$\Dec(\SK,\Ct)$:] on input a secret key $\SK$ and a ciphertext $\CT=(c_1,c_2,c_3)$, output $m=c_3/(c_1^x c_2^y)$.
\end{description}
This scheme fulfills the IND-CPA property under the Decision Linear assumption (see \cite{C:BonBoySha04} for details) and it is easy to verify that it fulfills the unique secret key property.



\begin{definition}[(Perfectly binding) Commitment Schemes]
\label{def:com}
A commitment scheme $\Com$ is a PPT algorithm that takes as input a string $x$ and randomness $r\in\zu^k$
and outputs $\com \from \Com(x; r)$. A perfectly binding commitment scheme must satisfy the following
properties:
\begin{itemize}
	\item{Perfectly Binding}: This property states that two different strings cannot have the same
commitment. More formally, $\forall x_1 \neq x_2$ and $r_1, r_2, \Com(x_1; r_1) \neq \Com(x_2; r_2)$.
\item{Computational Hiding}: For all strings $x_0$ and $x_1$ (of the same length), for all PPT
	adversaries $\A$ there exists a negligible function $\nu(\cdot)$ such that:
	$|\probb{r\in\zu^k}{\A(\Com(x_0;r)) = 1} - \probb{r\in\zu^k}{\A(\Com(x_1;r)) = 1)}| \le \nu(k). $

\end{itemize}
\end{definition}
\paragraph{NIWI proof systems.}\label{sec:niwi}
Next, we define (one-message) non-interactive witness indistinguishability (NIWI) proof systems \cite{C:GroOstSah06}. Groth \etal\ \cite{C:GroOstSah06} construct such NIWIs for all languages in $\NP$, and in particular for $\CircuitSat$.

\begin{definition}[Non-interactive Proof System] A non-interactive proof system for a language
	$L$ with a PPT relation $R$ is a tuple of algorithms $(\Prove, \Verify)$. $\Prove$ receives as input a statement $x$ and a witness $w$ and outputs a proof $\pi$. $\Verify$ receives as input a statement $x$ and a proof $\pi$ and outputs a symbol in $\{\OK,\bot\}$. The following
properties must hold:
\begin{itemize}
	\item{Perfect Completeness}: For every $(x, w) \in R$, it holds that\\
		$\Prob{\Verify(x, \Prove(x, w)) = \OK} = 1$,
where the probability is taken over the coins of $\Prove$ and
$\Verify$.
\item{Perfect Soundness}: For every adversary $\A$, it holds that:
$$\Prob{
	\begin{array}{lcl}
		\Verify(x, \pi) = \OK \wedge x \notin  L :\\
		(x, \pi) \from \A(1^k)\\				
	\end{array}
}=0.$$
\end{itemize}
\end{definition}
\begin{definition}[NIWI]\label{def:niwi} A non-interactive proof system $\NIWI=(\Prove, \Verify)$ for
	a language $L$ with a PPT relation $R$ is witness-indistinguishable (WI, in short) if for any triplet $(x, w_0, w_1)$ such
that $(x, w_0) \in R$ and $(x, w_1) \in R$, the distributions $\{\Prove(x, w_0)\}$ and $\{\Prove(x, w_1)\}$
are computationally indistinguishable.
\end{definition}


\section{Our Weakly Verifiable eVote}\label{sec:scheme}

In this section, we present our weakly verifiable eVote $\EVOTE$. This eVote fulfills the \wIND-Security and weak verifiability properties.
\begin{definition}[$\EVOTE$]  Let $\NIWIDec = (\ProveDec,\VerifyDec)$ be a NIWI proof system for the relation $\Rdec$, which we specify later.	Let $\PKE\allowbreak=\allowbreak(\PKE.\Setup,\allowbreak\PKE.\Enc,\allowbreak\PKE.\Dec)$ be a PKE scheme with perfect correctness and unique secret key (see Def. \ref{def:pke}).

We define as follows an $(N,\M,\Sigma,F)$-eVote
\begin{equation*}
\EVOTE^{N,\M,\Sigma,F,\PKE,\NIWIDec}=(\Setup,\Cast,\VerifyBallot,\EvalTally,\VerifyTally)
\end{equation*}

\begin{itemize}
	\item $\Setup(1^\lambda)$: on input the security parameter in unary, do the following.
		\begin{enumerate}
			\item For all $l\in[3]$, run $(\PKE.\PK_l,\PKE.\SK_l)=\PKE.\Setup(1^\lambda;s_l)$ with randomness $s_l$.
			\item Output $\PK\defeq(\PKE.\PK_1,\ldots,\PKE.\PK_3)$ and
				$\SK\defeq(\PKE.\SK_1,\PKE.\SK_2,s_1,s_2)$.\footnote{Actually, as the randomness for the setup of our PKE scheme uniquely determines the secret key, it would be sufficient to just include the $s_l$'s in $\SK$.}
		\end{enumerate}

\item $\Cast(\PK,j,v)$:
	on input the public key $\PK$, the voter index $j\in[N]$, and a vote $v$, do the following.
	\begin{enumerate}
		\item For all $l\in[3]$,  compute $\Ct_{j,l}\from\PKE.\Enc(\PKE.\PK_l,v)$.
		\item Output $\BT_j\defeq(\Ct_{j,1},\ldots,\CT_{j,3})$.
	\end{enumerate}

\item $\VerifyBallot(\PK,j,\BT)$:
	on input the public key $\PK$,
	the voter index $j\in[N]$, and a ballot $\BT$,
	output $\OK$ (i.e., accept any ballot, even invalid ones).

\item $\EvalTally(\PK,\SK,\BT_1,\ldots,\BT_N)$:
	on input the public key $\PK$, the secret key $\SK$, and a tuple of $N$ strings $(\BT_1,\ldots,\BT_N)$ that consists of either ballots cast by a voter or the special symbol $\bot$, do the following.
	\begin{enumerate}
		\item For all $j\in[N],l\in[2]$,
				$$m_j^{l}=
				\begin{cases} \bot &\mbox{if } \BT_j=\bot, \\
					\bot &\mbox{if } \BT_j\neq\bot \wedge \PKE.\Dec(\Ct_{j,l},\PKE.\SK_{l})\notin\M,\\
					\PKE.\Dec(\Ct_{j,l},\PKE.\SK_l)	& \mbox{otherwise. }
				\end{cases}
				$$
\item For all $l\in[2]$, compute $y_l= F(m_{1,l},\ldots,m_{N,l}).$
\item If $y_1=y_2$, then set $y=y_1$, else set $y=\bot$.

\item Consider the following relation $\Rdec$ in Fig.~\ref{fig:relationdecrypt}. Henceforth, if the indices $(i_1,i_2)$ in the witness of the relation $\Rdec$ fulfill $i_1=1$ and $i_2=2$ (resp. $i_1\neq 1$ or $i_2\neq 2$), the statement or the proof is in real mode (resp. trapdoor mode). Set the statement
    \begin{equation*}
    x\defeq (\BT_1,\ldots,\BT_N,\PKE.\PK_1,\ldots,\PKE.\PK_3,y)
    \end{equation*}
    and the witness
    \begin{equation*}
    w \defeq (\PKE.\SK_1,\PKE.\SK_2,s_1,s_2,i_1\defeq 1,i_2\defeq 2)
    \end{equation*}
    and compute a proof $\gamma \from \ProveDec(x,w)$.

\item Output $(y,\gamma)$.

\begin{figure}

  \begin{framed}

Relation $\Rdec(x,w)$:\\

Instance: $x\defeq(\BT_1,\ldots,\BT_N,\PKE.\PK_1,\ldots,\PKE.\PK_3, y).$ (Recall that a ballot is set to $\bot$ if the corresponding voter did not cast her vote.)\\

Witness: $w \defeq (\PKE.\SK_1',\PKE.\SK_2',s_1,s_2,i_1,i_2)$, where the $s_l$'s are the randomness used to generate the secret keys and public keys (which are known to the authority who set up the system).\\

$\Rdec(x, w) = 1$ if and only if the following condition holds.\\

$2$ of the secret keys corresponding to indices $\PKE.\PK_{i_1}, \PKE.\PK_{i_2}$ are constructed using honestly generated public and secret key
		pairs and are equal to $\PKE.\SK_1',\PKE.\SK_2'$; and either $y=\bot$ or for all $l\in[2]$, $y=F(m_1^l,\ldots,m_N^l)$ and for all $j\in[N]$, if $\BT_j\neq\bot$ then for $l\in[2]$, $\PKE.\SK_{i_l}$ decrypts ciphertext $\Ct_{j,i_l}$ in $\BT_j$ to $m_j^{i_l}\in\M$; and for all $l\in[2]$, $m_j^l=\bot$ if either $\BT_j=\bot$ or $\PKE.\SK_{i_l}$ decrypts $\Ct_{j,i_l}$ to a string $\notin \M$.\\


		Precisely, $\Rdec(x, w) = 1$ if and only if the following conditions hold. In the following, items (a) and (c) are not actually conditions that have to be checked but are steps needed to define  (note the use of ``$\defeq$'') the variables $\PKE.\PK_{i_l}$'s, $\PKE.\SK_{i_l}$'s and $m_j^{i_l}$'s that are used in the checks (b) and (d).\\

		\begin{enumerate}
			\item For all $l\in[2], (\PKE.\PK_{i_l},\PKE.\SK_{i_l})\defeq \PKE.\Setup(1^\lambda;s_l)$.
			\item For all $l\in[2], \PKE.\SK_{l}'=\PKE.\SK_{i_l}$.
			\item For all $j\in[N],l\in[2],$
				$$m_j^{i_l}\defeq
				\begin{cases} \bot &\mbox{if } \BT_j=\bot, \\
					\bot &\mbox{if } \BT_j\neq\bot \wedge \PKE.\Dec(\Ct_{j,i_l},\PKE.\SK_{i_l})\notin\M,\\
					\PKE.\Dec(\Ct_{j,i_l},\PKE.\SK_{i_l}) & \mbox{otherwise. }
				\end{cases}
				$$
			\item $(y=\bot)$ $\vee$ (for all $l\in[2]$, $y=F(m_1^{i_l},\ldots,m_N^{i_l})$).
		\end{enumerate}
		(Note that $\PKE.\SK_1'$ and $\PKE.\SK_2'$ do not necessarily have to correspond to the first two secret keys.)

\end{framed}
\caption{Relation $\Rdec$}
\label{fig:relationdecrypt}
\end{figure}

\medskip

\end{enumerate}
\item $\VerifyTally(\PK,\BT_1,\ldots,\BT_N,y,\gamma)$: on input the public key $\PK$, a tuple of $N$ strings that can be either ballots cast by a voter or the special symbol $\bot$, a tally $y$ and a proof $\gamma$,
		if $\gamma=\bot$ output $\bot$, else set
  \begin{equation*}
    x\defeq (\BT_1,\ldots,\BT_N,\PKE.\PK_1,\ldots,\PKE.\PK_3,y)
    \end{equation*}
 output $\VerifyDec(x,\gamma)$.
	
\end{itemize}
\end{definition}
Henceforth, for simplicity we omit the parameters of the scheme and we write just $\EVOTE$.
\subsection{Correctness and Weak Verifiability of the Construction}\label{sec:schemever}
\paragraph{\bf Correctness.}  The (perfect) correctness of $\EVOTE$ follows from the perfect correctness of $\PKE$ and the perfect completeness of $\NIWIDec$.
\paragraph{\bf Weak verifiability.}

\begin{theorem}\label{thm:weakver}
	For all $N>0$, all sets $\M,\Sigma\subset\zu^\star$, and all tally functions $F:(\M\cup\{\bot\})^N\rightarrow\Sigma\cup\{\bot\}$, if $\PKE$ is a perfectly correct PKE with unique secret key (cf. Def. \ref{def:pke}) and $\NIWIDec$ is a (one-message) NIWI (cf. Def. \ref{def:niwi}) for the relation $\Rdec$, then $\EVOTE^{N,\M,\Sigma,F,\PKE,\NIWIDec}$ satisfies the weak verifiability property (cf. Def. \ref{def:evote}).
\end{theorem}

\begin{proof}
	First, we prove that condition (1) of verifiability is satisfied. Since algorithm $\VerifyBallot$ accepts any ballot, even invalid ones, we have to prove that
	for all $\PK\in\zu^\star$, all $\BT_1,\ldots,\BT_N\in\zu^\star\cup \{\bot\}$, there exist $m_1,\ldots,m_N\in\M\cup\{\bot\}$ such that for all $y\neq\bot$ and all $\gamma$ in $\zu^\star$,	if $\VerifyTally(\PK,\allowbreak\BT_1,\allowbreak\ldots,\allowbreak\BT_N,\allowbreak y,\allowbreak \gamma)\allowbreak =\allowbreak 1$ then  $y=F(m_1,\ldots,m_N)$.

	Henceforth, w.l.o.g, we let $\PK$ and $\BT_1,\ldots,\BT_N$ be arbitrary strings.
	First, we prove the following claim.
	\begin{claim}
		Given $\PK$ and $(\BT_1,\ldots,\BT_N)$, for every two pairs $(y_0,\gamma_0)$ and $(y_1,\gamma_1)$ such that $y_0,y_1\neq\bot$, if $\VerifyTally(\PK,\allowbreak\BT_1,\allowbreak\ldots,\allowbreak\BT_N,\allowbreak y_0,\allowbreak \gamma_0)\allowbreak =\allowbreak \VerifyTally(\PK,\allowbreak\BT_1,\allowbreak\ldots,\allowbreak\BT_N,\allowbreak y_1,\allowbreak\gamma_1)\allowbreak =\allowbreak \OK$ then $y_0=y_1$.
\end{claim}
Let $y_0,\gamma_0,y_1,\gamma_1$ be arbitrary strings in $\zu^\star\cup\{\bot\}$ such that $y_0,y_1\neq\bot$.
Suppose that $\VerifyTally(\PK,\allowbreak\BT_1,\allowbreak\ldots,\allowbreak\BT_N,\allowbreak y_0,\allowbreak\gamma_0)\allowbreak =\allowbreak\VerifyTally(\PK,\allowbreak\BT_1,\allowbreak\ldots,\allowbreak\BT_N,\allowbreak y_1,\allowbreak\gamma_1)\allowbreak =\allowbreak\OK$.
	The perfect soundness of $\NIWIDec$ implies that, for all $b\in\zu$, the proof $\gamma_b$ is computed on input some witness
	$(\PKE.\SK_1'^b,\PKE.\SK_2'^b,s_1^b,s_2^b,i_1^b, i_2^b)$.

By the pigeon principle, there exists an index $i^\star$ such that one of the following cases holds.
\begin{enumerate}
\item{$i^\star=i_1^0=i_2^1$.}
	For all $b\in\zu$, let $(m_1^{i^\star,b},\ldots,m_N^{i^\star,b})$ be the messages guaranteed by condition (iii) of relation $\Rdec$ for proof $\gamma_b$.
	Condition (i) for proof $\gamma_0$ (resp. $\gamma_1$) implies that the secret key $\SK_1'^0$ (resp. $\SK_2'^1$) is honestly computed and thus, the unique secret key property and the fact that it fulfills $\PKE.\PK_{i_1^0}=\PKE.\PK_{i^\star}$ (resp. $\PKE.\PK_{i_2^1}=\PKE.\PK_{i^\star}$) imply that for all $j\in[N]$,
	$\PKE.\Dec(\Ct_{j,i^\star},\PKE.\SK_1'^0)=\PKE.\Dec(\Ct_{j,i^\star},\PKE.\SK_2'^1)$.
	
	Furthermore, condition (ii) and (iii) for proof $\gamma_0$ (resp. $\gamma_1$) imply that for all $j\in[N]$, either $m_j^{i^\star,0}=\bot$
	or $m_j^{i^\star,0}=\PKE.\Dec(\Ct_{j,i^\star},\PKE.\SK_1')\in\M$ (resp. either $m_j^{i^\star,1}=\bot$ or $m_j^{i^\star,1}=\PKE.\Dec(\Ct_{j,i^\star},\PKE.\SK_2'^1)\in\M$).
   	
	Hence, for all $j\in[N]$, $m_j^{i^\star,0}=m_j^{i^\star,1}\in\M\cup\{\bot\}$.
	Now, condition (iv) for proof $\gamma_0$ (resp. $\gamma_1$) implies that
	either $y_0=F(m_1^{i_1^0,0},\ldots,m_N^{i_1^0,0})$ or $y_0=\bot$
	(resp. either $y_1=F(m_1^{i_2^1,1},\ldots,m_N^{i_2^1,1})$ or $y_1=\bot$) and, as by hypothesis $y_0,y_1\neq\bot$, it holds that $y_0=y_1$.	(Here, the ``weakness'' of $\EVOTE$ arises, i.e., it cannot be proven (fully) verifiable because it could occur that, for example, $y_0\neq y_1, y_0=\bot$.)
\item{$i^\star=i_2^0=i_1^1$.}
	This case is identical to the first one, except that we replace $i_1^0$ with $i_2^0$ and $i_2^1$ with $i_1^1$.
\item{$i^\star=i_1^0=i_1^1$.} 	
	This case is identical to the first one, except that we replace $i_2^1$ with $i_1^1$.
\item{$i^\star=i_2^0=i_2^1$.} 	
	This case is identical to the first one, except that we replace $i_1^0$ with $i_2^0$.
\end{enumerate}
In all cases, we have that, if $\VerifyTally(\PK, \allowbreak\BT_1,\allowbreak\ldots,\allowbreak\BT_N,\allowbreak y_0,\allowbreak\gamma_0)\allowbreak=\allowbreak\VerifyTally(\PK,\allowbreak\BT_1,\allowbreak\ldots,\allowbreak\BT_N,\allowbreak y_1,\allowbreak\gamma_1)\allowbreak=\allowbreak\OK$ then $y_0\allowbreak =\allowbreak y_1$. In conclusion, the claim is proved.

From the previous claim, it follows that there exists a {\em unique} value $y^\star$ such that,
		for all $(y,\gamma)$ such that $y\neq\bot$,
		if $\VerifyTally(\PK,\BT_1,\allowbreak \ldots,\BT_N,y,\gamma)=\OK$
		then $y=y^\star$ (1). Moreover, it is easy to see that,
		for all $(y,\gamma)$,
		if $\VerifyTally(\PK,\allowbreak\BT_1,\allowbreak \ldots,\allowbreak\BT_N,\allowbreak y,\allowbreak \gamma)\allowbreak=\allowbreak\OK$, there exist messages $m_1,\allowbreak \ldots,\allowbreak m_N\in\M\cup\{\bot\}$ such that $y \allowbreak=\allowbreak F(m_1,\allowbreak \ldots,\allowbreak m_N)$ (2).

		Now, we have two mutually exclusive cases.
		\begin{itemize}
			\item For all $(y,\gamma)$ such that $y\neq\bot$,
					$\VerifyTally(\PK,\BT_1,\allowbreak \ldots,\BT_N,y,\gamma)=\bot$.
				Then, letting $m_1,\ldots,m_N$ in the statement of the theorem be arbitrary messages in $\M\cup\{\bot\}$, the statement is verified with respect to $\PK$ and $\BT_1,\ldots,\BT_N$.
			\item There exists $(y',\gamma)$ such that $y'\neq\bot$ and $\VerifyTally(\PK,\BT_1,\allowbreak \ldots,\BT_N,y',\gamma)\allowbreak=\allowbreak\OK$.
				In this case, (2) implies that there exist $m_1',\ldots,m_N'\in\M\cup\{\bot\}$ such that $y'=F(m_1',\ldots,m_N')$ (3).
				Hence, (1) and (3) together imply that $y^\star=F(m_1',\ldots,m_N')$ (4).

				Therefore, for all $(y,\gamma)$ such that $y\neq\bot$, if $\VerifyTally(\PK,\BT_1,\BT_N,y,\gamma)=\OK$ then (by (1)) $y=y^\star=$ (by (4)) $=F(m_1',\ldots,m_N')$.

				Then, for $m_1\defeq m_1',\ldots,m_N\defeq m_N'$, the statement of condition (1) of weak verifiability is verified with respect to
				$\PK$ and $\BT_1,\ldots,\BT_N$.

		\end{itemize}
		In both cases, for $m_1\defeq m_1',\ldots,m_N\defeq m_N'$, the statement of condition (1) of weak verifiability is verified with respect to $\PK$ and $\BT_1,\ldots,\BT_N$.

		As  $\PK$ and $\BT_1,\ldots,\BT_N$ are arbitrary strings, the statement of condition (1) of weak verifiability is proven.
	
		It is also easy to check that condition (2) of weak verifiability is satisfied. This follows straightforwardly from the perfect soundness of $\NIWIDec$. Thanks to $\NIWIDec$, the authority always proves that the public key of the PKE scheme is honestly generated. Therefore, by the perfect correctness of the PKE scheme, an honestly computed ballot for message $m$ for the $j$-th voter is decrypted to $m$ (because an honestly computed ballot, by definition, consists of three ciphertexts that encrypt the same message). Consequently, if the tally $y$ is different from $\bot$ (i.e., if the evaluation of the tally function is equal for all indices), then $y$ has to be compatible with $m$ at index $j$ (cf. Def.~\ref{def:restriction}).
\end{proof}


\subsection{Weak Privacy of the Construction}\label{sec:privacy}\label{sec:privacyweak}

\begin{theorem}\label{thm:priv}
	For all $N>0$, all sets $\M,\Sigma\subset\zu^\star$, and all tally functions $F:(\M\cup\{\bot\})^N\rightarrow\Sigma\cup\{\bot\}$, if $\PKE$ is a perfectly correct PKE scheme with unique secret key (cf.\ Def.~\ref{def:pke}) and $\NIWIDec$ is a (one-message) NIWI (cf. Def. \ref{def:niwi}) for the relation $\Rdec$, then $\EVOTE^{N,\M,\Sigma,F,\PKE,\NIWIDec}$ is \wIND-Secure (cf.\ Def.~\ref{def:evoteweakpriv}).
\end{theorem}
	\begin{proof}
		Let $\adv$ be a PPT adversary against the \wIND-Security property of $\EVOTE$. 
		We prove that $\Adv_{\adv}^{\EVOTE,\SecwGame}(1^\lambda)\le\nu(1^\lambda)$ for some negligible function $\nu(\lambda)$.

		We prove that by means of a series of hybrid experiments.
		We refer the reader to Table~\ref{table} for a pictorial explanation of the experiments, which are explained in Section~\ref{sec:sketchproofsweakscheme}. In the table, for simplicity, we omit the indices $k$ for the experiments $H_2^k$'s,$H_4^k$'s,$H_6^k$'s presented below. Therefore, hybrid experiment $H_2$ (resp. $H_4$, $H_6$) in the table corresponds to hybrid experiment $H_2^N$ (resp. $H_4^N,H_6^N$) below.

		\paragraph{{\bf Hybrid $H_1$.}} Experiment $H_1$ is equal to the experiment $\SecwGame^{N,\M,\Sigma,F,\EVOTE}_\adv$ except that the challenger sets $b\defeq 0$.

\paragraph{{\bf Hybrid $H_2^k,$} for $k=0,\ldots,N$.}
For all $k=0,\ldots,N$, experiment $H_2^k$ is identical to experiment $H_1$ except that, for all $j=1,\ldots,k$, the challenger computes $\Ct_{k,3}$ on input $m_{1,k}$.
Note that $H_2^0$ is identical to $H_1$.

\begin{claim}\label{clm:hybridtwo}
	For all $k=1,\ldots,N$, the advantage of $\adv$ in distinguishing $H_2^{k-1}$ from $H_2^k$ is negligible.
\end{claim}
\begin{proof}
	Suppose toward a contradiction that $\adv$ has instead non-negligible advantage $\epsilon(\lambda)$.
	We construct an adversary $\B$ that has advantage at most $\epsilon(\lambda)$ against the IND-CPA security of $\PKE$.

	$\B$ receives from the challenger of IND-CPA a public key $\pk$ and sets $\PK_3\defeq\pk$. For $l\in[2]$, $\B$ runs $\PKE.\Setup$ to compute $(\PKE.\PK_l,\PKE.\SK_l)$ and runs $\A$ on input $\PK\defeq(\PKE.\PK_1,\PKE.\PK_2,\PKE.\PK_3)$.

	$\A$ outputs two tuples $(m_{0,1},\ldots,m_{0,N})$ and $(m_{1,1},\ldots,m_{1,N})$, and a set $S$, which is empty for the \wIND-Security game. 	$\B$ returns $(m_{0,k},m_{1,k})$ as its pair of challenge messages to the IND-CPA challenger. The IND-CPA challenger sends $\B$ the challenge ciphertext $\ct^\star$.


	$\B$ computes $\BT_k\defeq(\Ct_{k,1},\Ct_{k,2},\ct^\star)$ by encrypting $m_{0,j}$ in $\Ct_{k,1}$ and $\Ct_{k,2}$.
	$\B$ can compute the ballots $\BT_j$ for all $j\in[N],j\neq k$ exactly as the challenger in both experiments would do.

	$\B$ computes $y$ as in the previous experiment (i.e., by running $\EvalTally$ on input $(\PK,\BT_1,\ldots,\BT_N)$) and uses the $2$ secret keys $(\PKE.\SK_1,\PKE.\SK_2)$ to compute a proof $\gamma$ exactly as the challenger in both experiments would do. $\B$ restarts $\A$ on input the computed ballots along with $(y,\gamma)$ and returns the output of $\A$.

	It is easy to see that, if $\ct^\star$ is an encryption of $m_{0,k}$, then $\B$ simulates experiment $H_2^{k-1}$, and, if
	$\ct^\star$ is an encryption of $m_{1,k}$, then $\B$ simulates experiment $H_2^{k}$. Therefore, $\B$ has probability $\epsilon(\lambda)$ of winning the IND-CPA game, which contradicts the assumption that the PKE scheme fulfills the IND-CPA property.
\end{proof}

\paragraph{{\bf Hybrid $H_3$}.} Experiment $H_3$ is identical to experiment $H_2^N$ except that the challenger computes the proof $\gamma$ on input a witness that contains indices $(1,3)$ and secret keys $\SK_1,\SK_3$ (precisely, with the randomness used to compute those secret keys, but henceforth, for simplicity, we omit this detail).
\begin{claim}\label{clm:hybridthree}
	The advantage of $\adv$ in distinguishing $H_2^{N}$ from $H_3$ is negligible.
\end{claim}
\begin{proof}
	This follows straightforwardly from the WI property of $\NIWIDec$. We note that both the randomness used to compute $\SK_1,\SK_2$ and the randomness used to compute $\SK_1,\SK_3$ constitute valid witnesses for the statement $(\BT_1, \allowbreak \ldots, \allowbreak \BT_N, \allowbreak \PKE.\PK_1, \allowbreak \ldots, \allowbreak \PKE.\PK_3, \allowbreak y)$.
	\end{proof}

	\paragraph{{\bf Hybrid $H_4^k,$} for $k=0,\ldots,N$.}
	For all $k=0,\ldots,N$, experiment $H_4^k$ is identical to experiment $H_3$, except that, for all $j=1,\ldots,k$, the challenger computes $\Ct_{k,2}$ on input $m_{1,k}$.
	Note that $H_4^0$ is identical to $H_3$.

\begin{claim}
	For all $k=1,\ldots,N$, the advantage of $\adv$ in distinguishing $H_4^{k-1}$ from $H_4^k$ is negligible.
\end{claim}
\begin{proof}
	The proof is identical to the one for Claim~\ref{clm:hybridtwo} except that the third index and the second index are swapped.
	\end{proof}

	\paragraph{{\bf Hybrid $H_5$}.} Experiment $H_5$ is identical to experiment $H_4^N$ except that the challenger computes the proof $\gamma$ on input a witness that contains indices $(2,3)$ and secret keys $\SK_2,\SK_3$.
\begin{claim}
	The advantage of $\adv$ in distinguishing $H_4^{N}$ from $H_5$ is negligible.
\end{claim}
\begin{proof}
	This follows straightforwardly from the WI property of $\NIWIDec$. We note that both the randomness used to compute $\SK_1,\SK_3$ and the randomness used to compute $\SK_2,\SK_3$ constitute valid witnesses for the statement $(\BT_1, \allowbreak \ldots, \allowbreak \BT_N, \allowbreak \PKE.\PK_1, \allowbreak \ldots, \allowbreak \PKE.\PK_3, \allowbreak y)$.
	\end{proof}
	\paragraph{{\bf Hybrid $H_6^k,$} for $k=0,\ldots,N$.}
	For all $k=0,\ldots,N$, experiment $H_6^k$ is identical to experiment $H_5$ except that, for all $j=1,\ldots,k$, the challenger computes $\Ct_{k,1}$ on input $m_{1,k}$.
Note that $H_6^0$ is identical to $H_5$.

\begin{claim}
	For all $k=1,\ldots,N$, the advantage of $\adv$ in distinguishing $H_6^{k-1}$ from $H_6^k$ is negligible.
\end{claim}
\begin{proof}
	The proof is identical to the one for Claim~\ref{clm:hybridtwo} except that the third index and the first index are swapped.
	\end{proof}
	\paragraph{{\bf Hybrid $H_7$}.} Experiment $H_7$ is identical to experiment $H_6^N$ except that the challenger computes the proof $\gamma$ on input a witness that contains indices $(1,2)$ and secret keys $(\SK_1,\SK_2)$.
\begin{claim}
	The advantage of $\adv$ in distinguishing $H_6^{N}$ from $H_7$ is negligible.
\end{claim}
\begin{proof}
	This follows straightforwardly from the WI property of $\NIWIDec$. We note that both the randomness used to compute $\SK_1,\SK_2$ and the randomness used to compute $\SK_2,\SK_3$ constitute valid witnesses for the statement $(\BT_1, \allowbreak \ldots, \allowbreak \BT_N, \allowbreak \PKE.\PK_1, \allowbreak \ldots, \allowbreak \PKE.\PK_3, \allowbreak y)$.
	\end{proof}
	\ignore{
	\paragraph{{\bf Hybrid $H_7',$}.} Experiment $H_7'$ is identical to experiment $H_7$ except that the tally $y$ is computed as in experiment $H_1$ (i.e., using $\EvalTally$).
\begin{claim}
	The advantage of $\adv$ in distinguishing $H_7$ from $H_7'$ is negligible.
\end{claim}
\begin{proof}
	The proof is symmetrical to the proof of Claim \ref{clm:hybridone} with the messages $m_{0,j}$'s swapped with the messages $m_{1,j}$'s.	
\end{proof}
}

Experiment $H_1$ (resp. $H_7$) is identical to experiment $\SecwGame^{N,\M,\Sigma,F,\EVOTE}_\adv$ except that the challenger sets $b=0$ (resp. $b=1$). Hence $\Adv_{\adv}^{\EVOTE,\SecwGame}(1^\lambda)$ equals at most the sum of the advantages of $\adv$ in distinguishing the previous hybrids. Since $N$ is a constant, such advantage is negligible and the theorem is proven.
	\end{proof}

	\begin{corollary} \label{cor:weak}
		\ifnum\fullversion=1
		If the Decision Linear assumption (see Section \ref{app:dlin}) holds, then there exists a weakly verifiable eVote.
		\else
		If the Decision Linear assumption (see Appendix \ref{app:dlin}) holds, then there exists a weakly verifiable eVote.
		\fi
	\end{corollary}
	\begin{proof}
		Boneh \etal\ \cite{C:BonBoySha04} show the existence of a PKE scheme with perfect correctness and unique secret key that fulfills the IND-CPA property under the Decision Linear assumption. Groth \etal\ \cite{C:GroOstSah06} show the existence of (one-message) NIWI proofs with perfect soundness for all languages in $\NP$ that is secure under the Decision Linear assumption. Then, because Theorem~\ref{thm:weakver} and Theorem~\ref{thm:priv} are proven, the corollary follows.
	\end{proof}





\section{Our (Fully) Verifiable eVote}\label{sec:fullyscheme}

In this Section, we present an eVote scheme $\EVOTEF$ that is \IND-Secure and (fully) verifiable.
\begin{definition}[$\EVOTEF$]
	Let $\PKE=(\PKE.\Setup,\PKE.\Enc,\PKE.\Dec)$ be a PKE scheme with perfect correctness and unique secret key (see Def. \ref{def:pke}). Let $\Com$ be a perfectly binding commitment scheme. Let $\NIWIDecf =(\ProveDecf,\VerifyDecf)$ and $\NIWIEncf = (\ProveEncf,\VerifyEncf)$ be two NIWI proof systems for the relations $\Rdecf$ and $\Rencf$, which we specify later.

We define as follows an $(N,\M,\Sigma,F)$-eVote
\begin{multline*}
\EVOTEF^{N,\M,\Sigma,F,\PKE,\Com,\NIWIEncf,\NIWIDecf}\\=(\Setupf,\Castf,\VerifyBallotf,\EvalTallyf,\VerifyTallyf)
\end{multline*}

\begin{itemize}
	\item $\Setupf(1^\lambda)$: on input the security parameter in unary, do the following.
		\begin{enumerate}
			\item Choose randomness $r\from\zu^\lambda$ and set $Z=\Com(1;r)$.
			\item For all $l\in[3]$, choose randomness $s_l\from\zu^\lambda$ and run $(\PKE.\PK_l,\PKE.\SK_l)=\PKE.\Setup(1^\lambda;s_l)$.
			\item Output $\PK\defeq(\PKE.\PK_1,\ldots,\PKE.\PK_3,Z)$ and
				$\SK\defeq(\PKE.\SK_1,\PKE.\SK_2,s_1,s_2,r)$.\footnote{Actually, as the randomness for the setup of our PKE scheme uniquely determines the secret key, it would be sufficient to just include the $s_l$'s in $\SK$.}
		\end{enumerate}	
\item $\Castf(\PK,j,v)$:
	on input the public key $\PK$, the voter index $j\in[N]$, and a vote $v$, do the following.
	\begin{enumerate}
			\item For all $l\in[3]$, choose randomness $r_l\from\zu^\lambda$ and compute $\Ct_{j,l}=\PKE.\Enc(\PKE.\PK_l,v;r_l)$.
		\item Consider the following relation $\Rencf$  in Fig.~\ref{fig:relationencrypt}. Run $\ProveEncf$ on input the statement
$(j,\Ct_1,\ldots,\Ct_3,\PKE.\PK_1,\ldots,\PKE.\PK_3, Z)$ and the witness $(v,r_1,\ldots,r_3)$ to compute a proof $\pi_j$. Output $\BT_j\defeq(\Ct_{j,1},\ldots,\CT_{j,3},\pi_j)$.

\medskip

\begin{figure}
  \begin{framed}
Relation $\Rencf(x,w)$:\\

Instance: $x\defeq(j,\Ct_1,\ldots,\Ct_3,\PKE.\PK_1,\ldots,\PKE.\PK_3, Z)$.\\

Witness : $w \defeq (m, r_1,\ldots,r_3, u)$, where the $r_l$'s are the randomness used to compute the ciphertexts $\CT_l$'s and $u$ is the randomness used to compute the commitment $Z$.\\

$\Rencf(x, w) = 1$ if and only if either of the following two conditions hold:\\

\begin{enumerate}
	\item{\bf Real mode.} All $3$ ciphertexts $(\Ct_1,\ldots,\Ct_3)$ encrypt the same string in $\M\cup\{\bot\}$.
		
		Precisely, for all $l\in[3]$, $\Ct_l=\PKE.\Enc(\PKE.\PK_l,m;r_l)$ and $m\in\M\cup\{\bot\}$.
		$$\text{OR}$$
	\item{\bf Trapdoor mode.} $Z$ is a commitment to $0$.

		Precisely, $Z=\Com(0;u)$.
\end{enumerate}
\end{framed}
\caption{Relation $\Rencf$}
\label{fig:relationencrypt}
\end{figure}


	\end{enumerate}

\item $\VerifyBallotf(\PK,j,\BT)$:
	on input the public key $\PK$,
	the voter index $j\in[N]$, and a ballot $\BT$,
	output $\VerifyEncf((j,\Ct_1,\ldots,\Ct_3,\PKE.\PK_1,\ldots,\PKE.\PK_3, Z),\pi)$.

\item $\EvalTallyf(\PK,\SK,\BT_1,\ldots,\BT_N)$:
	on input the public key $\PK$, the secret key $\SK$, and $N$ strings $(\BT_1,\ldots,\BT_N)$ that can be either ballots cast by a voter or the special symbol $\bot$, do the following.
	\begin{enumerate}
		\item For all $j\in[N]$, if $\VerifyBallotf(\PK,j,\BT_j)=\bot$, set $\BT_j=\bot$. If, for all $j\in[N]$, $\BT_j=\bot$,  then output $(y=\bot,\gamma=\bot)$.
		\item Else, for all $j\in[N],l\in[2]$,
				$$m_j^{l}=
				\begin{cases} \bot &\mbox{if } \BT_j=\bot, \\
					\bot &\mbox{if } \BT_j\neq\bot \wedge \PKE.\Dec(\Ct_{j,l},\PKE.\SK_{l})\notin\M,\\
					\PKE.\Dec(\Ct_{j,l},\PKE.\SK_l)	& \mbox{otherwise.}
				\end{cases}
				$$
\item For all $l\in[2]$, compute $y_l= F(m_{1,l},\ldots,m_{N,l}).$
\item If $y_1=y_2$ then set $y=y_1$.

\item Consider the following relation $\Rdecf$ in Fig.~\ref{fig:relationdecryptfull}. (The relation $\Rdecf$ is identical to the relation $\Rdec$ that is used in our weakly verifiable eVote. The only difference is that the ballots in the statement of $\Rdecf$ are replaced by $\bot$ if they are not accepted by the ballot verification algorithm. Henceforth, if the indices $(i_1,i_2)$ in the witness of the relation $\Rdec$ fulfill $i_1=1$ and $i_2=2$ (resp. $i_1\neq 1$ or $i_2\neq 2$), the statement or the proof is in real mode (resp. trapdoor mode).) Run $\ProveDecf$ on input the statement $(\BT_1,\allowbreak \ldots,\allowbreak \BT_N,\allowbreak \PKE.\PK_1, \allowbreak \ldots,\allowbreak \PKE.\PK_3,\allowbreak y)$ and the witness $(\PKE.\SK_1, \allowbreak \PKE.\SK_2, \allowbreak s_1,\allowbreak s_2, \allowbreak i_1 = 1, \allowbreak i_2 = 2)$ to compute a proof $\gamma$.
\item Output $(y, \allowbreak\gamma)$.


\begin{figure}

  \begin{framed}

Relation $\Rdecf(x,w)$:\\

Instance: $x\defeq(\BT_1,\ldots,\BT_N,\PKE.\PK_1,\ldots,\PKE.\PK_3, y).$ (Recall that a ballot is set to $\bot$ if either the corresponding voter did not cast her vote or her ballot is not accepted by the ballot verification algorithm.)\\

Witness: $w \defeq (\PKE.\SK_1',\PKE.\SK_2',s_1,s_2,i_1,i_2)$, where the $s_l$'s are the randomness used to generate the secret keys and public keys (which are known to the authority who set up the system).\\

$\Rdecf(x, w) = 1$ if and only if the following condition holds.\\

$2$ of the secret keys corresponding to indices $\PKE.\PK_{i_1}, \PKE.\PK_{i_2}$ are constructed using honestly generated public and secret key
		pairs and are equal to $\PKE.\SK_1',\PKE.\SK_2'$; and either $y=\bot$ or for all $l\in[2]$, $y=F(m_1^l,\ldots,m_N^l)$ and for all $j\in[N]$, if $\BT_j\neq\bot$ then for $l\in[2]$, $\PKE.\SK_{i_l}$ decrypts ciphertext $\Ct_{j,i_l}$ in $\BT_j$ to $m_j^{i_l}\in\M$; and for all $l\in[2]$, $m_j^l=\bot$ if either $\BT_j=\bot$ or $\PKE.\SK_{i_l}$ decrypts $\Ct_{j,i_l}$ to a string $\notin \M$.\\


		Precisely, $\Rdecf(x, w) = 1$ if and only if the following conditions hold. In the following, items (a) and (c) are not actually conditions that have to be checked but are steps needed to define  (note the use of ``$\defeq$'') the variables $\PKE.\PK_{i_l}$'s, $\PKE.\SK_{i_l}$'s and $m_j^{i_l}$'s that are used in the checks (b) and (d).\\

		\begin{enumerate}
			\item For all $l\in[2], (\PKE.\PK_{i_l},\PKE.\SK_{i_l})\defeq \PKE.\Setup(1^\lambda;s_l)$.
			\item For all $l\in[2], \PKE.\SK_{l}'=\PKE.\SK_{i_l}$.
			\item For all $j\in[N],l\in[2],$
				$$m_j^{i_l}\defeq
				\begin{cases} \bot &\mbox{if } \BT_j=\bot, \\
					\bot &\mbox{if } \BT_j\neq\bot \wedge \PKE.\Dec(\Ct_{j,i_l},\PKE.\SK_{i_l})\notin\M,\\
					\PKE.\Dec(\Ct_{j,i_l},\PKE.\SK_{i_l}) & \mbox{otherwise. }
				\end{cases}
				$$
			\item $(y=\bot)$ $\vee$ (for all $l\in[2]$, $y=F(m_1^{i_l},\ldots,m_N^{i_l})$).
		\end{enumerate}
		(Note that $\PKE.\SK_1'$ and $\PKE.\SK_2'$ do not necessarily have to correspond to the first two secret keys.)

\end{framed}
\caption{Relation $\Rdecf$}
\label{fig:relationdecryptfull}
\end{figure}

\end{enumerate}
\item $\VerifyTallyf(\PK,\BT_1,\ldots,\BT_N,y,\gamma)$: on input the public key $\PK$, $N$ strings that can be either ballots cast by a voter or the special symbol $\bot$, a tally $y$ and a proof $\gamma$ of tally correctness, do the following. If $y=\bot$ and all $\BT_j$'s are equal to $\bot$, output $\OK$. If $y=\bot$ but not all $\BT_j$'s are equal to $\bot$, output $\bot$.  Otherwise output the decision of $\VerifyDecf((\BT_1,\allowbreak\ldots,\allowbreak\BT_N,\allowbreak\PKE.\PK_1,\allowbreak\ldots,\allowbreak\PKE.\PK_3, \allowbreak y), \allowbreak \gamma)$, after having replaced $\BT_j$'s with $\bot$ when $\VerifyBallotf(\PK,\allowbreak j,\allowbreak\BT_j)\allowbreak =\allowbreak\bot$. Precisely, the algorithm does the following:
		\begin{enumerate}
			\item For all $j\in[N]$, if $\VerifyBallotf(\PK,j,\BT_j)=\bot$, set $\BT_j=\bot$.
			\item If $y\neq\bot$, then output $\VerifyDecf((\BT_1,\ldots,\BT_N,\PKE.\PK_1,\ldots,\PKE.\PK_3,y),\gamma)$.
			\item If $y=\bot$, then, if for all $j\in[N],\BT_j=\bot$, output $\OK$, else output $\bot$.
		
		\end{enumerate}

\end{itemize}
\end{definition}
Henceforth, for simplicity we omit the parameters of the scheme and we just write $\EVOTEF$.


\subsection{Correctness and (Full) Verifiability of the Construction}\label{sec:schemeverfull}

\paragraph{\bf Correctness.}  Condition (1) of (perfect) correctness of $\EVOTEF$ follows from the perfect correctness of the PKE scheme and the perfect completeness of $\NIWIDecf$ and $\NIWIEncf$.
Condition (2) follows analogously. We note the following.
For all  honestly computed $\PK$, $\PK=(\PK_1,\PK_2,\PK_3,Z)$ holds for some $\PK_1,\PK_2,\PK_3$ and $Z$. $Z$ is a commitment to $1$. Therefore, relation $\Rencf$ and the perfectly binding property of the commitment scheme imply that, if there exists a proof $\pi$ and a statement $x=(j,\Ct_1,\ldots,\Ct_3,\PK_1,\ldots,\allowbreak\PK_3,\allowbreak Z)$ such that $\VerifyBallotf$ accepts $(x,\allowbreak\pi)$, then it must be the case that $\Ct_1,\ldots,\Ct_3$ encrypt the same string in $\M\cup\{\bot\}$. For all $j\in[N]$, if $\BT_j$ is accepted by $\VerifyBallotf$, $\BT_j' = \BT_j$, else $\BT_j'= \bot$. Therefore, for all $\BT_1,\ldots,\BT_N$, if $(y,\gamma) =\EvalTallyf(\PK,\BT_1,\ldots,\BT_N)$, then $y=F(m_1,\allowbreak\ldots,\allowbreak m_N)$,
where, for all $j\in[N]$, if $\BT_j$ is accepted by $\VerifyBallotf$, $m_j$ is the string encrypted in the first two ciphertexts of $\BT_j$, else $m_j$ is $\bot$. Then, it is easy to see that $\VerifyTallyf(\PK,\BT_1,\ldots,\BT_N,y,\gamma)=\OK$.

\paragraph{\bf (Full) verifiability.}

\begin{theorem}\label{thm:fullver}
	For all $N>0$, all sets $\M,\Sigma\subset\zu^\star$, and all tally functions $F:(\M\cup\{\bot\})^N\rightarrow\Sigma\cup\{\bot\}$, if $\PKE$ is a perfectly correct PKE scheme with unique secret key (cf. Def.~\ref{def:pke}), $\Com$ is a PPT algorithm, and $\NIWIDecf$ and $\NIWIEncf$ are (one-message) NIWIs (cf. Def.~\ref{def:niwi}), for the relations $\Rdecf$ and $\Rencf$ respectively, then $\EVOTEF^{N,\M,\Sigma,F,\PKE,\Com,\NIWIEncf,\NIWIDecf}$  satisfies the (full) verifiability property (cf. Def.~\ref{def:evote}).
\end{theorem}

\begin{proof}
	We first prove that condition (1) of verifiability is satisfied.
	We have to prove that,
	for all $\PK \allowbreak \in \allowbreak \zu^\star$, and all $\BT_1,\allowbreak\ldots,\allowbreak\BT_N\in\zu^\star\cup \{\bot\}$
	such that, for all $j\in[N]$, either $\BT_j=\bot$ or $\VerifyBallotf(\PK,\allowbreak j,\allowbreak\BT_j)\allowbreak =\allowbreak\OK$,
	there exist $m_1,\ldots,m_N\in\M\cup\{\bot\}$ such that, for all $y,\gamma\in\zu^\star$,
	if $\VerifyTallyf(\PK,\allowbreak\BT_1,\allowbreak\ldots,\allowbreak\BT_N,\allowbreak y,\allowbreak \gamma)=1$ then  $y=F(m_1,\allowbreak\ldots,\allowbreak m_N)$.	Henceforth, w.l.o.g, we let $\PK$ and $\BT_1,\ldots,\BT_N$ be arbitrary strings such that, for all $j\in[N]$, either $\BT_j=\bot$ or $\VerifyBallotf(\PK,j,\BT_j)=\OK$.
	
	First, we prove the following claim.
	\begin{claim}
		Given $\PK$ and $(\BT_1,\ldots,\BT_N)$, for every two pairs $(y_0,\gamma_0)$ and $(y_1,\gamma_1)$,
		if $\VerifyTallyf(\PK,\allowbreak\BT_1,\allowbreak\ldots,\allowbreak\BT_N,\allowbreak y_0,\allowbreak\gamma_0)=\VerifyTallyf(\PK,\allowbreak\BT_1,\allowbreak\ldots,\allowbreak\BT_N,\allowbreak y_1,\allowbreak\gamma_1)\allowbreak=\allowbreak\OK$ then $y_0=y_1$.
\end{claim}
For every $(y_0,\gamma_0)$ and $(y_1,\gamma_1)$, we have two cases.
\begin{enumerate}
	\item{Either $y_0=\bot$ and $y_1\neq\bot$ or $y_1=\bot$ and $y_0\neq\bot$.}
		Suppose w.l.o.g. that $y_0=\bot$ and $y_1\neq\bot$. The other case (i.e., $y_1=\bot$ and $y_0\neq\bot$) is symmetrical.

		By construction, for all $(y,\gamma)$, it holds that (A)
		if $\BT_1=\cdots=\BT_N=\bot$, then
		$\VerifyTallyf(\PK,\BT_1,\ldots,\BT_N,y,\gamma)=\OK$ if and only if
		$y=\bot$ and (B)
		if, for some $j\in[N],\BT_j\neq\bot$, then
	       	$\VerifyTallyf(\PK,\BT_1,\ldots,\BT_N,\bot,\gamma)=\bot$.
	We now have two cases.
	\begin{enumerate}
		\item{$\BT_1=\cdots=\BT_N=\bot$.}
			Then we have that $\VerifyTallyf(\PK,\allowbreak\BT_1,\allowbreak\ldots,\allowbreak\BT_N,\allowbreak y_1,\allowbreak\gamma_1)\allowbreak=\allowbreak\OK$ and by (A) $y_1 \allowbreak = \allowbreak \bot$, which is a contradiction.
		\item{It is not the case that $\BT_1=\cdots=\BT_N=\bot$.} Then, by (B) we have that $\VerifyTallyf(\PK,\allowbreak\BT_1,\allowbreak\ldots,\allowbreak\BT_N,\allowbreak y_0,\allowbreak\gamma_0)\allowbreak=\allowbreak\bot$, which contradicts the fact that $(y_0,\gamma_0)$ is accepted.
	\end{enumerate}	
	
	\item{$y_0,y_1\neq\bot$.}

Let $y_0,\gamma_0,y_1,\gamma_1$ be arbitrary strings in $\zu^\star\cup\{\bot\}$ such that $y_0,y_1\neq\bot$.
Suppose that $\VerifyTallyf(\PK,\allowbreak\BT_1,\allowbreak\ldots,\allowbreak\BT_N,\allowbreak y_0,\allowbreak\gamma_0)\allowbreak =\allowbreak\VerifyTallyf(\PK,\allowbreak\BT_1,\allowbreak\ldots,\allowbreak\BT_N,\allowbreak y_1,\allowbreak\gamma_1)\allowbreak =\allowbreak\OK$.
	The perfect soundness of $\NIWIDecf$ implies that, for all $b\in\zu$, the proof $\gamma_b$ is computed on input some witness
	$(\PKE.\SK_1'^b,\PKE.\SK_2'^b,s_1^b,s_2^b,i_1^b, i_2^b)$.

By the pigeon principle, there exists an index $i^\star$ such that one of the following cases holds.
\begin{enumerate}
\item{$i^\star=i_1^0=i_2^1$.}
	For all $b\in\zu$, let $(m_1^{i^\star,b},\ldots,m_N^{i^\star,b})$ be the messages guaranteed by condition (iii) of relation $\Rdecf$ for proof $\gamma_b$.
	Condition (i) for proof $\gamma_0$ (resp. $\gamma_1$) implies that the secret key $\SK_1'^0$ (resp. $\SK_2'^1$) is honestly computed and thus, the unique secret key property and the fact that it fulfills $\PKE.\PK_{i_1^0}=\PKE.\PK_{i^\star}$ (resp. $\PKE.\PK_{i_2^1}=\PKE.\PK_{i^\star}$) imply that for all $j\in[N]$,
	$\PKE.\Dec(\Ct_{j,i^\star},\PKE.\SK_1'^0)=\PKE.\Dec(\Ct_{j,i^\star},\PKE.\SK_2'^1)$.
	
	Furthermore, condition (ii) and (iii) for proof $\gamma_0$ (resp. $\gamma_1$) imply that for all $j\in[N]$, either $m_j^{i^\star,0}=\bot$
	or $m_j^{i^\star,0}=\PKE.\Dec(\Ct_{j,i^\star},\PKE.\SK_1')\in\M$ (resp. either $m_j^{i^\star,1}=\bot$ or $m_j^{i^\star,1}=\PKE.\Dec(\Ct_{j,i^\star},\PKE.\SK_2'^1)\in\M$).
   	
	Hence, for all $j\in[N]$, $m_j^{i^\star,0}=m_j^{i^\star,1}\in\M\cup\{\bot\}$.
	Now, condition (iv) for proof $\gamma_0$ (resp. $\gamma_1$) implies that
	either $y_0=F(m_1^{i_1^0,0},\ldots,m_N^{i_1^0,0})$ or $y_0=\bot$
	(resp. either $y_1=F(m_1^{i_2^1,1},\ldots,m_N^{i_2^1,1})$ or $y_1=\bot$) and, as by hypothesis $y_0,y_1\neq\bot$, it holds that $y_0=y_1$.

\item{$i^\star=i_2^0=i_1^1$.}
	This case is identical to the first one, except that we replace $i_1^0$ with $i_2^0$ and $i_2^1$ with $i_1^1$.
\item{$i^\star=i_1^0=i_1^1$.} 	
	This case is identical to the first one, except that we replace $i_2^1$ with $i_1^1$.
\item{$i^\star=i_2^0=i_2^1$.} 	
	This case is identical to the first one, except that we replace $i_1^0$ with $i_2^0$.
\end{enumerate}
\end{enumerate}
In all cases,  if $\VerifyTallyf(\PK, \allowbreak\BT_1,\allowbreak\ldots,\allowbreak\BT_N,\allowbreak y_0,\allowbreak\gamma_0)\allowbreak=\allowbreak\VerifyTallyf(\PK,\allowbreak\BT_1,\allowbreak\ldots,\allowbreak\BT_N,\allowbreak y_1,\allowbreak\gamma_1)\allowbreak=\allowbreak\OK$ then $y_0\allowbreak =\allowbreak y_1$. In conclusion, the claim is proved.

From the previous claim, it follows that there exists a {\em unique} value $y^\star$ such that,
		for all $(y,\gamma)$ such that $y\neq\bot$,
		if $\VerifyTallyf(\PK,\BT_1,\allowbreak \ldots,\BT_N,y,\gamma)=\OK$
		then $y=y^\star$ (1). Moreover, it is easy to see that,
		for all $(y,\gamma)$,
		if $\VerifyTallyf(\PK,\allowbreak\BT_1,\allowbreak \ldots,\allowbreak\BT_N,\allowbreak y,\allowbreak \gamma)\allowbreak=\allowbreak\OK$, there exist messages $m_1,\allowbreak \ldots,\allowbreak m_N\in\M\cup\{\bot\}$ such that $y \allowbreak=\allowbreak F(m_1,\allowbreak \ldots,\allowbreak m_N)$ (2).

		Now, we have two mutually exclusive cases.
		\begin{itemize}
			\item For all $(y,\gamma)$ such that $y\neq\bot$,
					$\VerifyTallyf(\PK,\BT_1,\allowbreak \ldots,\BT_N,y,\gamma)=\bot$.
				Then, letting $m_1,\ldots,m_N$ in the statement of the theorem be arbitrary messages in $\M\cup\{\bot\}$, the statement is verified with respect to $\PK$ and $\BT_1,\ldots,\BT_N$.
			\item There exists $(y',\gamma)$ such that $y'\neq\bot$ and $\VerifyTallyf(\PK,\BT_1,\allowbreak \ldots,\BT_N,y',\allowbreak\gamma)\allowbreak=\allowbreak\OK$.
				In this case, (2) implies that there exist $m_1',\ldots,m_N'\in\M\cup\{\bot\}$ such that $y'=F(m_1',\ldots,m_N')$ (3).
				Hence, (1) and (3) together imply that $y^\star=F(m_1',\ldots,m_N')$ (4).

				Therefore, for all $(y,\gamma)$ such that $y\neq\bot$, if $\VerifyTallyf(\PK,\BT_1,\BT_N,y,\gamma)=\OK$ then (by (1)) $y=y^\star=$ (by (4)) $=F(m_1',\ldots,m_N')$.

				Then, for $m_1\defeq m_1',\ldots,m_N\defeq m_N'$, the statement of condition (1) of weak verifiability is verified with respect to
				$\PK$ and $\BT_1,\ldots,\BT_N$.

		\end{itemize}
		In both cases, for $m_1\defeq m_1',\ldots,m_N\defeq m_N'$, the statement of condition (1) of weak verifiability is verified with respect to $\PK$ and $\BT_1,\ldots,\BT_N$.

		As  $\PK$ and $\BT_1,\ldots,\BT_N$ are arbitrary strings, the statement of condition (1) of weak verifiability is proven.

	It is also easy to check that condition (2) of weak verifiability is satisfied. This follows straightforwardly from the perfect soundness of $\NIWIDecf$. Thanks to $\NIWIDecf$, the authority always proves that the public key of the PKE scheme is honestly generated. Therefore, by the perfect correctness of the PKE scheme, an honestly computed ballot for message $m$ for the $j$-th voter is decrypted to $m$ (because an honestly computed ballot, by definition, consists of three ciphertexts that encrypt the same message, and thus the value committed to in $Z$ is not relevant).  Consequently, if the tally $y$ is different from $\bot$ (i.e., if the evaluation of the tally function is equal for all indices), then $y$ has to be compatible with $m$ at index $j$ (cf. Def.~\ref{def:restriction}).

In essence, condition (2) is satisfied because the degree of freedom of the authority in creating a dishonest public key only allows it to set up the commitment dishonestly. This does not affect how honest ballots are decrypted and ``counted''.
\end{proof}
Note that, for the proof of the theorem above, the security of the commitment scheme $\Com$ is not needed, i.e., the theorem holds for any PPT algorithm $\Com$, even insecure ones.


\subsection{Privacy of the Construction}\label{sec:privacyfull}

\begin{theorem}\label{thm:privfull}
	For all $N>0$, all sets $\M,\Sigma\subset\zu^\star$, and all tally functions $F:(\M\cup\{\bot\})^N\rightarrow\Sigma\cup\{\bot\}$, if $\PKE$ is a perfectly correct PKE scheme with unique secret key (cf. Def. \ref{def:pke}), $\Com$ is a computationally hiding commitment scheme (cf. Def. \ref{def:com}), and $\NIWIDecf$ and $\NIWIEncf$  are (one-message) NIWIs (cf. Def. \ref{def:niwi}), respectively, for the relations $\Rdecf$ and $\Rencf$, then $\EVOTEF^{N,\M,\Sigma,F,\PKE,\Com,\NIWIEncf,\NIWIDecf}$ is \IND-Secure (cf. Def. \ref{def:evotepriv}).
\end{theorem}
	\begin{proof}
		Consider the following experiment $H_{\adv}^Z(1^\lambda)$ between a challenger and $\adv$ (henceforth, we often omit the parameters).
		\paragraph{{\bf Experiment $H^Z$.}} $H^Z$ is equal to the experiment $\SecGame^{N,\M,\Sigma,F,\EVOTEF}_\adv$ except that the challenger sets the commitment $Z$ in the public key to be a commitment to $0$ instead of $1$. We define the output of the experiment to be a bit that is $1$ if and only if all winning conditions are satisfied.
Then, consider the following claim.
		\begin{claim}\label{clm:expz}
			The probability $P_0$ that $\adv$ wins the experiment
			$\SecGame^{N,\M,\Sigma,F,\EVOTEF}_\adv$
			is negligibly different from the probability $P_1$ that $\adv$ wins game $H^Z$.
\end{claim}
\begin{proof}
Suppose towards a contradiction that the difference between $P_0$ and $P_1$ is some non-negligible function $\epsilon(\lambda)$.
We construct an adversary $\B$ that breaks the computationally hiding property of $\Com$ with non-negligible probability.

$\B$ receives as input a commitment $\com$ that is either a commitment to $0$ or to $1$.
For $l\in[3]$, $\B$ runs $\PKE.\Setup(1^\lambda)$ to compute $(\PKE.\PK_l,\PKE.\SK_l)$ and sets the public key $\PK = (\PKE.\PK_1,\ldots,\PKE.\PK_3,Z = \com)$. $\B$ follows the challenger of  $\SecGame^{N,\M,\Sigma,F,\EVOTEF}_\adv$ to compute the remaining messages that are sent to the adversary. Finally, $\B$ gets the output $b'$ from $\adv$. $\B$ outputs $1$ if and only if all winning conditions are satisfied.


By hypothesis, if $\com$ is a commitment to $1$, the probability that $\B$ outputs $1$ equals the probability that $\adv$ wins in
			$\SecGame^{N,\M,\Sigma,F,\EVOTEF}_\adv$,
and if $\com$ is a commitment to $0$, the probability that $\B$ outputs $1$ equals the probability that $\adv$ wins in $H^Z$. Thus, the advantage of $\B$ in breaking the computational hiding property of $\Com$ is $\epsilon(\lambda)$, which contradicts the assumption that the commitment scheme is computationally hiding.

	\end{proof}


	Before continuing with the proof, we would like to remark a subtle point. In the previous claim, we implicitly assumed that the adversary $\B$ is able to check all of the winning conditions efficiently.
	This is possible if $\M$ is efficiently enumerable and its cardinality, as well as the number of voters $N$, are {\em constant} in the security parameter. This could seem like resorting to ``complexity leveraging'' arguments. In fact, one could ask if our proof would break down if $N$ and $\M$ depend on the security parameter. However, the whole proof can be generalized to the case of $N$ and $|\M|$ polynomial in the security parameter by using the following observation. Let $A$ be the event that $\adv$ submits challenges that satisfy the winning condition. Then, if the probability that $\adv$ wins the $\SecGame^{N,\M,\Sigma,F,\EVOTEF}_\adv$ is non-negligible, then the event $A$ must occur with non-negligible probability and, conditioned on it, $\A$ wins with non-negligible probability as well. Therefore, the rest of the proof would follow analyzing the probability that $\adv$ wins in the next hybrid experiments conditioned under the occurrence of the event that, in such experiments, $\adv$ submit challenges satisfying the winning condition. As we will see now, a similar ``conditioning'' argument will be anyhow necessary for the rest of the proof.

	\ignore{Let $E^0$ be the event that in $\SecGame^{N,\M,\Sigma,F,\EVOTEF}_\adv$, $\adv$ submits as challenge
	two tuples $M_0\defeq(m_{0,1},\ldots,m_{0,N})$ and
	$M_1\defeq(m_{1,1},\ldots,m_{1,N})$, and a set $S\subset[N]$ such that there exists $j\in S$ such that $m_{0,j}=m_{1,j}$ and, letting $B\defeq m_{0,j}\defeq (\Ct_1,\ldots,\CT_3)$ (and supposing that it can be parsed in a such way), it holds that $\VerifyBallot(\PK,B)=\OK$ but there exist $i_1,i_2\in[3],i_1\neq i_2$ such that $\PKE.\Dec(\Ct_{i_1},\SK_{i_1})\neq\PKE.\Dec(\Ct_{i_2},\SK_{i_2})$. That is, the event that the adversary submits a message representing a possible dishonest ballot that passes the verification ballot test but such that the three ciphertexts of which it is made up cannot be decrypted to the same message $\in\M\cup\{\bot\}$.

	By the perfect soundness of $\NIWIEncf$ and the definition of relation $\Rencf$, it is straightforward to see that the probability that event $E^0$ occurs is $0$.
}
	Let $E^1$ be the event that, in experiment $H^Z,$ $\adv$ submits as challenge
	two tuples $M_0 = (m_{0,1},\ldots,m_{0,N})$ and
	$M_1 = (m_{1,1},\ldots,m_{1,N})$ and a set $S\subset[N]$ that fulfill the following condition: there exists $j\in S$ such that $m_{0,j}=m_{1,j}$ and, letting $B = m_{0,j} = (\Ct_1,\ldots,\CT_3)$ (suppose that $m_{0,j}$ can be parsed that way), it holds that $\VerifyBallotf(\PK,j,B)=\OK$ but there exist $i_1,i_2\in[3],i_1\neq i_2$ such that $\PKE.\Dec(\Ct_{i_1},\SK_{i_1})\neq\PKE.\Dec(\Ct_{i_2},\SK_{i_2})$.

	\begin{claim}\label{clm:eventeone}
			The probability that $E^1$ occurs is negligible.
\end{claim}
\begin{proof}
Suppose towards a contradiction that the probability of occurrence of $E^1$ be some non-negligible function $\epsilon(\lambda)$.
We construct an adversary $\B$ that breaks the computationally hiding property of $\Com$ with non-negligible probability.

$\B$ receives as input a commitment $\com$ that is either a commitment to $0$ or to $1$.
For $l\in[3]$, $\B$ runs $\PKE.\Setup(1^\lambda)$ to compute $(\PKE.\PK_l,\PKE.\SK_l)$ and sets the public key $\PK = (\PKE.\PK_1,\ldots,\PKE.\PK_3,Z = \com)$.  $\B$ follows the challenger of  $\SecGame^{N,\M,\Sigma,F,\EVOTEF}_\adv$ to compute the remaining messages that are sent to the adversary. $\B$ receives two tuples $M_0 = (m_{0,1},\ldots,m_{0,N})$ and
	$M_1 = (m_{1,1},\ldots,m_{1,N})$  and a set $S\subset[N]$ from the adversary.

	
	For all $j\in S$, $\B$ checks whether the following conditions are all satisfied: $m_{0,j}=m_{1,j}$ and, after setting $B = m_{0,j}$, $B$ can be parsed as $(\Ct_1,\ldots,\CT_3)$ and it holds that $\VerifyBallotf(\PK,j,B)=\OK$ but there exist $i_1,i_2\in[3],i_1\neq i_2$ such that $\PKE.\Dec(\Ct_{i_1},\SK_{i_1})\neq\PKE.\Dec(\Ct_{i_2},\SK_{i_2})$.
If for some $j\in S$ the conditions are satisfied, $\B$ outputs $0$, otherwise it outputs $1$.

If $\com$ is a commitment to $1$, the perfect soundness of $\NIWIEncf$ and the definition of relation $\Rencf$ guarantee that the conditions above are never satisfied for any $j\in S$. Therefore,  if $\com$ is a commitment to $1$ $\B$ outputs $1$ with probability $1$.

On the other hand, if $\com$ is a commitment to $0$, the probability that the conditions are satisfied for some $j\in[S]$ equals the probability of $E^1$. Therfore, $\B$ outputs $0$ with probability $\epsilon$ and $1$ with probability $1-\epsilon$. In conclusion, the advantage of $\B$ in breaking the computationally hiding property of $\Com$ is $\epsilon(\lambda)$, which contradicts the assumption that the commitment scheme is computationally hiding.
	\end{proof}

From Claim~\ref{clm:expz} and Claim~\ref{clm:eventeone}, we now know that, for some negligible function $\negl(\cdot)$, the following equations hold:
\begin{equation}
	\left|\Prob{\SecGame=1}-\Prob{H^Z=1}\right|\le \negl(\lambda),
\end{equation}
\begin{equation} \label{eqn:one}
\Prob{E^1}\le \negl(\lambda),
\end{equation}
\begin{multline} \label{eqn:three}
\Prob{H^Z=1}=\Prob{H^Z=1|E^1}\Prob{E^1}+\Prob{H^Z=1|\bar E^1}\Prob{\bar E^1} \le \\ \negl + \Prob{H^Z=1|\bar E^1}(1-\negl).
\end{multline}

(Here and henceforth, we omit the parameters, but it is meant that the experiments are parameterized by $\lambda$ and $\negl(\cdot)$.)

Thus, to show that $\Prob{\Priv=1}$ equals $1/2$ plus a negligible quantity, it is sufficient to show that $\Prob{H^Z=1|\bar E^1}$ equals $1/2$ plus a negligible quantity. We prove the latter by means of a series of hybrid experiments. The reader could still refer to Table~\ref{table} for a pictorial explanation of the hybrid experiments. However, the experiments in the table, though conceptually very similar, correspond to the security reduction for the weakly verifiable eVote. Moreover, in the following we analyze the behavior of the adversary conditioned on the occurrence of the event $\bar E^1$.
		\paragraph{{\bf Hybrid $H_1$.}} Experiment $H_1$ is equal to the experiment $H^Z$ except that the challenger sets $b = 0$.
	
\paragraph{{\bf Hybrid $H_2^k,$} for $k=0,\ldots,N$.}
For all $k=0,\ldots,N$, experiment $H_2^k$ is identical to experiment $H_1$ except that, for all $j=1,\ldots,k$ such that $j\notin S$, the challenger computes $\Ct_{k,3}$ on input $m_{1,k}$.
Note that $H_2^0$ is identical to $H_1$.

\begin{claim}\label{clm:hybridtwofull}
	For all $k=1,\ldots,N$, $\left|\Prob{H_2^{k-1}=1|\bar E^1}-\Prob{H_2^k=1|\bar E^1}\right|$ is negligible.
\end{claim}
\begin{proof}
	Suppose toward a contradiction that the difference between such probabilities is non-negligible function $\epsilon(\lambda)$.
	We construct an adversary $\B$ that has advantage at most $\epsilon(\lambda)$ against the IND-CPA security of $\PKE$.

	$\B$ receives from the challenger of IND-CPA a public key $\pk$ and sets $\PK_3 = \pk$. For $l\in[2]$, $\B$ runs $\PKE.\Setup$ to compute $(\PKE.\PK_l,\PKE.\SK_l)$, computes $Z\from\Com(0)$ and runs $\A$ on input $\PK = (\PKE.\PK_1,\PKE.\PK_2,\PKE.\PK_3,Z)$.

	$\A$ outputs two tuples $(m_{0,1},\ldots,m_{0,N})$ and $(m_{1,1},\ldots,m_{1,N})$ and a set $S$. If $k\in S$, $\B$ sends $(0,0)$ as its pair of challenge messages to the IND-CPA challenger, which returns the challenge ciphertext $\ct^\star$ to $\B$.
	If $k\notin S$, $\B$ sends $(m_{0,k},m_{1,k})$ as its pair of challenge messages to the IND-CPA challenger, which returns  the challenge ciphertext $\ct^\star$ to $\B$.

	If $k\in S$, $\B$ sets $\BT_j$ as the challenger in the real experiment would do, else $\B$ computes $\BT_k = (\Ct_{k,1},\Ct_{k,2},\ct^\star)$ by computing $\Ct_{k,1}$ and $\Ct_{k,2}$ on input $m_{0,j}$.
	For all $j\in[N] (j\neq k)$, $\B$ computes the ballots $\BT_j$  exactly as the challenger in both experiments would do.
	$\B$ computes $y$ using $\EvalTallyf$ and uses the $2$ secret keys $\PKE.\SK_1,\PKE.\SK_2$ to compute a proof $\gamma$ exactly as the challenger in both experiments would do.
	$\B$ sends $\A$ the computed ballots along with $(y,\gamma)$ and returns the output of $\A$.
	
	It is easy to see that, if $\ct^\star$ is an encryption of $m_{0,k}$ and if $k\notin S$, then $\B$ simulates experiment $H_2^{k-1}$ and if
	$\ct^\star$ is an encryption of $m_{1,k}$ and $k\notin S$, then $\B$ simulates experiment $H_2^{k}$.
	If $k\in S$ the advantage of $\adv$ is $0$.

	Therefore, $\B$ has non-negligible probability of winning the IND-CPA game, which contradicts the assumption that the PKE scheme fulfills the IND-CPA property.
\end{proof}

\paragraph{{\bf Hybrid $H_3$}.} Experiment $H_3$ is identical to experiment $H_2^N$ except that the challenger computes the proof $\gamma$ on input a witness that contains indices $(1,3)$ and secret keys $(\SK_1,\SK_3)$ (precisely, the witness contains the randomness used to compute those secret keys, but henceforth, for simplicity, we omit this detail).
\begin{claim}\label{clm:hybridthreefull}
	$\left|\Prob{H_2^{N}=1|\bar E^1}-\Prob{H_3=1|\bar E^1}\right|$ is negligible.
\end{claim}
\begin{proof}
	The proof follows from the WI property of $\NIWIDecf$. We observe that both the randomness used to compute $(\SK_1,\SK_2)$ and the randomness used to compute $(\SK_1,\SK_3)$ constitute valid witnesses for the statement $(\BT_1, \allowbreak \ldots, \allowbreak \BT_N, \allowbreak \PKE.\PK_1, \allowbreak \ldots, \allowbreak \PKE.\PK_3, \allowbreak y)$. Additionally, we observe that, if event $\bar E^1$ occurs, any ballot in the set $S$ is in both experiments either replaced by $\bot$, if $\VerifyBallotf$ refuses it, or decrypted to the same value. Consequently, the tally is identical in both experiments.
	\end{proof}

	\paragraph{{\bf Hybrid $H_4^k,$} for $k=0,\ldots,N$.}
	For all $k=0,\ldots,N$, experiment $H_4^k$ is identical to experiment $H_3$ except that, for all $j=1,\ldots,k$ such that $j\notin S$, the challenger computes $\Ct_{k,2}$ on input $m_{1,k}$.
Note that $H_4^0$ is identical to $H_3$.

\begin{claim}
	For all $k=1,\ldots,N$, $\left|\Prob{H_4^{k-1}=1|\bar E^1}-\Prob{H_4^k=1|\bar E^1}\right|$ is negligible.
\end{claim}
\begin{proof}
	The proof is identical to the one for Claim~\ref{clm:hybridtwofull} except that the third index and the second index are swapped.
	\end{proof}

	\paragraph{{\bf Hybrid $H_5$}.} Experiment $H_5$ is identical to experiment $H_4^N$ except that the challenger computes the proof $\gamma$ on input a witness that contains indices $(2,3)$ and secret keys $(\SK_2,\SK_3)$.
\begin{claim}
	$\left|\Prob{H_4^{N}=1|\bar E^1}-\Prob{H_5=1|\bar E^1}\right|$ is negligible.
\end{claim}
\begin{proof}
	This follows straightforwardly from the WI property of $\NIWIDecf$. We observe that both the randomness used to compute $(\SK_1,\SK_3)$ and the randomness used to compute $(\SK_2,\SK_3)$ constitute valid witnesses for the statement  $(\BT_1, \allowbreak \ldots, \allowbreak \BT_N, \allowbreak \PKE.\PK_1, \allowbreak \ldots, \allowbreak \PKE.\PK_3, \allowbreak y)$. Additionally, we observe that, if event $\bar E^1$ occurs, any ballot in the set $S$ is in both experiments either replaced by $\bot$, if $\VerifyBallotf$ refuses it, or decrypted to the same value. Consequently, the tally is identical in both experiments.
	\end{proof}
	\paragraph{{\bf Hybrid $H_6^k,$} for $k=0,\ldots,N$.}
For all $k=0,\ldots,N$, experiment $H_6^k$ is identical to experiment $H_5$ except that, for all $j=1,\ldots,k$ such that $j\notin S$, the challenger computes $\Ct_{k,1}$ on input $m_{1,k}$.
Note that $H_6^0$ is identical to $H_5$.
\begin{claim}
	For all $k=1,\ldots,N$, $\left|\Prob{H_6^{k-1}=1|\bar E^1}-\Prob{H_6^k=1|\bar E^1}\right|$ is negligible.
\end{claim}
\begin{proof}
	The proof is identical to the one for Claim~\ref{clm:hybridtwofull} except that the third index and the first index are swapped.
	\end{proof}
	\paragraph{{\bf Hybrid $H_7$}.} Experiment $H_7$ is identical to experiment $H_6^N$ except that the challenger sets $b=1$ (so that the winning condition be computed differently) and computes the proof $\gamma$ on input a witness that contains indices $(1,2)$ and secret keys $(\SK_1,\SK_2)$.
	\begin{claim}\label{clm:lasthybrid}
	$\left|\Prob{H_6^{N}=1|\bar E^1}-\Prob{H_7=0|\bar E^1}\right|$ is negligible.
\end{claim}
\begin{proof}
	The proof follows straightforwardly from the WI property of $\NIWIDecf$. We observe that both the randomness used to compute $(\SK_1,\SK_2)$ and the randomness used to compute $(\SK_2,\SK_3)$ constitute valid witnesses for the statement  $(\BT_1, \allowbreak \ldots, \allowbreak \BT_N, \allowbreak \PKE.\PK_1, \allowbreak \ldots, \allowbreak \PKE.\PK_3, \allowbreak y)$. Additionally, we observe that, if event $\bar E^1$ occurs, any ballot in the set $S$ is in both experiments either replaced by $\bot$, if $\VerifyBallotf$ refuses it, or decrypted to the same value. Consequently, the tally is identical in both experiments.

	Note that according to the proof received, an adversary against NIWI can emulate experiment $H_6^N$ or $H_7$, and return the output of $\adv$.
      In the first case, the probability that $\adv$ outputs $0$ is exactly $\Prob{H_6^{N}=1|\bar E^1}$ because the winning condition is computed with respect to $b=0$, whereas in the second case it is $\Prob{H_7=0|\bar E^1}$ because the winning condition is computed with respect to $b=1$.
	\end{proof}

	Now, consider Equation~\ref{eqn:four} in Fig.~\ref{fig:equation}.
\begin{figure}
\begin{framed}
\tiny
	\begin{equation}\label{eqn:four}
	\begin{split}
	\Prob{H^Z=1|\bar E^1}=\\
	\Prob{H^Z=1|\bar E^1 \wedge b=0}\Prob{b=0}+\Prob{H^Z=1|\bar E^1 \wedge b=1}\Prob{b=1}=&\\
	=1/2\cdot\left(\Prob{H^Z=1|\bar E^1 \wedge b=0}+\Prob{H^Z=1|\bar E^1 \wedge b=1}\right)=&\\
	(\text{since $H_1$ is identically distributed to $H^Z$ with bit $b=0$ and $H_7$ to $H^Z$ with $b=1$})&\\
	=1/2\cdot\left(\Prob{H_1=1|\bar E^1}+\Prob{H_7=1|\bar E^1}\right)=&\\
	=1/2+1/2\cdot\left(\Prob{H_1=1|\bar E^1}-\Prob{H_7=0|\bar E^1}\right)=\\
	(\text{since $H_1$ (resp. $H_3,H_5$) is identically distributed to $H_2^0$ (resp. $H_4^0, H_6^0$)})&\\
	=1/2+1/2\cdot(
		\sum_{k=0}^{N-1}(\Prob{H_2^k=1|\bar E^1}-\Prob{H_2^{k+1}=1|\bar E^1})+
		(\Prob{H_2^N=1|\bar E^1}-\Prob{H_4^0=1|\bar E^1})+&\\
		\sum_{k=0}^{N-1}(\Prob{H_4^k=1|\bar E^1}-\Prob{H_4^{k+1}=1|\bar E^1})
		(\Prob{H_4^N=1|\bar E^1}-\Prob{H_6^0=1|\bar E^1})+&\\
		\sum_{k=0}^{N-1}(\Prob{H_6^k=1|\bar E^1}-\Prob{H_6^{k+1}=1|\bar E^1})
		(\Prob{H_6^N=1|\bar E^1}-\Prob{H_7=0|\bar E^1}))\le&\\
	1\le/2+1/2\cdot|(
		\sum_{k=0}^{N-1}(\Prob{H_2^k=1|\bar E^1}-\Prob{H_2^{k+1}=1|\bar E^1})+
		(\Prob{H_2^N=1|\bar E^1}-\Prob{H_4^0=1|\bar E^1})+&\\
		\sum_{k=0}^{N-1}(\Prob{H_4^k=1|\bar E^1}-\Prob{H_4^{k+1}=1|\bar E^1})
		(\Prob{H_4^N=1|\bar E^1}-\Prob{H_6^0=1|\bar E^1})+&\\
		\sum_{k=0}^{N-1}(\Prob{H_6^k=1|\bar E^1}-\Prob{H_6^{k+1}=1|\bar E^1})
		(\Prob{H_6^N=1|\bar E^1}-\Prob{H_7=0|\bar E^1}))|\le&\\
		(\text{by the triangle inequality})&\\
		\le1/2+1/2\cdot(
		\sum_{k=0}^{N-1}|\Prob{H_2^k=1|\bar E^1}-\Prob{H_2^{k+1}=1|\bar E^1}|+
		|(\Prob{H_2^N=1|\bar E^1}-\Prob{H_4^0=1|\bar E^1})|+&\\
		\sum_{k=0}^{N-1}|\Prob{H_4^k=1|\bar E^1}-\Prob{H_4^{k+1}=1|\bar E^1}|
		|\Prob{H_4^N=1|\bar E^1}-\Prob{H_6^0=1|\bar E^1}|+&\\
		\sum_{k=0}^{N-1}|\Prob{H_6^k=1|\bar E^1}-\Prob{H_6^{k+1}=1|\bar E^1}|
		|\Prob{H_6^N=1|\bar E^1}-\Prob{H_7=0|\bar E^1}|)\le&\\
	(\text{by Claims \ref{clm:hybridtwofull}\ -\ \ref{clm:lasthybrid}})&\\
	\le3k\cdot\negl,
\text{where $\negl$ is the sum of the negligible functions guaranteed by Claims \ref{clm:hybridtwofull}\ -\ \ref{clm:lasthybrid}}.
\end{split}
\end{equation}
\end{framed}
\caption{Equation~\ref{eqn:four}}
\label{fig:equation}
\end{figure}
Finally, Claim \ref{clm:expz} and equations \ref{eqn:one},\ref{eqn:three} and \ref{eqn:four} imply that $\Prob{\SecGame=1}\le \nu$ for some negligible function $\nu$ and the theorem is proven.
	\end{proof}

\begin{corollary} \label{cor:full}
	\ifnum\fullversion=1
	If the Decision Linear assumption (see Section \ref{app:dlin}) holds, then there exists a (fully) verifiable eVote.
	\else
	If the Decision Linear assumption (see Appendix \ref{app:dlin}) holds, then there exists a (fully) verifiable eVote.
	\fi
	\end{corollary}
	\begin{proof}
		Boneh \etal\ \cite{C:BonBoySha04} show the existence of a PKE with perfect correctness and unique secret key that fulfills the IND-CPA property under the Decision Linear assumption. Groth \etal\ \cite{C:GroOstSah06} show the existence of (one-message) NIWI (with perfect soundness) for all languages in $\NP$ and of statistically binding commitments. Both constructions are secure under the Decision Linear assumption.	Then, because Theorem~\ref{thm:fullver} and Theorem~\ref{thm:privfull} are proven, the corollary follows.
	\end{proof}




\section{Future Directions}\label{sec:future}
Our work opens up new directions in e-voting and generally in cryptography. We discuss some of them.
\begin{itemize}
\item{\bf Efficiency.} In our work, in order to compute the NIWI proofs of Groth \etal\ \cite{C:GroOstSah06} for $\CircuitSat$, we need to represent the computation as a Boolean circuit and, though this can be done in polynomial time, it can be inefficient in practice. An important objective is to sidestep the reduction to circuits by employing a more direct approach. A possibility would be to explore the achievability of our results from variants of Groth-Sahai NIWIs~\cite{EC:GroSah08}. The NIWI of Groth-Sahai, as it stands, is formulated in the CRS model but it is worthy to study in which settings it can be instantiated without CRS.

Another important direction is to improve the efficiency of verification. It would be desirable that the cost for  verifiers be sub-linear in the number of voters. The verifiability guarantees attained would then be computational but hopefully it could be possible to avoid trust assumptions. A possibility would be to employ variants of succinct arguments (see~\cite{PHD:Bitansky14} for a survey).
\item{\bf Receipt-freeness.} Perfect verifiability and perfect correctness seem incompatible with receipt-freeness~\cite{STOC:BenTui94,EC:SakKil95,AC:MicHor96a,C:MorNao06,DelKreRya09,BeleniosRF}, but we think that it should be possible to define a statistical variant of verifiability that could coexist with some form of receipt-freeness. Another possibility could be to resort to some voting server trusted for receipt-freeness but not for privacy that re-randomizes the ballots, as done in BeleniosRF of Chaidos, Cortier, Fuchsbauer and Galindo \cite{BeleniosRF}.
\item{\bf Other applications of our techniques.} We think that our techniques could be of wide applicability to other settings. For instance, Camenisch and Shoup~\cite{C:CamSho03} put forth the concept of verifiable encryption (that in some sense could be also viewed as a special case of verifiable functional encryption~\cite{EPRINT:BGJS16}) and present numerous applications of it, such as key escrow, optimistic fair exchange, publicly verifiable secret and signature sharing, universally composable commitments, group signatures, and confirmer signatures. We believe that our techniques can be employed profitably to improve their results with the aim of removing the need of trust assumptions.
\end{itemize}


\fi
\section{Acknowledgments}

Vincenzo Iovino thanks Saikrishna Badrinarayanan and Aayush Jain for helpful discussions about verifiability, and Peter B. R\o nne thanks Steve Kremer for suggestions.  

Vincenzo Iovino is supported by the Luxembourg National Research Fund (FNR grant no. 7884937). Further, this work is also supported by the INTER-Sequoia project from the Luxembourg National Research Fund, which is joint with the ANR project SEQUOIA ANR-14-CE28-0030-01.


\newcommand{\etalchar}[1]{$^{#1}$}


\begin{thebibliography}{BDSG{\etalchar{+}}13}

\bibitem[Adi08]{Helios}
Ben Adida.
\newblock Helios: Web-based open-audit voting.
\newblock In {\em USENIX Security Symposium}, volume~17, pages 335--348, 2008.

\bibitem[BBS04]{C:BonBoySha04}
Dan Boneh, Xavier Boyen, and Hovav Shacham.
\newblock Short group signatures.
\newblock In Matthew Franklin, editor, {\em Advances in Cryptology --
  {CRYPTO}~2004}, volume 3152 of {\em Lecture Notes in Computer Science}, pages
  41--55. Springer, August 2004.

\bibitem[BCG{\etalchar{+}}15]{BCGPW15}
David Bernhard, V{\'e}ronique Cortier, David Galindo, Olivier Pereira, and
  Bogdan Warinschi.
\newblock Sok: A comprehensive analysis of game-based ballot privacy
  definitions.
\newblock In {\em 2015 IEEE Symposium on Security and Privacy}, pages 499--516.
  IEEE, 2015.

\bibitem[BDPA11]{Keccak}
G.~Bertoni, J.~Daemen, M.~Peeters, and G.~Van Assche.
\newblock The {{\sc Keccak}} reference, 2011.
\newblock \url{http://keccak.noekeon.org/}.

\bibitem[BDSG{\etalchar{+}}13]{TCC:BDGJKLW13}
Nir Bitansky, Dana Dachman-Soled, Sanjam Garg, Abhishek Jain, Yael~Tauman
  Kalai, Adriana L{\'o}pez-Alt, and Daniel Wichs.
\newblock Why ``fiat-shamir for proofs'' lacks a proof.
\newblock In {\em Theory of Cryptography: 10th Theory of Cryptography
  Conference, TCC 2013, Tokyo, Japan, March 3-6, 2013.}, pages 182--201.
  Springer, 2013.

\bibitem[Ben87]{thesis:Benaloh}
J.~Benaloh.
\newblock {\em Verifiable secret-ballot elections}.
\newblock PhD thesis, Yale University, 1987.

\bibitem[BF03]{BonFra03}
Dan Boneh and Matthew~K. Franklin.
\newblock Identity based encryption from the {Weil} pairing.
\newblock {\em {SIAM} Journal on Computing}, 32(3):586--615, 2003.

\bibitem[BFM88]{STOC:BluFelMic88}
Manuel Blum, Paul Feldman, and Silvio Micali.
\newblock Non-interactive zero-knowledge and its applications (extended
  abstract).
\newblock In {\em 20th Annual {ACM} Symposium on Theory of Computing}, pages
  103--112. {ACM} Press, May 1988.

\bibitem[BFS16]{EPRINT:BelFucSca16}
Mihir Bellare, Georg Fuchsbauer, and Alessandra Scafuro.
\newblock {NIZKs} with an untrusted {CRS}: Security in the face of parameter
  subversion.
\newblock Cryptology ePrint Archive, Report 2016/372, 2016.
\newblock \url{http://eprint.iacr.org/2016/372} {. To appear in ASIACRYPT
  2016}.

\bibitem[BGJS16]{EPRINT:BGJS16}
Saikrishna Badrinarayanan, Vipul Goyal, Aayush Jain, and Amit Sahai.
\newblock Verifiable functional encryption.
\newblock Cryptology ePrint Archive, Report 2016/629, 2016.
\newblock \url{http://eprint.iacr.org/2016/629}. To appear in ASIACRYPT 2016.

\bibitem[Bit14]{PHD:Bitansky14}
Nir Bitansky.
\newblock {\em Getting inside the Adversary’s Head: New Directions in
  Non-Black-Box Knowledge Extraction}.
\newblock PhD thesis, Tel Aviv University, 2014.

\bibitem[BOV03]{C:BarOngVad03}
Boaz Barak, Shien~Jin Ong, and Salil~P. Vadhan.
\newblock Derandomization in cryptography.
\newblock In Dan Boneh, editor, {\em Advances in Cryptology -- {CRYPTO}~2003},
  volume 2729 of {\em Lecture Notes in Computer Science}, pages 299--315.
  Springer, August 2003.

\bibitem[BP15]{TCC:BitPan15}
Nir Bitansky and Omer Paneth.
\newblock Zaps and non-interactive witness indistinguishability from
  indistinguishability obfuscation.
\newblock In {\em Theory of Cryptography Conference}, pages 401--427. Springer,
  2015.

\bibitem[BR93]{CCS:BelRog93}
Mihir Bellare and Phillip Rogaway.
\newblock Random oracles are practical: {A} paradigm for designing efficient
  protocols.
\newblock In V.~Ashby, editor, {\em ACM CCS 93: 1st Conference on Computer and
  Communications Security}, pages 62--73. {ACM} Press, November 1993.

\bibitem[BSW11]{TCC:BonSahWat11}
Dan Boneh, Amit Sahai, and Brent Waters.
\newblock Functional encryption: Definitions and challenges.
\newblock In Yuval Ishai, editor, {\em TCC~2011: 8th Theory of Cryptography
  Conference}, volume 6597 of {\em Lecture Notes in Computer Science}, pages
  253--273. Springer, March 2011.

\bibitem[BT94]{STOC:BenTui94}
Josh~Cohen Benaloh and Dwight Tuinstra.
\newblock Receipt-free secret-ballot elections (extended abstract).
\newblock In {\em 26th Annual {ACM} Symposium on Theory of Computing}, pages
  544--553. {ACM} Press, May 1994.

\bibitem[CCC{\etalchar{+}}09]{DBLP:journals/tifs/ChaumCCEPRRSSV09}
David Chaum, Richard Carback, Jeremy Clark, Aleksander Essex, Stefan
  Popoveniuc, Ronald~L. Rivest, Peter Y.~A. Ryan, Emily Shen, Alan~T. Sherman,
  and Poorvi~L. Vora.
\newblock Scantegrity {II:} end-to-end verifiability by voters of optical scan
  elections through confirmation codes.
\newblock {\em {IEEE} Trans. Information Forensics and Security},
  4(4):611--627, 2009.

\bibitem[CCFG15]{BeleniosRF}
Pyrros Chaidos, V{\'e}ronique Cortier, Georg Fuchsbauer, and David Galindo.
\newblock Beleniosrf: A non-interactive receipt-free electronic voting scheme.
\newblock Cryptology ePrint Archive, Report 2015/629, 2015.
\newblock \url{http://eprint.iacr.org/2015/629}. To appear in ACM CCS 2016.

\bibitem[CG15]{PKC:ChaGro15}
Pyrros Chaidos and Jens Groth.
\newblock Making sigma-protocols non-interactive without random oracles.
\newblock In {\em Public-Key Cryptography - {PKC} 2015 - 18th {IACR}
  International Conference on Practice and Theory in Public-Key Cryptography,
  Gaithersburg, MD, USA, March 30 - April 1, 2015, Proceedings}, pages
  650--670, 2015.

\bibitem[CGGI14]{ESORICS:CGGI14}
V{\'e}ronique Cortier, David Galindo, St{\'e}phane Glondu, and Malika
  Izabach{\`e}ne.
\newblock Election verifiability for helios under weaker trust assumptions.
\newblock In Miroslaw Kutylowski and Jaideep Vaidya, editors, {\em
  ESORICS~2014: 19th European Symposium on Research in Computer Security, Part
  II}, volume 8713 of {\em Lecture Notes in Computer Science}, pages 327--344.
  Springer, September 2014.

\bibitem[CGH98]{STOC:CanGolHal98}
Ran Canetti, Oded Goldreich, and Shai Halevi.
\newblock The random oracle methodology, revisited (preliminary version).
\newblock In {\em 30th Annual {ACM} Symposium on Theory of Computing}, pages
  209--218. {ACM} Press, May 1998.

\bibitem[CGK{\etalchar{+}}16]{EPRINT:CGKMT16}
V{\'e}ronique Cortier, David Galindo, Ralf Kuesters, Johannes Mueller, and
  Tomasz Truderung.
\newblock Verifiability notions for e-voting protocols.
\newblock Cryptology ePrint Archive, Report 2016/287, 2016.
\newblock \url{http://eprint.iacr.org/2016/287}.

\bibitem[CGS97]{EC:CraGenSch97}
Ronald Cramer, Rosario Gennaro, and Berry Schoenmakers.
\newblock A secure and optimally efficient multi-authority election scheme.
\newblock In Walter Fumy, editor, {\em Advances in Cryptology --
  {EUROCRYPT}'97}, volume 1233 of {\em Lecture Notes in Computer Science},
  pages 103--118. Springer, May 1997.

\bibitem[Cha81]{CACM:Cha81}
David~L. Chaum.
\newblock Untraceable electronic mail, return addresses, and digital
  pseudonyms.
\newblock {\em Communications of the ACM}, 24(2):84--90, 1981.

\bibitem[CHK04]{EC:CanHalKat04}
Ran Canetti, Shai Halevi, and Jonathan Katz.
\newblock Chosen-ciphertext security from identity-based encryption.
\newblock In Christian Cachin and Jan Camenisch, editors, {\em Advances in
  Cryptology -- {EUROCRYPT}~2004}, volume 3027 of {\em Lecture Notes in
  Computer Science}, pages 207--222. Springer, May 2004.

\bibitem[CKLM13]{PKC:CKLM13}
Melissa Chase, Markulf Kohlweiss, Anna Lysyanskaya, and Sarah Meiklejohn.
\newblock Verifiable elections that scale for free.
\newblock In Kaoru Kurosawa and Goichiro Hanaoka, editors, {\em PKC~2013: 16th
  International Workshop on Theory and Practice in Public Key Cryptography},
  volume 7778 of {\em Lecture Notes in Computer Science}, pages 479--496.
  Springer, February~/~March 2013.

\bibitem[CPSV16]{TCC:CPSV16}
Michele Ciampi, Giuseppe Persiano, Luisa Siniscalchi, and Ivan Visconti.
\newblock A transform for {NIZK} almost as efficient and general as the
  fiat-shamir transform without programmable random oracles.
\newblock In {\em Theory of Cryptography - 13th International Conference, {TCC}
  2016-A, Tel Aviv, Israel, January 10-13, 2016, Proceedings, Part {II}}, pages
  83--111, 2016.

\bibitem[CS03a]{C:CamSho03}
Jan Camenisch and Victor Shoup.
\newblock Practical verifiable encryption and decryption of discrete
  logarithms.
\newblock In Dan Boneh, editor, {\em Advances in Cryptology -- {CRYPTO}~2003},
  volume 2729 of {\em Lecture Notes in Computer Science}, pages 126--144.
  Springer, August 2003.

\bibitem[CS03b]{CraSho03}
Ronald Cramer and Victor Shoup.
\newblock Design and analysis of practical public-key encryption schemes secure
  against adaptive chosen ciphertext attack.
\newblock {\em {SIAM} Journal on Computing}, 33(1):167--226, 2003.

\bibitem[CS10]{EPRINT:CorSmy10}
Veronique Cortier and Ben Smyth.
\newblock Attacking and fixing helios: An analysis of ballot secrecy.
\newblock Cryptology ePrint Archive, Report 2010/625, 2010.
\newblock \url{http://eprint.iacr.org/2010/625}.

\bibitem[CZZ{\etalchar{+}}15]{CZZDMPDKR15}
Nikos Chondros, Bingsheng Zhang, Thomas Zacharias, Panos Diamantopoulos,
  Stathis Maneas, Christos Patsonakis, Alex Delis, Aggelos Kiayias, and Mema
  Roussopoulos.
\newblock A distributed, end-to-end verifiable, internet voting system.
\newblock {\em CoRR}, abs/1507.06812, 2015.

\bibitem[DDO{\etalchar{+}}01]{C:SCOPS01}
Alfredo {De Santis}, Giovanni {Di Crescenzo}, Rafail Ostrovsky, Giuseppe
  Persiano, and Amit Sahai.
\newblock Robust non-interactive zero knowledge.
\newblock In {\em Advances in Cryptology - {CRYPTO} 2001, 21st Annual
  International Cryptology Conference, Santa Barbara, California, USA, August
  19-23, 2001, Proceedings}, pages 566--598, 2001.

\bibitem[DFN06]{TCC:DamFazNic06}
Ivan Damg{\aa}rd, Nelly Fazio, and Antonio Nicolosi.
\newblock Non-interactive zero-knowledge from homomorphic encryption.
\newblock In Shai Halevi and Tal Rabin, editors, {\em TCC~2006: 3rd Theory of
  Cryptography Conference}, volume 3876 of {\em Lecture Notes in Computer
  Science}, pages 41--59. Springer, March 2006.

\bibitem[DH76]{DifHel76}
Whitfield Diffie and Martin~E. Hellman.
\newblock New directions in cryptography.
\newblock {\em {IEEE} Transactions on Information Theory}, 22(6):644--654,
  1976.

\bibitem[DJ01]{PKC:DamJur01}
Ivan Damg{\aa}rd and Mats Jurik.
\newblock A generalisation, a simplification and some applications of
  {Paillier}'s probabilistic public-key system.
\newblock In Kwangjo Kim, editor, {\em PKC~2001: 4th International Workshop on
  Theory and Practice in Public Key Cryptography}, volume 1992 of {\em Lecture
  Notes in Computer Science}, pages 119--136. Springer, February 2001.

\bibitem[DJ03]{ACISP:DamJur03}
Ivan Damg{\r a}rd and Mads Jurik.
\newblock A length-flexible threshold cryptosystem with applications.
\newblock In Reihaneh Safavi-Naini and Jennifer Seberry, editors, {\em ACISP
  03: 8th Australasian Conference on Information Security and Privacy}, volume
  2727 of {\em Lecture Notes in Computer Science}, pages 350--364. Springer,
  July 2003.

\bibitem[DKR09]{DelKreRya09}
St{\'{e}}phanie Delaune, Steve Kremer, and Mark Ryan.
\newblock Verifying privacy-type properties of electronic voting protocols.
\newblock {\em Journal of Computer Security}, 17(4):435--487, 2009.

\bibitem[DMP88]{C:DesMicPer87}
Alfredo {De Santis}, Silvio Micali, and Giuseppe Persiano.
\newblock Non-interactive zero-knowledge proof systems.
\newblock In Carl Pomerance, editor, {\em Advances in Cryptology --
  {CRYPTO}'87}, volume 293 of {\em Lecture Notes in Computer Science}, pages
  52--72. Springer, August 1988.

\bibitem[DN00]{FOCS:DwoNao00}
Cynthia Dwork and Moni Naor.
\newblock Zaps and their applications.
\newblock In {\em 41st Annual Symposium on Foundations of Computer Science},
  pages 283--293. {IEEE} Computer Society Press, November 2000.

\bibitem[{Fed}12]{Streebog}
{Federal Agency on Technical Regulation and Metrology}.
\newblock Gost r 34.11-2012: Streebog hash function, 2012.
\newblock \url{https://www.streebog.net}.

\bibitem[FLS90]{FOCS:FeiLapSha90}
Uriel Feige, Dror Lapidot, and Adi Shamir.
\newblock Multiple non-interactive zero knowledge proofs based on a single
  random string (extended abstract).
\newblock In {\em 31st Annual Symposium on Foundations of Computer Science},
  pages 308--317. {IEEE} Computer Society Press, October 1990.

\bibitem[FS87]{C:FiaSha86}
Amos Fiat and Adi Shamir.
\newblock How to prove yourself: {Practical} solutions to identification and
  signature problems.
\newblock In Andrew~M. Odlyzko, editor, {\em Advances in Cryptology --
  {CRYPTO}'86}, volume 263 of {\em Lecture Notes in Computer Science}, pages
  186--194. Springer, August 1987.

\bibitem[GGG{\etalchar{+}}14]{EC:GGGJKL14}
Shafi Goldwasser, S.~Dov Gordon, Vipul Goyal, Abhishek Jain, Jonathan Katz,
  Feng-Hao Liu, Amit Sahai, Elaine Shi, and Hong-Sheng Zhou.
\newblock Multi-input functional encryption.
\newblock In Phong~Q. Nguyen and Elisabeth Oswald, editors, {\em Advances in
  Cryptology -- {EUROCRYPT}~2014}, volume 8441 of {\em Lecture Notes in
  Computer Science}, pages 578--602. Springer, May 2014.

\bibitem[GGH{\etalchar{+}}13]{FOCS:GGHRSW13}
Sanjam Garg, Craig Gentry, Shai Halevi, Mariana Raykova, Amit Sahai, and Brent
  Waters.
\newblock Candidate indistinguishability obfuscation and functional encryption
  for all circuits.
\newblock In {\em 54th Annual Symposium on Foundations of Computer Science},
  pages 40--49. {IEEE} Computer Society Press, October 2013.

\bibitem[GGHZ16]{TCC:GGHZ16}
Sanjam Garg, Craig Gentry, Shai Halevi, and Mark Zhandry.
\newblock Functional encryption without obfuscation.
\newblock In Eyal Kushilevitz and Tal Malkin, editors, {\em Theory of
  Cryptography: 13th International Conference, TCC 2016-A, Tel Aviv, Israel,
  January 10-13, 2016, Proceedings, Part II}, pages 480--511. Springer, 2016.

\bibitem[GH07]{AC:GreHoh07}
Matthew Green and Susan Hohenberger.
\newblock Blind identity-based encryption and simulatable oblivious transfer.
\newblock In Kaoru Kurosawa, editor, {\em Advances in Cryptology --
  {ASIACRYPT}~2007}, volume 4833 of {\em Lecture Notes in Computer Science},
  pages 265--282. Springer, December 2007.

\bibitem[GIR16]{GiuIovRon16}
Rosario Giustolisi, Vincenzo Iovino, and Peter {R\o nne}.
\newblock On the possibility of non-interactive voting in the public-key
  setting.
\newblock In {\em Financial Cryptography and Data Security - FC 2016
  International Workshops, BITCOIN, VOTING, and WAHC, Christ Church, Barbados,
  February 26, 2016, Revised Selected Papers}, 2016.

\bibitem[GK03]{FOCS:GolKal03}
Shafi Goldwasser and Yael~Tauman Kalai.
\newblock On the (in)security of the {Fiat}-{Shamir} paradigm.
\newblock In {\em 44th Annual Symposium on Foundations of Computer Science},
  pages 102--115. {IEEE} Computer Society Press, October 2003.

\bibitem[GKP{\etalchar{+}}13]{STOC:GKPVZ13}
Shafi Goldwasser, Yael~Tauman Kalai, Raluca~A. Popa, Vinod Vaikuntanathan, and
  Nickolai Zeldovich.
\newblock Reusable garbled circuits and succinct functional encryption.
\newblock In Dan Boneh, Tim Roughgarden, and Joan Feigenbaum, editors, {\em
  45th Annual {ACM} Symposium on Theory of Computing}, pages 555--564. {ACM}
  Press, June 2013.

\bibitem[GM84]{GolMic84}
Shafi Goldwasser and Silvio Micali.
\newblock Probabilistic encryption.
\newblock {\em Journal of Computer and System Sciences}, 28(2):270--299, 1984.

\bibitem[GO14]{JC:GroOst14}
Jens Groth and Rafail Ostrovsky.
\newblock Cryptography in the multi-string model.
\newblock {\em Journal of Cryptology}, 27(3):506--543, July 2014.

\bibitem[Gol01]{Goldreich01}
Oded Goldreich.
\newblock {\em Foundations of Cryptography: Basic Techniques}, volume~1.
\newblock Cambridge University Press, Cambridge, UK, 2001.

\bibitem[GOS06]{C:GroOstSah06}
Jens Groth, Rafail Ostrovsky, and Amit Sahai.
\newblock Non-interactive zaps and new techniques for {NIZK}.
\newblock In Cynthia Dwork, editor, {\em Advances in Cryptology --
  {CRYPTO}~2006}, volume 4117 of {\em Lecture Notes in Computer Science}, pages
  97--111. Springer, August 2006.

\bibitem[Gro04]{FC:Gro04}
Jens Groth.
\newblock Efficient maximal privacy in boardroom voting and anonymous
  broadcast.
\newblock In {\em International Conference on Financial Cryptography}, pages
  90--104. Springer, 2004.

\bibitem[GS08]{EC:GroSah08}
Jens Groth and Amit Sahai.
\newblock Efficient non-interactive proof systems for bilinear groups.
\newblock In Nigel~P. Smart, editor, {\em Advances in Cryptology --
  {EUROCRYPT}~2008}, volume 4965 of {\em Lecture Notes in Computer Science},
  pages 415--432. Springer, April 2008.

\bibitem[GVW12]{C:GorVaiWee12}
Sergey Gorbunov, Vinod Vaikuntanathan, and Hoeteck Wee.
\newblock Functional encryption with bounded collusions via multi-party
  computation.
\newblock In Reihaneh Safavi-Naini and Ran Canetti, editors, {\em Advances in
  Cryptology -- {CRYPTO}~2012}, volume 7417 of {\em Lecture Notes in Computer
  Science}, pages 162--179. Springer, August 2012.

\bibitem[HRZ10]{HRZ10}
Feng Hao, Peter Y.~A. Ryan, and Piotr Zielinski.
\newblock Anonymous voting by two-round public discussion.
\newblock {\em {IET} Information Security}, 4(2):62--67, 2010.

\bibitem[JCJ10]{JCJ10}
Ari Juels, Dario Catalano, and Markus Jakobsson.
\newblock Coercion-resistant electronic elections.
\newblock In {\em Towards Trustworthy Elections}, pages 37--63. Springer, 2010.

\bibitem[Jou04]{JC:Joux04}
Antoine Joux.
\newblock A one round protocol for tripartite {Diffie}-{Hellman}.
\newblock {\em Journal of Cryptology}, 17(4):263--276, September 2004.

\bibitem[Kal06]{PHD:Kalai06}
Yael~Tauman Kalai.
\newblock {\em Attacks on the Fiat-Shamir paradigm and program obfuscation}.
\newblock PhD thesis, Massachusetts Institute of Technology, 2006.

\bibitem[KRS10]{KreRyaSmy10}
Steve Kremer, Mark Ryan, and Ben Smyth.
\newblock Election verifiability in electronic voting protocols.
\newblock In {\em European Symposium on Research in Computer Security}, pages
  389--404. Springer, 2010.

\bibitem[KSRH12]{EV:KSRH12}
Dalia Khader, Ben Smyth, Peter Y.~A. Ryan, and Feng Hao.
\newblock A fair and robust voting system by broadcast.
\newblock In {\em 5th International Conference on Electronic Voting 2012,
  {(EVOTE} 2012), Co-organized by the Council of Europe, Gesellschaft f{\"{u}}r
  Informatik and E-Voting.CC, July 11-14, 2012, Castle Hofen, Bregenz,
  Austria}, pages 285--299, 2012.

\bibitem[KY02]{PKC:KiaYun02}
Aggelos Kiayias and Moti Yung.
\newblock Self-tallying elections and perfect ballot secrecy.
\newblock In David Naccache and Pascal Paillier, editors, {\em PKC~2002: 5th
  International Workshop on Theory and Practice in Public Key Cryptography},
  volume 2274 of {\em Lecture Notes in Computer Science}, pages 141--158.
  Springer, February 2002.

\bibitem[KZZ15]{EC:KiaZacZha15}
Aggelos Kiayias, Thomas Zacharias, and Bingsheng Zhang.
\newblock End-to-end verifiable elections in the standard model.
\newblock In {\em Advances in Cryptology - {EUROCRYPT} 2015 - 34th Annual
  International Conference on the Theory and Applications of Cryptographic
  Techniques, Sofia, Bulgaria, April 26-30, 2015, Proceedings, Part {II}},
  pages 468--498, 2015.

\bibitem[Lin15]{TCC:Lin15}
Yehuda Lindell.
\newblock An efficient transform from sigma protocols to {NIZK} with a {CRS}
  and non-programmable random oracle.
\newblock In {\em Theory of Cryptography - 12th Theory of Cryptography
  Conference, {TCC} 2015, Warsaw, Poland, March 23-25, 2015, Proceedings, Part
  {I}}, pages 93--109, 2015.

\bibitem[Lip05]{Lipmaa05}
Helger Lipmaa.
\newblock Secure electronic voting protocols.
\newblock In Hossein Bidgoli, editor, {\em Handbook of Information Security,
  Volume 2, Information Warfare, Social, Legal, and International Issues and
  Security Foundations}, pages 647--657. John Wiley \& Sons, Inc., 2005.
\newblock Electronic edition available at
  \url{http://kodu.ut.ee/~lipmaa/papers/voting4hb.pdf}.

\bibitem[LP09]{JC:LinPin09}
Yehuda Lindell and Benny Pinkas.
\newblock A proof of security of {Yao}'s protocol for two-party computation.
\newblock {\em Journal of Cryptology}, 22(2):161--188, April 2009.

\bibitem[MH96]{AC:MicHor96a}
Markus Michels and Patrick Horster.
\newblock Some remarks on a receipt-free and universally verifiable mix-type
  voting scheme.
\newblock In Kwangjo Kim and Tsutomu Matsumoto, editors, {\em Advances in
  Cryptology -- {ASIACRYPT}'96}, volume 1163 of {\em Lecture Notes in Computer
  Science}, pages 125--132. Springer, November 1996.

\bibitem[MN06]{C:MorNao06}
Tal Moran and Moni Naor.
\newblock Receipt-free universally-verifiable voting with everlasting privacy.
\newblock In Cynthia Dwork, editor, {\em Advances in Cryptology --
  {CRYPTO}~2006}, volume 4117 of {\em Lecture Notes in Computer Science}, pages
  373--392. Springer, August 2006.

\bibitem[Nao03]{C:Naor03}
Moni Naor.
\newblock On cryptographic assumptions and challenges (invited talk).
\newblock In Dan Boneh, editor, {\em Advances in Cryptology -- {CRYPTO}~2003},
  volume 2729 of {\em Lecture Notes in Computer Science}, pages 96--109.
  Springer, August 2003.

\bibitem[NY90]{STOC:NaoYun90}
Moni Naor and Moti Yung.
\newblock Public-key cryptosystems provably secure against chosen ciphertext
  attacks.
\newblock In {\em 22nd Annual {ACM} Symposium on Theory of Computing}, pages
  427--437. {ACM} Press, May 1990.

\bibitem[Riv06]{threeballot}
Ronald~L. Rivest.
\newblock The threeballot voting system, 2006.

\bibitem[RR06]{Randell06votingtechnologies}
Brian Randell and Peter Y.~A. Ryan.
\newblock Voting technologies and trust.
\newblock In {\em IEEE Security and Privacy}, pages 50--56, 2006.

\bibitem[RRI16]{selene}
Peter Y.~A. Ryan, Peter~B. R{\o}nne, and Vincenzo Iovino.
\newblock Selene: Voting with transparent verifiability and
  coercion-mitigation.
\newblock In {\em Financial Cryptography and Data Security - {FC} 2016
  International Workshops, BITCOIN, VOTING, and WAHC, Christ Church, Barbados,
  February 26, 2016, Revised Selected Papers}, pages 176--192, 2016.

\bibitem[RS92]{C:RacSim91}
Charles Rackoff and Daniel~R. Simon.
\newblock Non-interactive zero-knowledge proof of knowledge and chosen
  ciphertext attack.
\newblock In Joan Feigenbaum, editor, {\em Advances in Cryptology --
  {CRYPTO}'91}, volume 576 of {\em Lecture Notes in Computer Science}, pages
  433--444. Springer, August 1992.

\bibitem[RS06]{ryan006}
Peter Y.~A. Ryan and S.~A. Schneider.
\newblock Pr\^{e}t \`{a} voter with re-encryption mixes.
\newblock Technical Report CS-TR-956, University of Newcastle, 2006.

\bibitem[RT09]{Ryan09prettygood}
Peter Y.~A. Ryan and Vanessa Teague.
\newblock Pretty good democracy.
\newblock In {\em IN: WORKSHOP ON SECURITY PROTOCOLS}, 2009.

\bibitem[SK95]{EC:SakKil95}
Kazue Sako and Joe Kilian.
\newblock Receipt-free mix-type voting scheme - a practical solution to the
  implementation of a voting booth.
\newblock In Louis~C. Guillou and Jean-Jacques Quisquater, editors, {\em
  Advances in Cryptology -- {EUROCRYPT}'95}, volume 921 of {\em Lecture Notes
  in Computer Science}, pages 393--403. Springer, May 1995.

\bibitem[SS10]{CCS:SahSey10}
Amit Sahai and Hakan Seyalioglu.
\newblock Worry-free encryption: functional encryption with public keys.
\newblock In Ehab Al-Shaer, Angelos~D. Keromytis, and Vitaly Shmatikov,
  editors, {\em ACM CCS 10: 17th Conference on Computer and Communications
  Security}, pages 463--472. {ACM} Press, October 2010.

\bibitem[Yao86]{FOCS:Yao86}
Andrew Chi-Chih Yao.
\newblock How to generate and exchange secrets (extended abstract).
\newblock In {\em 27th Annual Symposium on Foundations of Computer Science},
  pages 162--167. {IEEE} Computer Society Press, October 1986.

\end{thebibliography}
\ifnum\fullversion=1
\else
\appendix{\bf \Large{Supplementary material.}}
\fi
\ifnum\fullversion=1
\else
\section{Definitions}\label{sec:def}
\paragraph{\bf Notation.}
A {\em negligible} function $\negl(k)$ is a function that is smaller than the inverse of any polynomial in $k$ (from a certain point and on).
We denote by $[n]$ the set of numbers $\{1,\ldots,n\}$.
If $S$ is a finite set, we denote by $a\from S$ the process of setting $a$ equal to a uniformly chosen element of $S$.
With a slight abuse of notation, we assume the existence of a special symbol $\bot$ that does not belong to $\zu^\star$.

If $A$ is an algorithm, then $A(x_1,x_2,\ldots)$ denotes the probability distribution of the output of $A$ when $A$ is run on input $(x_1,x_2,\ldots)$ and randomly chosen coin tosses. Instead, $A(x_1,x_2,\ldots; r)$ denotes 	the output of $A$ when run on input $(x_1,x_2,\ldots)$ and (sufficiently long) coin tosses $r$. All algorithms, unless explicitly noted, are probabilistic polynomial time (PPT) and all adversaries are modeled by non-uniform PPT algorithms.

If $A$ is a PPT algorithm, we say that $y\in A(x)$ iff there exists a random value $r$ such that $y=A(x;r)$; in that case, we say that $y$ is in the range of $A(x)$.
If $E$ is an event in a probability space, $\bar E$ denotes its complement.

The following definition is used in the definition of verifiability. Essentially, it states that a tally $y$ is compatible with votes $z_1,\ldots,z_k$ if the latter values are in its pre-image.
\begin{definition}\label{def:restriction}
	Given a function $F(x_1,\ldots,x_n):A^n\rightarrow B$, we say that a value $y\in B$ is compatible with $z_1,\ldots,z_k\in A$ at indices $i_1,\ldots,i_k\in[N]$ if $y$ is in the range of the restriction $F_{|C_{z_1,\ldots,z_k,i_1,\ldots,i_k}}$ of $F$ to $C_{z_1,\ldots,z_n,i_1,\ldots,i_n}\defeq\{(x_1,\ldots,x_n)| \forall j\in[k], x_{i_j}=z_j\}$.
\end{definition}

\subsection{E-Voting Schemes}\label{sec:defevote}

An e-voting scheme (eVote, in short) is parameterized by the tuple $(N,\M,\Sigma,F)$. The natural number $N>0$ is the {\em number of voters}. The set $\M$ is the {\em domain of valid votes}. The set $\Sigma$ is the {\em range of possible results}. The function  $F:(\M\cup\{\bot\})^N\rightarrow\Sigma\cup\{\bot\}$ is the {\em tally function}.
We allow the tally function to take as input the special symbol $\bot$, which denotes either an abstention, an invalid ballot or a blank vote\footnote{We note that our tally function can be made more general by assigning different symbols to an abstention, to an invalid ballot and to a blank vote.}, and to output $\bot$ to indicate an error. We require that the tally function outputs an error on input a sequence of strings iff all the strings are equal to $\bot$. Formally, the tally function is defined as follows.
\begin{definition}[Tally function]\label{def:tally}
	A function $F$ is a tally function if there exists a natural number $N>1$, and sets $\M,\Sigma\subset\zu^\star$ such that the domain of $F$ is $\M\cup\{\bot\}$, the range is $\Sigma\cup\{\bot\}$ and for all strings $m_1,\ldots,m_N\in\M\cup\{\bot\},$ it holds that $F(m_1,\ldots,m_N)=\bot$ iff $m_1=\bot,\ldots,m_N=\bot$.
\end{definition}

Before defining formally an eVote, we explain how its algorithms are used to conduct an election.
\paragraph{The voting ceremony.}
The voting ceremony occurs as follows.
\begin{itemize}
	\item{Setup phase.} An authority (also called voting authority or election authority) uses  algorithm $\Setup$ to compute a public key $\PK$ and a secret key $\SK$.
	\item{Voting phase.} Each of the $N$ voters runs an algorithm $\Cast$ on input the voter identifier $j\in[N]$, the public key $\PK$ and a vote $v\in \M$ to compute a ballot $\BT$. The voter sends $\BT$ to an append-only public bulletin board (PBB).
	\item{Tallying phase.} The well-formedness of each ballot $\BT$ published in the PBB can be publicly verified by means of an algorithm $\VerifyBallot$. If the ballot is invalid,  a new row in which the ballot is replaced by $\bot$ is appended to the PBB. Later, only the new row is used. If a voter did not cast a vote, $\bot$  is appended to the PBB.

		The authority runs {\em evaluation tally} algorithm $\EvalTally$ on input the public key, the secret key, and $N$ strings that represent either ballots or $\bot$ symbols appended to the PBB. $\EvalTally$ outputs the tally, i.e., the result of the election, and a  proof of tally correctness. The tally equals the special symbol $\bot$ to indicate an error.
\item{Verification phase.}
Algorithm $\VerifyTally$ takes as input the public key, a tuple of $N$ strings that represent either ballots or the special symbol $\bot$, the tally and the proof of tally correctness.  $\VerifyTally$ outputs a value in $\{\OK,\bot\}$.

Each participant, not necessarily a voter, can verify the correctness of the result of the election as follows.
First, verify whether the ballots cast by the voters are valid using the $\VerifyBallot$ algorithm. Check whether the authority replaced with $\bot$ only the invalid ballots. Assign $\bot$ to any voter who did not cast her vote. After that, run the $\VerifyTally$ algorithm on input the public key, the $N$ strings that represent either ballots or the special symbol $\bot$, the tally and the proof of tally correctness.
\end{itemize}

\begin{definition}[E-voting Scheme]\label{def:evote}
	A $(N,\M,\Sigma,F)$-{\em e-voting scheme} $\EVOTE$ for number of voters $N>1$, domain of valid votes $\M$, range of possible results $\Sigma$ and tally function $F:(\M\cup\{\bot\})^N\rightarrow\Sigma\cup\{\bot\}$ is a tuple
\begin{equation*}
\EVOTE\defeq(\Setup,\Cast,\VerifyBallot, \EvalTally,\VerifyTally)
\end{equation*}
 of $5$ PPT algorithms, where $\VerifyBallot$ and $\VerifyTally$ are deterministic, that fulfill the following syntax:
\begin{enumerate}
\item $\Setup(1^\lambda)$: on input the security parameter in unary, it
	outputs the {{\em public key}} $\PK$ and the {\em secret key} $\SK$.

\item $\Cast(\PK,j,v)$: on input the public key $\PK$, the voter identifier $j\in[N]$, and a vote $v \in \M$, it outputs a {\em ballot} $\BT$.
\item $\VerifyBallot(\PK,j,\BT)$: on input the public key $\PK$, the voter identifier $j\in[N]$ and a ballot $\BT$, it outputs a value in $\{\OK, \bot\}$.

\item $\EvalTally(\PK,\SK,\BT_1,\ldots,\BT_N)$: on input the public key $\PK$, the secret key $\SK$, and
	$N$ strings that are either ballots or the special symbol $\bot$, it  outputs the tally $y \in \Sigma \cup \{\bot\}$ and a proof $\gamma$ of tally correctness.

\item $\VerifyTally(\PK,\BT_1,\ldots,\BT_N,y,\gamma)$: on input the public key $\PK$, $N$ strings that are either ballots or the special symbol $\bot$, a tally $y\in\zu^\star\cup\{\bot\}$ and a proof $\gamma$ of tally correctness, it outputs a value in $\{\OK,\bot\}$.
\end{enumerate}
\end{definition}

An eVote must satisfy the following correctness, verifiability, and privacy properties. We also define the weak verifiability and weak privacy properties. A weakly verifiable eVote must satisfy correctness, weak verifiability and weak privacy.
\subsubsection{Correctness and verifiability}

\begin{itemize}
	\item{\em (Perfect) Correctness.}
		We require the following conditions (1) and (2) to hold.
		\begin{enumerate}
			\item
				Let $\Abst$ be a special symbol not in $\M\cup\{\bot\}$ that denotes that a voter did not cast her vote.\footnote{In the following definition, we need $\Abst$ to differentiate the case of a voter who did not cast a vote at all ($\Abst$) from the case of a voter who casts $\bot$ as her own vote but wishes to preserve the anonymity of her choice. However, in both cases, correctness guarantees that the result of the election equals the output of the tally function, and the input to the tally function is $\bot$ both when a voter casts $\bot$ and when a voter does not cast any vote.}
For all $\PK\in\Setup(1^\lambda)$,
all $m_1,\ldots,m_N\in \M\cup\{\bot,\Abst \}$,
all $(\BT_j)_{j=1}^N$ such that for all $j\in[N]$, $\BT_j=\bot$ if $m_j=\Abst$, $\BT_j\in\Cast(\PK,j,m_j)$ if $m_j\in\M$ and $\BT_j\in\Cast(\PK,j,\bot)$ otherwise, the following two conditions (a) and (b) hold:
\begin{enumerate}
	\item For all $j\in[N],$ if $m_j\neq\Abst$ then $\VerifyBallot(\PK,j,\BT_j)=\OK$.
	\item if $(y,\gamma)\defeq\EvalTally(\PK,\BT_1,\ldots,\BT_N)$ then it holds that:\\ $y=F(m_1,\ldots,m_N)$ and $\VerifyTally(\PK,\BT_1,\ldots,\BT_N,y,\gamma)=\OK$.

\end{enumerate}
	\item For all $\PK\in\Setup(1^\lambda)$, $\BT_1,\ldots,\BT_N\in\zu^\star\cup \{\bot\}$,
		if $S\defeq\{j|\ \BT_j\neq\bot \wedge \VerifyBallot(\PK,j,\BT_j)=\bot\}$ and $\BT_1',\ldots,\BT_N'$ are such that for all $j\in[N],$ $\BT_j'=\BT_j$ if $j\notin S$ and $\BT'_j=\bot$ otherwise, it holds that:\\
	If $(y,\gamma)\defeq\EvalTally(\PK,\BT'_1,\ldots,\BT'_N)$ then $\VerifyTally(\PK, \allowbreak\BT'_1, \allowbreak \ldots, \allowbreak \BT'_N, \allowbreak y, \allowbreak \gamma) \allowbreak = \allowbreak \OK$.
\end{enumerate}
\item{\em Weak verifiability.}
		We require the following conditions (1) and (2) to hold.
	\begin{enumerate}
		\item For all $\PK\in\zu^\star,\BT_1,\ldots,\BT_N\in\zu^\star\cup \{\bot\}$, there exist $m_1,\ldots,m_N\in\M\cup\{\bot\}$ such that for all $y\neq\bot$ and $\gamma$ in $\zu^\star$, if $S\defeq\{j|\ \BT_j\neq\bot \wedge \VerifyBallot(\PK,j,\BT_j)=\bot\}$ and $\BT_1',\ldots,\BT_N'$ are such that for all $j\in[N],$ $\BT_j'=\BT_j$ if $j\notin S$ and $\BT'_j=\bot$ otherwise, it holds that:\\
	if $\VerifyTally(\PK,\BT_1',\ldots,\BT_N',y,\gamma)=\OK$ then  $y=F(m_1,\ldots,m_N)$.
\item For all $\PK\in\zu^\lambda$, all $k\in[N]$, $i_1,\ldots,i_k\in[N]$, all $m_{i_1},\ldots,m_{i_k}\in\M\cup\{\bot\}$, all $\BT_1,\ldots,\BT_N\in\zu^\star\cup \{\bot\}$ such that for all $j\in[k]$, $\BT_j\in\Cast(\PK,m_{i_j})$ and $\VerifyBallot(\PK,\BT_j)=\OK$,
		if $S\defeq\{j|\ \BT_j\neq\bot \wedge \VerifyBallot(\PK,j,\BT_j)=\bot\}$ and $\BT_1',\ldots,\BT_N'$ are such that for all $j\in[N],$ $\BT_j'=\BT_j$ if $j\notin S$ and $\BT'_j=\bot$ otherwise, it holds that:\\
		if there exist $y\in\zu^\star$ and $\gamma\in\zu^\star$ such that $\VerifyTally(\PK,\allowbreak\BT_1',\allowbreak\ldots,\allowbreak\BT_N',\allowbreak y,\allowbreak\gamma)\allowbreak = \allowbreak \OK$, then $y$ is compatible with $m_{i_1},\allowbreak \ldots,\allowbreak m_{i_k}$ at indices $i_1,\allowbreak \ldots,\allowbreak i_k$.
\end{enumerate}
\item{\em Verifiability.}
		We require the following conditions (1) and (2) to hold.
	\begin{enumerate}
		\item For all $\PK\in\zu^\star,\BT_1,\ldots,\BT_N\in\zu^\star\cup \{\bot\}$, there exist $m_1,\allowbreak \ldots,\allowbreak m_N \allowbreak \in \allowbreak \M\cup\{\bot\}$ such that for all $y\in\zu^\star\cup\{\bot\}$ and $\gamma$ in $\zu^\star$, if $S\defeq\{j|\ \BT_j\neq\bot \wedge \VerifyBallot(\PK,j,\BT_j)=\bot\}$ and $\BT_1',\ldots,\BT_N'$ are such that for all $j\in[N],$ $\BT_j'=\BT_j$ if $j\notin S$ and $\BT'_j=\bot$ otherwise, it holds that:\\
	if $\VerifyTally(\PK,\BT_1',\ldots,\BT_N',y,\gamma)=\OK$ then  $y=F(m_1,\ldots,m_N)$.
\item For all $\PK\in\zu^\lambda$, all $k\in[N]$, $i_1,\ldots,i_k\in[N]$, all $m_{i_1},\ldots,m_{i_k}\in\M\cup\{\bot\}$, all $\BT_1,\ldots,\BT_N\in\zu^\star\cup \{\bot\}$ such that for all $j\in[k]$, $\BT_j\in\Cast(\PK,m_{i_j})$ and $\VerifyBallot(\PK,\BT_j)=\OK$,
		if $S\defeq\{j|\ \BT_j\neq\bot \wedge \VerifyBallot(\PK,j,\BT_j)=\bot\}$ and $\BT_1',\ldots,\BT_N'$ are such that for all $j\in[N],$ $\BT_j'=\BT_j$ if $j\notin S$ and $\BT'_j=\bot$ otherwise, it holds that:\\
		if there exist $y\in\zu^\star\cup\{\bot\}$ and $\gamma\in\zu^\star$ such that $\VerifyTally(\PK,\allowbreak\BT_1',\allowbreak\ldots,\allowbreak\BT_N',\allowbreak y,\allowbreak\gamma)\allowbreak=\allowbreak\OK$, then $y$ is compatible with $m_{i_1},\allowbreak\ldots,\allowbreak m_{i_k}$ at indices $i_1,\allowbreak \ldots,\allowbreak i_k$.
\end{enumerate}

\medskip

Note that the difference between condition (2) of verifiability and condition (2) of weak verifiability lies in the fact that, in the latter, $y$ cannot equal $\bot$, whereas, in the former, the condition has to hold even for $y=\bot$. In this work, we use the terms verifiability and full verifiability interchangeably to differentiate them from weak verifiability.
\end{itemize}
\subsubsection{Privacy}

We define {\em privacy} in the style of indistinguishability-based security.
Privacy for a $(N,\M,\Sigma,F)$-eVote
\begin{equation*}
\EVOTE\allowbreak\defeq\allowbreak(\Setup,\allowbreak\Cast,\allowbreak\VerifyBallot,\allowbreak\EvalTally,\allowbreak\VerifyTally)
\end{equation*}
 is formalized by means of the game
$\SecGame^{N,\M,\Sigma,F,\EVOTE}_\adv$
between a stateful adversary
$\adv$ and a {\em challenger} $\C$. We describe the game in Fig.~\ref{definitionprivacy}.

\begin{figure}
\small
\begin{center}
\begin{framed}
$\SecGame^{N,\M,\Sigma,F,\EVOTE}_\adv(1^\lambda)$
\begin{itemize}
\item{Setup phase.} $\C$ generates
	$(\PK,\SK)\from\Setup(1^\lambda)$, chooses a random bit $b\from\zu$ and runs
$\adv$ on input $\PK$.

\item{Query phase.} $\adv$ outputs two tuples
	$M_0\defeq(m_{0,1},\ldots,m_{0,N})$ and
	$M_1\defeq(m_{1,1},\ldots,m_{1,N})$, and a set $S\subset[N]$. (The set $S$ contains the indices of the strings in the tuples that are possibly dishonest ballots. The strings in the tuples whose indices are not in $S$ are supposed to be votes to be given as input to the $\Cast$ algorithm.)
\item{Challenge phase.} The challenger does the following.
	For all $j\in[N]$, if $j\in S$, then set $\BT_j\defeq m_{b,j}$, else set $\BT_j\from\Cast(\PK,j,m_{b,j})$.
	For all $j\in S,$ if $\VerifyBallot(\PK,j,\BT_j)=\bot$, set $\BT_j\defeq\bot$.	Compute $(y,\gamma) \from \EvalTally(\PK,\allowbreak\SK,\allowbreak\BT_1,\allowbreak\ldots, \allowbreak\BT_N)$ and return $(\BT_1,\ldots,\BT_N,y,\gamma)$ to the adversary.

\item{Output.} At some point the adversary outputs its guess $b'$.

\item{Winning condition.}
	The adversary wins the game if all the following conditions hold:
\begin{enumerate}
	\item $b'=b$.
	\item For all $j\in S, m_{0,j}=m_{1,j}$. (That is, if the adversary submits a dishonest ballot, it has to be the same in both tuples.)
\item For all $d_1,\ldots,d_N\in\M\cup\{\bot\}$, for all $j\in [N]$, let $m'_{0,j}\defeq m'_{1,j}\defeq d_j$ if $j\in S$, and for all $b\in\zu$ let $m'_{b,j}\defeq m_{b,j}$ if $m_{b,j}\in\M$ and $m_{b,j}'\defeq \bot$ if $m_{b,j}\notin\M$. Then, $F(m_{0,1}',\ldots,m_{0,N}')=F(m_{1,1}',\ldots,m_{1,N}')$.
		
		(That is, the tally function outputs the same result on input both tuples, even if the ballots corresponding to indices in $S$ are replaced by arbitrary messages in $\M\cup\{\bot\}$.)
\end{enumerate}
\end{itemize}
\end{framed}
\end{center}
\caption{Definition of privacy}
\label{definitionprivacy}
\end{figure}

The advantage of adversary $\A$ in the above game
is defined as
$$
\Adv_{\adv}^{\EVOTE,\Priv}(1^\lambda)\defeq
|\prob{\SecGame^{N,\M,\Sigma,F,\EVOTE}_\adv(1^\lambda)=1}-1/2|
$$

\begin{definition}\label{def:evotepriv}
An $\EVOTE$ for parameters $(N,\M,\Sigma,F)$ is {\em private} or \IND-Secure
if the advantage of all PPT adversaries
$\adv$
is at most negligible in $\lambda$ in
the above game.
\end{definition}
\begin{definition}\label{def:evoteweakpriv}
An $\EVOTE$ for parameters $(N,\M,\Sigma,F)$ is {\em weakly private} or \wIND-Secure
if the advantage of all PPT adversaries
$\adv$
is at most negligible in $\lambda$ in
a game $\SecwGame^{N,\M,\Sigma,F,\EVOTE}_\adv(1^\lambda)$ identical to the one above except that $\adv$ is required to output an empty set $S$, i.e., $\adv$ cannot submit dishonest ballots.
\end{definition}

\begin{remark}
We make some remarks on the previous definitions.
\begin{itemize}
	\item
	Our definitions suppose that algorithm $\VerifyBallot$ is run on input each ballot before running algorithm $\VerifyTally$. The ballots that are input to $\VerifyTally$ are replaced by $\bot$ if they were not accepted by $\VerifyBallot$. Another possibility would be to let $\VerifyTally$ do this task itself.

	\item We require that $\VerifyBallot$ and $\VerifyTally$ be deterministic algorithms. Alternatively, they can be defined as PPT, but then definitions of weak verifiability and verifiability would have to be changed accordingly to hold with probability $1$ over the random coins of the algorithms.
	\item Our definition is parameterized by the number of voters $N$. It is possible to define a more restricted eVote that may possibly be ``unbounded''. Note that our definition is more general and, for instance, takes into account e-voting schemes in which the public key is of size proportional to the number of voters.
	\item Both condition (2) of verifiability and condition (2) of weak verifiability lie in some sense between correctness and verifiability as they state a requirement about honest voters.
	\item In our weakly verifiable construction in Section~\ref{sec:scheme}, algorithm $\VerifyBallot$ could be completely discarded because it accepts any ballot. Both for the sake of generality (there could exist some weakly verifiable eVote that makes a non-trivial use of $\VerifyBallot$) and to avoid overburdening the presentation, we use the same syntax for weakly verifiable eVotes and for verifiable eVotes.
	\item For the necessity of condition (2) of correctness, we refer the reader to the discussion in Section~\ref{sec:strongcorrectness}.
	\item As shown in~\cite{BCGPW15}, the definition of ``Benaloh'' (recall that we restate it using modern terminology) is subject to attacks when instantiated with specific tally functions like the majority. Nonetheless, ours is strengthened to withstand such attacks. This is done by adding the $3$-rd winning condition.
\end{itemize}
	\end{remark}


\section{Building Blocks}\label{sec:buildingblocks} \label{app:dlin}\label{sec:dlin}Our constructions use perfectly binding commitment schemes,  (one-message) non-interactive witness-indistinguishable proof systems with perfect soundness for $\NP$ \cite{C:GroOstSah06} (see also \cite{FOCS:FeiLapSha90,FOCS:DwoNao00,FOCS:DwoNao00,C:BarOngVad03,TCC:BitPan15}) and IND-CPA public key encryption with perfect correctness and unique secret key. In this section, we recall the definitions of those primitives.

\begin{definition}[IND-CPA secure PKE with perfect correctness and unique secret key]\label{def:pke}
An IND-CPA (or semantically) secure Public Key Encryption (PKE) scheme consists of three PPT
algorithms $(\Setup, \Enc, \Dec)$ defined as follows.
\begin{itemize}
	\item{$\Setup(1^\lambda)$}: On input $1^\lambda$, it outputs public key $\PK$ and decryption key $\SK$.
	\item{$\Enc(m, \PK)$}: On input message $m$ and the public key, it outputs ciphertext $\CT$.
	\item{$\Dec(\CT, \SK)$}: On input ciphertext $\CT$ and the decryption key, it outputs $m$.
	\end{itemize}
	The PKE scheme is said to be IND-CPA (or semantically) secure if for any PPT adversary $\A$, there exists
	a negligible function $\nu(\cdot)$ such that the following is satisfied for any two messages $m_0, m_1$ and for
$b \in \{0, 1\}$:
\begin{equation*}
|\Prob{\A(1^\lambda, \Enc(m_0, \PK)) = b} - \Prob{\A(1^\lambda, \Enc(m_1, \PK)) = b}| \le \nu(\lambda).
\end{equation*}

{\em Perfect correctness} requires that, for all pairs $(\PK,\SK)\in\Setup$, for all messages $m$ in the message space and all ciphertexts $\Ct$ output by $\Enc(\PK,m)$, $\Dec(\CT,\SK)=m$ must hold. {\em Unique secret key} requires that, for all $\PK$, there exists at most {\em one} $\SK$ such that $(\PK,\SK)\in\Setup(1^\lambda)$.

The Decision Linear Encryption scheme~\cite{C:BonBoySha04} fulfills those properties. It is secure under the Decision Linear Assumption~\cite{C:BonBoySha04}. We recall them next.
\end{definition}



\newcommand{\BilinearSetup}{\ensuremath{\mathcal{G}}}

\newcommand{\myvar}[1]{\ensuremath{\mathit{#1}}}
\newcommand{\securityparameter}{\myvar{k}}
\newcommand{\Gp}{\mathbb{G}_p}
\newcommand{\Ga}{\mathbb{G}}
\newcommand{\Gb}{\mathbb{\tilde{G}}}
\newcommand{\Gt}{{\mathbb{G}_{t}}}
\newcommand{\p}{\myvar{p}}
\newcommand{\q}{\myvar{q}}
\newcommand{\Zp}{\mathbb{Z}_p}
\newcommand{\Zq}{\mathbb{Z}_q}
\newcommand{\g}{\myvar{g}}
\newcommand{\ga}{\myvar{g}}
\newcommand{\gb}{\myvar{\tilde{g}}}
\newcommand{\gt}{\myvar{g_t}}
\newcommand{\h}{\myvar{h}}
\newcommand{\hb}{\myvar{\tilde{h}}}
\newcommand{\hbig}{\myvar{H}}
\newcommand{\grp}{\myvar{grp}}

\paragraph{Bilinear Groups.} We assume the existence of a PPT algorithm $\BilinearSetup(1^\lambda)$, the {\em bilinear group generator}, that outputs a {\em pairing group setup} $(\p,\Ga,\allowbreak \Gt,\e,\ga)$, where $\Ga$ and $\Gt$ are multiplicative groups of prime order $\p$  and $\e: \Ga \times \Ga \rightarrow \Gt$ is a {\em bilinear map} satisfying the following three properties: (1) bilinearity, i.e., $\e(\ga^x,\ga^y)=e(\ga,\ga)^{xy}$; (2) non-degeneracy, i.e., for all generators $\ga \in \Ga$, $\e(\ga,\ga)$ generates $\Gt$; (3) efficiency, i.e., $\e$ can be computed in polynomial time.

\begin{assm}[Decision Linear Assumption for $\BilinearSetup$.\cite{C:BonBoySha04}] Let the tuple $(p,\allowbreak\Ga,\allowbreak\Gt,\allowbreak\e,\allowbreak\ga)$ be a pairing group setup output by  $\BilinearSetup$ as defined above, and let $g_1$, $g_2$ and $g_3$ be generators of $\Ga$. Given $(g_1, g_2, g_3, g_1^a, g_2^b, g_3^c)$, where $a$ and $b$ are picked randomly from $\mathbb{Z}_p$, the Decision Linear (DLIN) assumption is to decide whether $c=a+b\ \mathrm{mod}\ p$. Precisely, the advantage of an adversary $\mathcal{A}$ in solving the Decision Linear assumption is given by:
\begin{align*}
&\bigl|\mathrm{Pr}\ [\mathcal{A}(\mathbb{G},p,g_1,g_2,g_3,g_1^a,g_2^b,g_3^{a+b})=1\mid (\p,\Ga,\allowbreak \Gt,\e,\ga)\from\BilinearSetup(1^\lambda); \\& (g_1,g_2,g_3) \leftarrow\Ga; (a,b) \leftarrow \mathbb{Z}_p] - \\& \mathrm{Pr}\ [\mathcal{A}(\mathbb{G},p,g_1,g_2,g_3,g_1^a,g_2^b,g_3^{c})=1\mid (\p,\Ga,\allowbreak \Gt,\e,\ga)\from\BilinearSetup(1^\lambda); \\& (g_1,g_2,g_3) \leftarrow\Ga; (a,b,c) \leftarrow \mathbb{Z}_p]\bigr|
\end{align*}
The Decision Linear assumption states that the advantage of $\mathcal{A}$ is negligible in $\lambda$. Boneh \etal\ \cite{C:BonBoySha04} provide a bilinear group generator $\BilinearSetup$ for which such assumption is conjectured to hold.
\end{assm}
\paragraph{Decision Linear Encryption Scheme.} Consider the following PKE scheme described by a setup algorithm $\Setup$, an encryption algorithm $\Enc$ and a decryption algorithm $\Dec$.
\begin{description}
	\item[$\Setup(1^\lambda)$:] pick
	$(\p,\Ga,\allowbreak \Gt,\e,\ga)\from\BilinearSetup(1^\lambda)$, pick randomly $g_3 \leftarrow\Ga$ and $(x,y) \leftarrow \mathbb{Z}_p$. Compute $g_1 = g_3^{1/x}$ and $g_2 = g_3^{1/y}$. Output the public key $\PK=(\mathbb{G},\allowbreak p,\allowbreak g_1,\allowbreak g_2,\allowbreak g_3)$ and the secret key $\SK\allowbreak=\allowbreak(\PK,\allowbreak x,\allowbreak y)$.
\item[$\Enc(\PK,m)$:] on input a public key $\PK$ and a message $m \in\Ga$, pick random $(a,b) \leftarrow \mathbb{Z}_p$. Output a ciphertext $\CT=(g_1^a,g_2^b,m \cdot g_3^{a+b})$.
\item[$\Dec(\SK,\Ct)$:] on input a secret key $\SK$ and a ciphertext $\CT=(c_1,c_2,c_3)$, output $m=c_3/(c_1^x c_2^y)$.
\end{description}
This scheme fulfills the IND-CPA property under the Decision Linear assumption (see \cite{C:BonBoySha04} for details) and it is easy to verify that it fulfills the unique secret key property.



\begin{definition}[(Perfectly binding) Commitment Schemes]
\label{def:com}
A commitment scheme $\Com$ is a PPT algorithm that takes as input a string $x$ and randomness $r\in\zu^k$
and outputs $\com \from \Com(x; r)$. A perfectly binding commitment scheme must satisfy the following
properties:
\begin{itemize}
	\item{Perfectly Binding}: This property states that two different strings cannot have the same
commitment. More formally, $\forall x_1 \neq x_2$ and $r_1, r_2, \Com(x_1; r_1) \neq \Com(x_2; r_2)$.
\item{Computational Hiding}: For all strings $x_0$ and $x_1$ (of the same length), for all PPT
	adversaries $\A$ there exists a negligible function $\nu(\cdot)$ such that:
	$|\probb{r\in\zu^k}{\A(\Com(x_0;r)) = 1} - \probb{r\in\zu^k}{\A(\Com(x_1;r)) = 1)}| \le \nu(k). $

\end{itemize}
\end{definition}
\paragraph{NIWI proof systems.}\label{sec:niwi}
Next, we define (one-message) non-interactive witness indistinguishability (NIWI) proof systems \cite{C:GroOstSah06}. Groth \etal\ \cite{C:GroOstSah06} construct such NIWIs for all languages in $\NP$, and in particular for $\CircuitSat$.

\begin{definition}[Non-interactive Proof System] A non-interactive proof system for a language
	$L$ with a PPT relation $R$ is a tuple of algorithms $(\Prove, \Verify)$. $\Prove$ receives as input a statement $x$ and a witness $w$ and outputs a proof $\pi$. $\Verify$ receives as input a statement $x$ and a proof $\pi$ and outputs a symbol in $\{\OK,\bot\}$. The following
properties must hold:
\begin{itemize}
	\item{Perfect Completeness}: For every $(x, w) \in R$, it holds that\\
		$\Prob{\Verify(x, \Prove(x, w)) = \OK} = 1$,
where the probability is taken over the coins of $\Prove$ and
$\Verify$.
\item{Perfect Soundness}: For every adversary $\A$, it holds that:
$$\Prob{
	\begin{array}{lcl}
		\Verify(x, \pi) = \OK \wedge x \notin  L :\\
		(x, \pi) \from \A(1^k)\\				
	\end{array}
}=0.$$
\end{itemize}
\end{definition}
\begin{definition}[NIWI]\label{def:niwi} A non-interactive proof system $\NIWI=(\Prove, \Verify)$ for
	a language $L$ with a PPT relation $R$ is witness-indistinguishable (WI, in short) if for any triplet $(x, w_0, w_1)$ such
that $(x, w_0) \in R$ and $(x, w_1) \in R$, the distributions $\{\Prove(x, w_0)\}$ and $\{\Prove(x, w_1)\}$
are computationally indistinguishable.
\end{definition}


\section{Our Weakly Verifiable eVote}\label{sec:scheme}

In this section, we present our weakly verifiable eVote $\EVOTE$. This eVote fulfills the \wIND-Security and weak verifiability properties.
\begin{definition}[$\EVOTE$]  Let $\NIWIDec = (\ProveDec,\VerifyDec)$ be a NIWI proof system for the relation $\Rdec$, which we specify later.	Let $\PKE\allowbreak=\allowbreak(\PKE.\Setup,\allowbreak\PKE.\Enc,\allowbreak\PKE.\Dec)$ be a PKE scheme with perfect correctness and unique secret key (see Def. \ref{def:pke}).

We define as follows an $(N,\M,\Sigma,F)$-eVote
\begin{equation*}
\EVOTE^{N,\M,\Sigma,F,\PKE,\NIWIDec}=(\Setup,\Cast,\VerifyBallot,\EvalTally,\VerifyTally)
\end{equation*}

\begin{itemize}
	\item $\Setup(1^\lambda)$: on input the security parameter in unary, do the following.
		\begin{enumerate}
			\item For all $l\in[3]$, run $(\PKE.\PK_l,\PKE.\SK_l)=\PKE.\Setup(1^\lambda;s_l)$ with randomness $s_l$.
			\item Output $\PK\defeq(\PKE.\PK_1,\ldots,\PKE.\PK_3)$ and
				$\SK\defeq(\PKE.\SK_1,\PKE.\SK_2,s_1,s_2)$.\footnote{Actually, as the randomness for the setup of our PKE scheme uniquely determines the secret key, it would be sufficient to just include the $s_l$'s in $\SK$.}
		\end{enumerate}

\item $\Cast(\PK,j,v)$:
	on input the public key $\PK$, the voter index $j\in[N]$, and a vote $v$, do the following.
	\begin{enumerate}
		\item For all $l\in[3]$,  compute $\Ct_{j,l}\from\PKE.\Enc(\PKE.\PK_l,v)$.
		\item Output $\BT_j\defeq(\Ct_{j,1},\ldots,\CT_{j,3})$.
	\end{enumerate}

\item $\VerifyBallot(\PK,j,\BT)$:
	on input the public key $\PK$,
	the voter index $j\in[N]$, and a ballot $\BT$,
	output $\OK$ (i.e., accept any ballot, even invalid ones).

\item $\EvalTally(\PK,\SK,\BT_1,\ldots,\BT_N)$:
	on input the public key $\PK$, the secret key $\SK$, and a tuple of $N$ strings $(\BT_1,\ldots,\BT_N)$ that consists of either ballots cast by a voter or the special symbol $\bot$, do the following.
	\begin{enumerate}
		\item For all $j\in[N],l\in[2]$,
				$$m_j^{l}=
				\begin{cases} \bot &\mbox{if } \BT_j=\bot, \\
					\bot &\mbox{if } \BT_j\neq\bot \wedge \PKE.\Dec(\Ct_{j,l},\PKE.\SK_{l})\notin\M,\\
					\PKE.\Dec(\Ct_{j,l},\PKE.\SK_l)	& \mbox{otherwise. }
				\end{cases}
				$$
\item For all $l\in[2]$, compute $y_l= F(m_{1,l},\ldots,m_{N,l}).$
\item If $y_1=y_2$, then set $y=y_1$, else set $y=\bot$.

\item Consider the following relation $\Rdec$ in Fig.~\ref{fig:relationdecrypt}. Henceforth, if the indices $(i_1,i_2)$ in the witness of the relation $\Rdec$ fulfill $i_1=1$ and $i_2=2$ (resp. $i_1\neq 1$ or $i_2\neq 2$), the statement or the proof is in real mode (resp. trapdoor mode). Set the statement
    \begin{equation*}
    x\defeq (\BT_1,\ldots,\BT_N,\PKE.\PK_1,\ldots,\PKE.\PK_3,y)
    \end{equation*}
    and the witness
    \begin{equation*}
    w \defeq (\PKE.\SK_1,\PKE.\SK_2,s_1,s_2,i_1\defeq 1,i_2\defeq 2)
    \end{equation*}
    and compute a proof $\gamma \from \ProveDec(x,w)$.

\item Output $(y,\gamma)$.

\begin{figure}

  \begin{framed}

Relation $\Rdec(x,w)$:\\

Instance: $x\defeq(\BT_1,\ldots,\BT_N,\PKE.\PK_1,\ldots,\PKE.\PK_3, y).$ (Recall that a ballot is set to $\bot$ if the corresponding voter did not cast her vote.)\\

Witness: $w \defeq (\PKE.\SK_1',\PKE.\SK_2',s_1,s_2,i_1,i_2)$, where the $s_l$'s are the randomness used to generate the secret keys and public keys (which are known to the authority who set up the system).\\

$\Rdec(x, w) = 1$ if and only if the following condition holds.\\

$2$ of the secret keys corresponding to indices $\PKE.\PK_{i_1}, \PKE.\PK_{i_2}$ are constructed using honestly generated public and secret key
		pairs and are equal to $\PKE.\SK_1',\PKE.\SK_2'$; and either $y=\bot$ or for all $l\in[2]$, $y=F(m_1^l,\ldots,m_N^l)$ and for all $j\in[N]$, if $\BT_j\neq\bot$ then for $l\in[2]$, $\PKE.\SK_{i_l}$ decrypts ciphertext $\Ct_{j,i_l}$ in $\BT_j$ to $m_j^{i_l}\in\M$; and for all $l\in[2]$, $m_j^l=\bot$ if either $\BT_j=\bot$ or $\PKE.\SK_{i_l}$ decrypts $\Ct_{j,i_l}$ to a string $\notin \M$.\\


		Precisely, $\Rdec(x, w) = 1$ if and only if the following conditions hold. In the following, items (a) and (c) are not actually conditions that have to be checked but are steps needed to define  (note the use of ``$\defeq$'') the variables $\PKE.\PK_{i_l}$'s, $\PKE.\SK_{i_l}$'s and $m_j^{i_l}$'s that are used in the checks (b) and (d).\\

		\begin{enumerate}
			\item For all $l\in[2], (\PKE.\PK_{i_l},\PKE.\SK_{i_l})\defeq \PKE.\Setup(1^\lambda;s_l)$.
			\item For all $l\in[2], \PKE.\SK_{l}'=\PKE.\SK_{i_l}$.
			\item For all $j\in[N],l\in[2],$
				$$m_j^{i_l}\defeq
				\begin{cases} \bot &\mbox{if } \BT_j=\bot, \\
					\bot &\mbox{if } \BT_j\neq\bot \wedge \PKE.\Dec(\Ct_{j,i_l},\PKE.\SK_{i_l})\notin\M,\\
					\PKE.\Dec(\Ct_{j,i_l},\PKE.\SK_{i_l}) & \mbox{otherwise. }
				\end{cases}
				$$
			\item $(y=\bot)$ $\vee$ (for all $l\in[2]$, $y=F(m_1^{i_l},\ldots,m_N^{i_l})$).
		\end{enumerate}
		(Note that $\PKE.\SK_1'$ and $\PKE.\SK_2'$ do not necessarily have to correspond to the first two secret keys.)

\end{framed}
\caption{Relation $\Rdec$}
\label{fig:relationdecrypt}
\end{figure}

\medskip

\end{enumerate}
\item $\VerifyTally(\PK,\BT_1,\ldots,\BT_N,y,\gamma)$: on input the public key $\PK$, a tuple of $N$ strings that can be either ballots cast by a voter or the special symbol $\bot$, a tally $y$ and a proof $\gamma$,
		if $\gamma=\bot$ output $\bot$, else set
  \begin{equation*}
    x\defeq (\BT_1,\ldots,\BT_N,\PKE.\PK_1,\ldots,\PKE.\PK_3,y)
    \end{equation*}
 output $\VerifyDec(x,\gamma)$.
	
\end{itemize}
\end{definition}
Henceforth, for simplicity we omit the parameters of the scheme and we write just $\EVOTE$.
\subsection{Correctness and Weak Verifiability of the Construction}\label{sec:schemever}
\paragraph{\bf Correctness.}  The (perfect) correctness of $\EVOTE$ follows from the perfect correctness of $\PKE$ and the perfect completeness of $\NIWIDec$.
\paragraph{\bf Weak verifiability.}

\begin{theorem}\label{thm:weakver}
	For all $N>0$, all sets $\M,\Sigma\subset\zu^\star$, and all tally functions $F:(\M\cup\{\bot\})^N\rightarrow\Sigma\cup\{\bot\}$, if $\PKE$ is a perfectly correct PKE with unique secret key (cf. Def. \ref{def:pke}) and $\NIWIDec$ is a (one-message) NIWI (cf. Def. \ref{def:niwi}) for the relation $\Rdec$, then $\EVOTE^{N,\M,\Sigma,F,\PKE,\NIWIDec}$ satisfies the weak verifiability property (cf. Def. \ref{def:evote}).
\end{theorem}

\begin{proof}
	First, we prove that condition (1) of verifiability is satisfied. Since algorithm $\VerifyBallot$ accepts any ballot, even invalid ones, we have to prove that
	for all $\PK\in\zu^\star$, all $\BT_1,\ldots,\BT_N\in\zu^\star\cup \{\bot\}$, there exist $m_1,\ldots,m_N\in\M\cup\{\bot\}$ such that for all $y\neq\bot$ and all $\gamma$ in $\zu^\star$,	if $\VerifyTally(\PK,\allowbreak\BT_1,\allowbreak\ldots,\allowbreak\BT_N,\allowbreak y,\allowbreak \gamma)\allowbreak =\allowbreak 1$ then  $y=F(m_1,\ldots,m_N)$.

	Henceforth, w.l.o.g, we let $\PK$ and $\BT_1,\ldots,\BT_N$ be arbitrary strings.
	First, we prove the following claim.
	\begin{claim}
		Given $\PK$ and $(\BT_1,\ldots,\BT_N)$, for every two pairs $(y_0,\gamma_0)$ and $(y_1,\gamma_1)$ such that $y_0,y_1\neq\bot$, if $\VerifyTally(\PK,\allowbreak\BT_1,\allowbreak\ldots,\allowbreak\BT_N,\allowbreak y_0,\allowbreak \gamma_0)\allowbreak =\allowbreak \VerifyTally(\PK,\allowbreak\BT_1,\allowbreak\ldots,\allowbreak\BT_N,\allowbreak y_1,\allowbreak\gamma_1)\allowbreak =\allowbreak \OK$ then $y_0=y_1$.
\end{claim}
Let $y_0,\gamma_0,y_1,\gamma_1$ be arbitrary strings in $\zu^\star\cup\{\bot\}$ such that $y_0,y_1\neq\bot$.
Suppose that $\VerifyTally(\PK,\allowbreak\BT_1,\allowbreak\ldots,\allowbreak\BT_N,\allowbreak y_0,\allowbreak\gamma_0)\allowbreak =\allowbreak\VerifyTally(\PK,\allowbreak\BT_1,\allowbreak\ldots,\allowbreak\BT_N,\allowbreak y_1,\allowbreak\gamma_1)\allowbreak =\allowbreak\OK$.
	The perfect soundness of $\NIWIDec$ implies that, for all $b\in\zu$, the proof $\gamma_b$ is computed on input some witness
	$(\PKE.\SK_1'^b,\PKE.\SK_2'^b,s_1^b,s_2^b,i_1^b, i_2^b)$.

By the pigeon principle, there exists an index $i^\star$ such that one of the following cases holds.
\begin{enumerate}
\item{$i^\star=i_1^0=i_2^1$.}
	For all $b\in\zu$, let $(m_1^{i^\star,b},\ldots,m_N^{i^\star,b})$ be the messages guaranteed by condition (iii) of relation $\Rdec$ for proof $\gamma_b$.
	Condition (i) for proof $\gamma_0$ (resp. $\gamma_1$) implies that the secret key $\SK_1'^0$ (resp. $\SK_2'^1$) is honestly computed and thus, the unique secret key property and the fact that it fulfills $\PKE.\PK_{i_1^0}=\PKE.\PK_{i^\star}$ (resp. $\PKE.\PK_{i_2^1}=\PKE.\PK_{i^\star}$) imply that for all $j\in[N]$,
	$\PKE.\Dec(\Ct_{j,i^\star},\PKE.\SK_1'^0)=\PKE.\Dec(\Ct_{j,i^\star},\PKE.\SK_2'^1)$.
	
	Furthermore, condition (ii) and (iii) for proof $\gamma_0$ (resp. $\gamma_1$) imply that for all $j\in[N]$, either $m_j^{i^\star,0}=\bot$
	or $m_j^{i^\star,0}=\PKE.\Dec(\Ct_{j,i^\star},\PKE.\SK_1')\in\M$ (resp. either $m_j^{i^\star,1}=\bot$ or $m_j^{i^\star,1}=\PKE.\Dec(\Ct_{j,i^\star},\PKE.\SK_2'^1)\in\M$).
   	
	Hence, for all $j\in[N]$, $m_j^{i^\star,0}=m_j^{i^\star,1}\in\M\cup\{\bot\}$.
	Now, condition (iv) for proof $\gamma_0$ (resp. $\gamma_1$) implies that
	either $y_0=F(m_1^{i_1^0,0},\ldots,m_N^{i_1^0,0})$ or $y_0=\bot$
	(resp. either $y_1=F(m_1^{i_2^1,1},\ldots,m_N^{i_2^1,1})$ or $y_1=\bot$) and, as by hypothesis $y_0,y_1\neq\bot$, it holds that $y_0=y_1$.	(Here, the ``weakness'' of $\EVOTE$ arises, i.e., it cannot be proven (fully) verifiable because it could occur that, for example, $y_0\neq y_1, y_0=\bot$.)
\item{$i^\star=i_2^0=i_1^1$.}
	This case is identical to the first one, except that we replace $i_1^0$ with $i_2^0$ and $i_2^1$ with $i_1^1$.
\item{$i^\star=i_1^0=i_1^1$.} 	
	This case is identical to the first one, except that we replace $i_2^1$ with $i_1^1$.
\item{$i^\star=i_2^0=i_2^1$.} 	
	This case is identical to the first one, except that we replace $i_1^0$ with $i_2^0$.
\end{enumerate}
In all cases, we have that, if $\VerifyTally(\PK, \allowbreak\BT_1,\allowbreak\ldots,\allowbreak\BT_N,\allowbreak y_0,\allowbreak\gamma_0)\allowbreak=\allowbreak\VerifyTally(\PK,\allowbreak\BT_1,\allowbreak\ldots,\allowbreak\BT_N,\allowbreak y_1,\allowbreak\gamma_1)\allowbreak=\allowbreak\OK$ then $y_0\allowbreak =\allowbreak y_1$. In conclusion, the claim is proved.

From the previous claim, it follows that there exists a {\em unique} value $y^\star$ such that,
		for all $(y,\gamma)$ such that $y\neq\bot$,
		if $\VerifyTally(\PK,\BT_1,\allowbreak \ldots,\BT_N,y,\gamma)=\OK$
		then $y=y^\star$ (1). Moreover, it is easy to see that,
		for all $(y,\gamma)$,
		if $\VerifyTally(\PK,\allowbreak\BT_1,\allowbreak \ldots,\allowbreak\BT_N,\allowbreak y,\allowbreak \gamma)\allowbreak=\allowbreak\OK$, there exist messages $m_1,\allowbreak \ldots,\allowbreak m_N\in\M\cup\{\bot\}$ such that $y \allowbreak=\allowbreak F(m_1,\allowbreak \ldots,\allowbreak m_N)$ (2).

		Now, we have two mutually exclusive cases.
		\begin{itemize}
			\item For all $(y,\gamma)$ such that $y\neq\bot$,
					$\VerifyTally(\PK,\BT_1,\allowbreak \ldots,\BT_N,y,\gamma)=\bot$.
				Then, letting $m_1,\ldots,m_N$ in the statement of the theorem be arbitrary messages in $\M\cup\{\bot\}$, the statement is verified with respect to $\PK$ and $\BT_1,\ldots,\BT_N$.
			\item There exists $(y',\gamma)$ such that $y'\neq\bot$ and $\VerifyTally(\PK,\BT_1,\allowbreak \ldots,\BT_N,y',\gamma)\allowbreak=\allowbreak\OK$.
				In this case, (2) implies that there exist $m_1',\ldots,m_N'\in\M\cup\{\bot\}$ such that $y'=F(m_1',\ldots,m_N')$ (3).
				Hence, (1) and (3) together imply that $y^\star=F(m_1',\ldots,m_N')$ (4).

				Therefore, for all $(y,\gamma)$ such that $y\neq\bot$, if $\VerifyTally(\PK,\BT_1,\BT_N,y,\gamma)=\OK$ then (by (1)) $y=y^\star=$ (by (4)) $=F(m_1',\ldots,m_N')$.

				Then, for $m_1\defeq m_1',\ldots,m_N\defeq m_N'$, the statement of condition (1) of weak verifiability is verified with respect to
				$\PK$ and $\BT_1,\ldots,\BT_N$.

		\end{itemize}
		In both cases, for $m_1\defeq m_1',\ldots,m_N\defeq m_N'$, the statement of condition (1) of weak verifiability is verified with respect to $\PK$ and $\BT_1,\ldots,\BT_N$.

		As  $\PK$ and $\BT_1,\ldots,\BT_N$ are arbitrary strings, the statement of condition (1) of weak verifiability is proven.
	
		It is also easy to check that condition (2) of weak verifiability is satisfied. This follows straightforwardly from the perfect soundness of $\NIWIDec$. Thanks to $\NIWIDec$, the authority always proves that the public key of the PKE scheme is honestly generated. Therefore, by the perfect correctness of the PKE scheme, an honestly computed ballot for message $m$ for the $j$-th voter is decrypted to $m$ (because an honestly computed ballot, by definition, consists of three ciphertexts that encrypt the same message). Consequently, if the tally $y$ is different from $\bot$ (i.e., if the evaluation of the tally function is equal for all indices), then $y$ has to be compatible with $m$ at index $j$ (cf. Def.~\ref{def:restriction}).
\end{proof}


\subsection{Weak Privacy of the Construction}\label{sec:privacy}\label{sec:privacyweak}

\begin{theorem}\label{thm:priv}
	For all $N>0$, all sets $\M,\Sigma\subset\zu^\star$, and all tally functions $F:(\M\cup\{\bot\})^N\rightarrow\Sigma\cup\{\bot\}$, if $\PKE$ is a perfectly correct PKE scheme with unique secret key (cf.\ Def.~\ref{def:pke}) and $\NIWIDec$ is a (one-message) NIWI (cf. Def. \ref{def:niwi}) for the relation $\Rdec$, then $\EVOTE^{N,\M,\Sigma,F,\PKE,\NIWIDec}$ is \wIND-Secure (cf.\ Def.~\ref{def:evoteweakpriv}).
\end{theorem}
	\begin{proof}
		Let $\adv$ be a PPT adversary against the \wIND-Security property of $\EVOTE$. 
		We prove that $\Adv_{\adv}^{\EVOTE,\SecwGame}(1^\lambda)\le\nu(1^\lambda)$ for some negligible function $\nu(\lambda)$.

		We prove that by means of a series of hybrid experiments.
		We refer the reader to Table~\ref{table} for a pictorial explanation of the experiments, which are explained in Section~\ref{sec:sketchproofsweakscheme}. In the table, for simplicity, we omit the indices $k$ for the experiments $H_2^k$'s,$H_4^k$'s,$H_6^k$'s presented below. Therefore, hybrid experiment $H_2$ (resp. $H_4$, $H_6$) in the table corresponds to hybrid experiment $H_2^N$ (resp. $H_4^N,H_6^N$) below.

		\paragraph{{\bf Hybrid $H_1$.}} Experiment $H_1$ is equal to the experiment $\SecwGame^{N,\M,\Sigma,F,\EVOTE}_\adv$ except that the challenger sets $b\defeq 0$.

\paragraph{{\bf Hybrid $H_2^k,$} for $k=0,\ldots,N$.}
For all $k=0,\ldots,N$, experiment $H_2^k$ is identical to experiment $H_1$ except that, for all $j=1,\ldots,k$, the challenger computes $\Ct_{k,3}$ on input $m_{1,k}$.
Note that $H_2^0$ is identical to $H_1$.

\begin{claim}\label{clm:hybridtwo}
	For all $k=1,\ldots,N$, the advantage of $\adv$ in distinguishing $H_2^{k-1}$ from $H_2^k$ is negligible.
\end{claim}
\begin{proof}
	Suppose toward a contradiction that $\adv$ has instead non-negligible advantage $\epsilon(\lambda)$.
	We construct an adversary $\B$ that has advantage at most $\epsilon(\lambda)$ against the IND-CPA security of $\PKE$.

	$\B$ receives from the challenger of IND-CPA a public key $\pk$ and sets $\PK_3\defeq\pk$. For $l\in[2]$, $\B$ runs $\PKE.\Setup$ to compute $(\PKE.\PK_l,\PKE.\SK_l)$ and runs $\A$ on input $\PK\defeq(\PKE.\PK_1,\PKE.\PK_2,\PKE.\PK_3)$.

	$\A$ outputs two tuples $(m_{0,1},\ldots,m_{0,N})$ and $(m_{1,1},\ldots,m_{1,N})$, and a set $S$, which is empty for the \wIND-Security game. 	$\B$ returns $(m_{0,k},m_{1,k})$ as its pair of challenge messages to the IND-CPA challenger. The IND-CPA challenger sends $\B$ the challenge ciphertext $\ct^\star$.


	$\B$ computes $\BT_k\defeq(\Ct_{k,1},\Ct_{k,2},\ct^\star)$ by encrypting $m_{0,j}$ in $\Ct_{k,1}$ and $\Ct_{k,2}$.
	$\B$ can compute the ballots $\BT_j$ for all $j\in[N],j\neq k$ exactly as the challenger in both experiments would do.

	$\B$ computes $y$ as in the previous experiment (i.e., by running $\EvalTally$ on input $(\PK,\BT_1,\ldots,\BT_N)$) and uses the $2$ secret keys $(\PKE.\SK_1,\PKE.\SK_2)$ to compute a proof $\gamma$ exactly as the challenger in both experiments would do. $\B$ restarts $\A$ on input the computed ballots along with $(y,\gamma)$ and returns the output of $\A$.

	It is easy to see that, if $\ct^\star$ is an encryption of $m_{0,k}$, then $\B$ simulates experiment $H_2^{k-1}$, and, if
	$\ct^\star$ is an encryption of $m_{1,k}$, then $\B$ simulates experiment $H_2^{k}$. Therefore, $\B$ has probability $\epsilon(\lambda)$ of winning the IND-CPA game, which contradicts the assumption that the PKE scheme fulfills the IND-CPA property.
\end{proof}

\paragraph{{\bf Hybrid $H_3$}.} Experiment $H_3$ is identical to experiment $H_2^N$ except that the challenger computes the proof $\gamma$ on input a witness that contains indices $(1,3)$ and secret keys $\SK_1,\SK_3$ (precisely, with the randomness used to compute those secret keys, but henceforth, for simplicity, we omit this detail).
\begin{claim}\label{clm:hybridthree}
	The advantage of $\adv$ in distinguishing $H_2^{N}$ from $H_3$ is negligible.
\end{claim}
\begin{proof}
	This follows straightforwardly from the WI property of $\NIWIDec$. We note that both the randomness used to compute $\SK_1,\SK_2$ and the randomness used to compute $\SK_1,\SK_3$ constitute valid witnesses for the statement $(\BT_1, \allowbreak \ldots, \allowbreak \BT_N, \allowbreak \PKE.\PK_1, \allowbreak \ldots, \allowbreak \PKE.\PK_3, \allowbreak y)$.
	\end{proof}

	\paragraph{{\bf Hybrid $H_4^k,$} for $k=0,\ldots,N$.}
	For all $k=0,\ldots,N$, experiment $H_4^k$ is identical to experiment $H_3$, except that, for all $j=1,\ldots,k$, the challenger computes $\Ct_{k,2}$ on input $m_{1,k}$.
	Note that $H_4^0$ is identical to $H_3$.

\begin{claim}
	For all $k=1,\ldots,N$, the advantage of $\adv$ in distinguishing $H_4^{k-1}$ from $H_4^k$ is negligible.
\end{claim}
\begin{proof}
	The proof is identical to the one for Claim~\ref{clm:hybridtwo} except that the third index and the second index are swapped.
	\end{proof}

	\paragraph{{\bf Hybrid $H_5$}.} Experiment $H_5$ is identical to experiment $H_4^N$ except that the challenger computes the proof $\gamma$ on input a witness that contains indices $(2,3)$ and secret keys $\SK_2,\SK_3$.
\begin{claim}
	The advantage of $\adv$ in distinguishing $H_4^{N}$ from $H_5$ is negligible.
\end{claim}
\begin{proof}
	This follows straightforwardly from the WI property of $\NIWIDec$. We note that both the randomness used to compute $\SK_1,\SK_3$ and the randomness used to compute $\SK_2,\SK_3$ constitute valid witnesses for the statement $(\BT_1, \allowbreak \ldots, \allowbreak \BT_N, \allowbreak \PKE.\PK_1, \allowbreak \ldots, \allowbreak \PKE.\PK_3, \allowbreak y)$.
	\end{proof}
	\paragraph{{\bf Hybrid $H_6^k,$} for $k=0,\ldots,N$.}
	For all $k=0,\ldots,N$, experiment $H_6^k$ is identical to experiment $H_5$ except that, for all $j=1,\ldots,k$, the challenger computes $\Ct_{k,1}$ on input $m_{1,k}$.
Note that $H_6^0$ is identical to $H_5$.

\begin{claim}
	For all $k=1,\ldots,N$, the advantage of $\adv$ in distinguishing $H_6^{k-1}$ from $H_6^k$ is negligible.
\end{claim}
\begin{proof}
	The proof is identical to the one for Claim~\ref{clm:hybridtwo} except that the third index and the first index are swapped.
	\end{proof}
	\paragraph{{\bf Hybrid $H_7$}.} Experiment $H_7$ is identical to experiment $H_6^N$ except that the challenger computes the proof $\gamma$ on input a witness that contains indices $(1,2)$ and secret keys $(\SK_1,\SK_2)$.
\begin{claim}
	The advantage of $\adv$ in distinguishing $H_6^{N}$ from $H_7$ is negligible.
\end{claim}
\begin{proof}
	This follows straightforwardly from the WI property of $\NIWIDec$. We note that both the randomness used to compute $\SK_1,\SK_2$ and the randomness used to compute $\SK_2,\SK_3$ constitute valid witnesses for the statement $(\BT_1, \allowbreak \ldots, \allowbreak \BT_N, \allowbreak \PKE.\PK_1, \allowbreak \ldots, \allowbreak \PKE.\PK_3, \allowbreak y)$.
	\end{proof}
	\ignore{
	\paragraph{{\bf Hybrid $H_7',$}.} Experiment $H_7'$ is identical to experiment $H_7$ except that the tally $y$ is computed as in experiment $H_1$ (i.e., using $\EvalTally$).
\begin{claim}
	The advantage of $\adv$ in distinguishing $H_7$ from $H_7'$ is negligible.
\end{claim}
\begin{proof}
	The proof is symmetrical to the proof of Claim \ref{clm:hybridone} with the messages $m_{0,j}$'s swapped with the messages $m_{1,j}$'s.	
\end{proof}
}

Experiment $H_1$ (resp. $H_7$) is identical to experiment $\SecwGame^{N,\M,\Sigma,F,\EVOTE}_\adv$ except that the challenger sets $b=0$ (resp. $b=1$). Hence $\Adv_{\adv}^{\EVOTE,\SecwGame}(1^\lambda)$ equals at most the sum of the advantages of $\adv$ in distinguishing the previous hybrids. Since $N$ is a constant, such advantage is negligible and the theorem is proven.
	\end{proof}

	\begin{corollary} \label{cor:weak}
		\ifnum\fullversion=1
		If the Decision Linear assumption (see Section \ref{app:dlin}) holds, then there exists a weakly verifiable eVote.
		\else
		If the Decision Linear assumption (see Appendix \ref{app:dlin}) holds, then there exists a weakly verifiable eVote.
		\fi
	\end{corollary}
	\begin{proof}
		Boneh \etal\ \cite{C:BonBoySha04} show the existence of a PKE scheme with perfect correctness and unique secret key that fulfills the IND-CPA property under the Decision Linear assumption. Groth \etal\ \cite{C:GroOstSah06} show the existence of (one-message) NIWI proofs with perfect soundness for all languages in $\NP$ that is secure under the Decision Linear assumption. Then, because Theorem~\ref{thm:weakver} and Theorem~\ref{thm:priv} are proven, the corollary follows.
	\end{proof}





\section{Our (Fully) Verifiable eVote}\label{sec:fullyscheme}

In this Section, we present an eVote scheme $\EVOTEF$ that is \IND-Secure and (fully) verifiable.
\begin{definition}[$\EVOTEF$]
	Let $\PKE=(\PKE.\Setup,\PKE.\Enc,\PKE.\Dec)$ be a PKE scheme with perfect correctness and unique secret key (see Def. \ref{def:pke}). Let $\Com$ be a perfectly binding commitment scheme. Let $\NIWIDecf =(\ProveDecf,\VerifyDecf)$ and $\NIWIEncf = (\ProveEncf,\VerifyEncf)$ be two NIWI proof systems for the relations $\Rdecf$ and $\Rencf$, which we specify later.

We define as follows an $(N,\M,\Sigma,F)$-eVote
\begin{multline*}
\EVOTEF^{N,\M,\Sigma,F,\PKE,\Com,\NIWIEncf,\NIWIDecf}\\=(\Setupf,\Castf,\VerifyBallotf,\EvalTallyf,\VerifyTallyf)
\end{multline*}

\begin{itemize}
	\item $\Setupf(1^\lambda)$: on input the security parameter in unary, do the following.
		\begin{enumerate}
			\item Choose randomness $r\from\zu^\lambda$ and set $Z=\Com(1;r)$.
			\item For all $l\in[3]$, choose randomness $s_l\from\zu^\lambda$ and run $(\PKE.\PK_l,\PKE.\SK_l)=\PKE.\Setup(1^\lambda;s_l)$.
			\item Output $\PK\defeq(\PKE.\PK_1,\ldots,\PKE.\PK_3,Z)$ and
				$\SK\defeq(\PKE.\SK_1,\PKE.\SK_2,s_1,s_2,r)$.\footnote{Actually, as the randomness for the setup of our PKE scheme uniquely determines the secret key, it would be sufficient to just include the $s_l$'s in $\SK$.}
		\end{enumerate}	
\item $\Castf(\PK,j,v)$:
	on input the public key $\PK$, the voter index $j\in[N]$, and a vote $v$, do the following.
	\begin{enumerate}
			\item For all $l\in[3]$, choose randomness $r_l\from\zu^\lambda$ and compute $\Ct_{j,l}=\PKE.\Enc(\PKE.\PK_l,v;r_l)$.
		\item Consider the following relation $\Rencf$  in Fig.~\ref{fig:relationencrypt}. Run $\ProveEncf$ on input the statement
$(j,\Ct_1,\ldots,\Ct_3,\PKE.\PK_1,\ldots,\PKE.\PK_3, Z)$ and the witness $(v,r_1,\ldots,r_3)$ to compute a proof $\pi_j$. Output $\BT_j\defeq(\Ct_{j,1},\ldots,\CT_{j,3},\pi_j)$.

\medskip

\begin{figure}
  \begin{framed}
Relation $\Rencf(x,w)$:\\

Instance: $x\defeq(j,\Ct_1,\ldots,\Ct_3,\PKE.\PK_1,\ldots,\PKE.\PK_3, Z)$.\\

Witness : $w \defeq (m, r_1,\ldots,r_3, u)$, where the $r_l$'s are the randomness used to compute the ciphertexts $\CT_l$'s and $u$ is the randomness used to compute the commitment $Z$.\\

$\Rencf(x, w) = 1$ if and only if either of the following two conditions hold:\\

\begin{enumerate}
	\item{\bf Real mode.} All $3$ ciphertexts $(\Ct_1,\ldots,\Ct_3)$ encrypt the same string in $\M\cup\{\bot\}$.
		
		Precisely, for all $l\in[3]$, $\Ct_l=\PKE.\Enc(\PKE.\PK_l,m;r_l)$ and $m\in\M\cup\{\bot\}$.
		$$\text{OR}$$
	\item{\bf Trapdoor mode.} $Z$ is a commitment to $0$.

		Precisely, $Z=\Com(0;u)$.
\end{enumerate}
\end{framed}
\caption{Relation $\Rencf$}
\label{fig:relationencrypt}
\end{figure}


	\end{enumerate}

\item $\VerifyBallotf(\PK,j,\BT)$:
	on input the public key $\PK$,
	the voter index $j\in[N]$, and a ballot $\BT$,
	output $\VerifyEncf((j,\Ct_1,\ldots,\Ct_3,\PKE.\PK_1,\ldots,\PKE.\PK_3, Z),\pi)$.

\item $\EvalTallyf(\PK,\SK,\BT_1,\ldots,\BT_N)$:
	on input the public key $\PK$, the secret key $\SK$, and $N$ strings $(\BT_1,\ldots,\BT_N)$ that can be either ballots cast by a voter or the special symbol $\bot$, do the following.
	\begin{enumerate}
		\item For all $j\in[N]$, if $\VerifyBallotf(\PK,j,\BT_j)=\bot$, set $\BT_j=\bot$. If, for all $j\in[N]$, $\BT_j=\bot$,  then output $(y=\bot,\gamma=\bot)$.
		\item Else, for all $j\in[N],l\in[2]$,
				$$m_j^{l}=
				\begin{cases} \bot &\mbox{if } \BT_j=\bot, \\
					\bot &\mbox{if } \BT_j\neq\bot \wedge \PKE.\Dec(\Ct_{j,l},\PKE.\SK_{l})\notin\M,\\
					\PKE.\Dec(\Ct_{j,l},\PKE.\SK_l)	& \mbox{otherwise.}
				\end{cases}
				$$
\item For all $l\in[2]$, compute $y_l= F(m_{1,l},\ldots,m_{N,l}).$
\item If $y_1=y_2$ then set $y=y_1$.

\item Consider the following relation $\Rdecf$ in Fig.~\ref{fig:relationdecryptfull}. (The relation $\Rdecf$ is identical to the relation $\Rdec$ that is used in our weakly verifiable eVote. The only difference is that the ballots in the statement of $\Rdecf$ are replaced by $\bot$ if they are not accepted by the ballot verification algorithm. Henceforth, if the indices $(i_1,i_2)$ in the witness of the relation $\Rdec$ fulfill $i_1=1$ and $i_2=2$ (resp. $i_1\neq 1$ or $i_2\neq 2$), the statement or the proof is in real mode (resp. trapdoor mode).) Run $\ProveDecf$ on input the statement $(\BT_1,\allowbreak \ldots,\allowbreak \BT_N,\allowbreak \PKE.\PK_1, \allowbreak \ldots,\allowbreak \PKE.\PK_3,\allowbreak y)$ and the witness $(\PKE.\SK_1, \allowbreak \PKE.\SK_2, \allowbreak s_1,\allowbreak s_2, \allowbreak i_1 = 1, \allowbreak i_2 = 2)$ to compute a proof $\gamma$.
\item Output $(y, \allowbreak\gamma)$.


\begin{figure}

  \begin{framed}

Relation $\Rdecf(x,w)$:\\

Instance: $x\defeq(\BT_1,\ldots,\BT_N,\PKE.\PK_1,\ldots,\PKE.\PK_3, y).$ (Recall that a ballot is set to $\bot$ if either the corresponding voter did not cast her vote or her ballot is not accepted by the ballot verification algorithm.)\\

Witness: $w \defeq (\PKE.\SK_1',\PKE.\SK_2',s_1,s_2,i_1,i_2)$, where the $s_l$'s are the randomness used to generate the secret keys and public keys (which are known to the authority who set up the system).\\

$\Rdecf(x, w) = 1$ if and only if the following condition holds.\\

$2$ of the secret keys corresponding to indices $\PKE.\PK_{i_1}, \PKE.\PK_{i_2}$ are constructed using honestly generated public and secret key
		pairs and are equal to $\PKE.\SK_1',\PKE.\SK_2'$; and either $y=\bot$ or for all $l\in[2]$, $y=F(m_1^l,\ldots,m_N^l)$ and for all $j\in[N]$, if $\BT_j\neq\bot$ then for $l\in[2]$, $\PKE.\SK_{i_l}$ decrypts ciphertext $\Ct_{j,i_l}$ in $\BT_j$ to $m_j^{i_l}\in\M$; and for all $l\in[2]$, $m_j^l=\bot$ if either $\BT_j=\bot$ or $\PKE.\SK_{i_l}$ decrypts $\Ct_{j,i_l}$ to a string $\notin \M$.\\


		Precisely, $\Rdecf(x, w) = 1$ if and only if the following conditions hold. In the following, items (a) and (c) are not actually conditions that have to be checked but are steps needed to define  (note the use of ``$\defeq$'') the variables $\PKE.\PK_{i_l}$'s, $\PKE.\SK_{i_l}$'s and $m_j^{i_l}$'s that are used in the checks (b) and (d).\\

		\begin{enumerate}
			\item For all $l\in[2], (\PKE.\PK_{i_l},\PKE.\SK_{i_l})\defeq \PKE.\Setup(1^\lambda;s_l)$.
			\item For all $l\in[2], \PKE.\SK_{l}'=\PKE.\SK_{i_l}$.
			\item For all $j\in[N],l\in[2],$
				$$m_j^{i_l}\defeq
				\begin{cases} \bot &\mbox{if } \BT_j=\bot, \\
					\bot &\mbox{if } \BT_j\neq\bot \wedge \PKE.\Dec(\Ct_{j,i_l},\PKE.\SK_{i_l})\notin\M,\\
					\PKE.\Dec(\Ct_{j,i_l},\PKE.\SK_{i_l}) & \mbox{otherwise. }
				\end{cases}
				$$
			\item $(y=\bot)$ $\vee$ (for all $l\in[2]$, $y=F(m_1^{i_l},\ldots,m_N^{i_l})$).
		\end{enumerate}
		(Note that $\PKE.\SK_1'$ and $\PKE.\SK_2'$ do not necessarily have to correspond to the first two secret keys.)

\end{framed}
\caption{Relation $\Rdecf$}
\label{fig:relationdecryptfull}
\end{figure}

\end{enumerate}
\item $\VerifyTallyf(\PK,\BT_1,\ldots,\BT_N,y,\gamma)$: on input the public key $\PK$, $N$ strings that can be either ballots cast by a voter or the special symbol $\bot$, a tally $y$ and a proof $\gamma$ of tally correctness, do the following. If $y=\bot$ and all $\BT_j$'s are equal to $\bot$, output $\OK$. If $y=\bot$ but not all $\BT_j$'s are equal to $\bot$, output $\bot$.  Otherwise output the decision of $\VerifyDecf((\BT_1,\allowbreak\ldots,\allowbreak\BT_N,\allowbreak\PKE.\PK_1,\allowbreak\ldots,\allowbreak\PKE.\PK_3, \allowbreak y), \allowbreak \gamma)$, after having replaced $\BT_j$'s with $\bot$ when $\VerifyBallotf(\PK,\allowbreak j,\allowbreak\BT_j)\allowbreak =\allowbreak\bot$. Precisely, the algorithm does the following:
		\begin{enumerate}
			\item For all $j\in[N]$, if $\VerifyBallotf(\PK,j,\BT_j)=\bot$, set $\BT_j=\bot$.
			\item If $y\neq\bot$, then output $\VerifyDecf((\BT_1,\ldots,\BT_N,\PKE.\PK_1,\ldots,\PKE.\PK_3,y),\gamma)$.
			\item If $y=\bot$, then, if for all $j\in[N],\BT_j=\bot$, output $\OK$, else output $\bot$.
		
		\end{enumerate}

\end{itemize}
\end{definition}
Henceforth, for simplicity we omit the parameters of the scheme and we just write $\EVOTEF$.


\subsection{Correctness and (Full) Verifiability of the Construction}\label{sec:schemeverfull}

\paragraph{\bf Correctness.}  Condition (1) of (perfect) correctness of $\EVOTEF$ follows from the perfect correctness of the PKE scheme and the perfect completeness of $\NIWIDecf$ and $\NIWIEncf$.
Condition (2) follows analogously. We note the following.
For all  honestly computed $\PK$, $\PK=(\PK_1,\PK_2,\PK_3,Z)$ holds for some $\PK_1,\PK_2,\PK_3$ and $Z$. $Z$ is a commitment to $1$. Therefore, relation $\Rencf$ and the perfectly binding property of the commitment scheme imply that, if there exists a proof $\pi$ and a statement $x=(j,\Ct_1,\ldots,\Ct_3,\PK_1,\ldots,\allowbreak\PK_3,\allowbreak Z)$ such that $\VerifyBallotf$ accepts $(x,\allowbreak\pi)$, then it must be the case that $\Ct_1,\ldots,\Ct_3$ encrypt the same string in $\M\cup\{\bot\}$. For all $j\in[N]$, if $\BT_j$ is accepted by $\VerifyBallotf$, $\BT_j' = \BT_j$, else $\BT_j'= \bot$. Therefore, for all $\BT_1,\ldots,\BT_N$, if $(y,\gamma) =\EvalTallyf(\PK,\BT_1,\ldots,\BT_N)$, then $y=F(m_1,\allowbreak\ldots,\allowbreak m_N)$,
where, for all $j\in[N]$, if $\BT_j$ is accepted by $\VerifyBallotf$, $m_j$ is the string encrypted in the first two ciphertexts of $\BT_j$, else $m_j$ is $\bot$. Then, it is easy to see that $\VerifyTallyf(\PK,\BT_1,\ldots,\BT_N,y,\gamma)=\OK$.

\paragraph{\bf (Full) verifiability.}

\begin{theorem}\label{thm:fullver}
	For all $N>0$, all sets $\M,\Sigma\subset\zu^\star$, and all tally functions $F:(\M\cup\{\bot\})^N\rightarrow\Sigma\cup\{\bot\}$, if $\PKE$ is a perfectly correct PKE scheme with unique secret key (cf. Def.~\ref{def:pke}), $\Com$ is a PPT algorithm, and $\NIWIDecf$ and $\NIWIEncf$ are (one-message) NIWIs (cf. Def.~\ref{def:niwi}), for the relations $\Rdecf$ and $\Rencf$ respectively, then $\EVOTEF^{N,\M,\Sigma,F,\PKE,\Com,\NIWIEncf,\NIWIDecf}$  satisfies the (full) verifiability property (cf. Def.~\ref{def:evote}).
\end{theorem}

\begin{proof}
	We first prove that condition (1) of verifiability is satisfied.
	We have to prove that,
	for all $\PK \allowbreak \in \allowbreak \zu^\star$, and all $\BT_1,\allowbreak\ldots,\allowbreak\BT_N\in\zu^\star\cup \{\bot\}$
	such that, for all $j\in[N]$, either $\BT_j=\bot$ or $\VerifyBallotf(\PK,\allowbreak j,\allowbreak\BT_j)\allowbreak =\allowbreak\OK$,
	there exist $m_1,\ldots,m_N\in\M\cup\{\bot\}$ such that, for all $y,\gamma\in\zu^\star$,
	if $\VerifyTallyf(\PK,\allowbreak\BT_1,\allowbreak\ldots,\allowbreak\BT_N,\allowbreak y,\allowbreak \gamma)=1$ then  $y=F(m_1,\allowbreak\ldots,\allowbreak m_N)$.	Henceforth, w.l.o.g, we let $\PK$ and $\BT_1,\ldots,\BT_N$ be arbitrary strings such that, for all $j\in[N]$, either $\BT_j=\bot$ or $\VerifyBallotf(\PK,j,\BT_j)=\OK$.
	
	First, we prove the following claim.
	\begin{claim}
		Given $\PK$ and $(\BT_1,\ldots,\BT_N)$, for every two pairs $(y_0,\gamma_0)$ and $(y_1,\gamma_1)$,
		if $\VerifyTallyf(\PK,\allowbreak\BT_1,\allowbreak\ldots,\allowbreak\BT_N,\allowbreak y_0,\allowbreak\gamma_0)=\VerifyTallyf(\PK,\allowbreak\BT_1,\allowbreak\ldots,\allowbreak\BT_N,\allowbreak y_1,\allowbreak\gamma_1)\allowbreak=\allowbreak\OK$ then $y_0=y_1$.
\end{claim}
For every $(y_0,\gamma_0)$ and $(y_1,\gamma_1)$, we have two cases.
\begin{enumerate}
	\item{Either $y_0=\bot$ and $y_1\neq\bot$ or $y_1=\bot$ and $y_0\neq\bot$.}
		Suppose w.l.o.g. that $y_0=\bot$ and $y_1\neq\bot$. The other case (i.e., $y_1=\bot$ and $y_0\neq\bot$) is symmetrical.

		By construction, for all $(y,\gamma)$, it holds that (A)
		if $\BT_1=\cdots=\BT_N=\bot$, then
		$\VerifyTallyf(\PK,\BT_1,\ldots,\BT_N,y,\gamma)=\OK$ if and only if
		$y=\bot$ and (B)
		if, for some $j\in[N],\BT_j\neq\bot$, then
	       	$\VerifyTallyf(\PK,\BT_1,\ldots,\BT_N,\bot,\gamma)=\bot$.
	We now have two cases.
	\begin{enumerate}
		\item{$\BT_1=\cdots=\BT_N=\bot$.}
			Then we have that $\VerifyTallyf(\PK,\allowbreak\BT_1,\allowbreak\ldots,\allowbreak\BT_N,\allowbreak y_1,\allowbreak\gamma_1)\allowbreak=\allowbreak\OK$ and by (A) $y_1 \allowbreak = \allowbreak \bot$, which is a contradiction.
		\item{It is not the case that $\BT_1=\cdots=\BT_N=\bot$.} Then, by (B) we have that $\VerifyTallyf(\PK,\allowbreak\BT_1,\allowbreak\ldots,\allowbreak\BT_N,\allowbreak y_0,\allowbreak\gamma_0)\allowbreak=\allowbreak\bot$, which contradicts the fact that $(y_0,\gamma_0)$ is accepted.
	\end{enumerate}	
	
	\item{$y_0,y_1\neq\bot$.}

Let $y_0,\gamma_0,y_1,\gamma_1$ be arbitrary strings in $\zu^\star\cup\{\bot\}$ such that $y_0,y_1\neq\bot$.
Suppose that $\VerifyTallyf(\PK,\allowbreak\BT_1,\allowbreak\ldots,\allowbreak\BT_N,\allowbreak y_0,\allowbreak\gamma_0)\allowbreak =\allowbreak\VerifyTallyf(\PK,\allowbreak\BT_1,\allowbreak\ldots,\allowbreak\BT_N,\allowbreak y_1,\allowbreak\gamma_1)\allowbreak =\allowbreak\OK$.
	The perfect soundness of $\NIWIDecf$ implies that, for all $b\in\zu$, the proof $\gamma_b$ is computed on input some witness
	$(\PKE.\SK_1'^b,\PKE.\SK_2'^b,s_1^b,s_2^b,i_1^b, i_2^b)$.

By the pigeon principle, there exists an index $i^\star$ such that one of the following cases holds.
\begin{enumerate}
\item{$i^\star=i_1^0=i_2^1$.}
	For all $b\in\zu$, let $(m_1^{i^\star,b},\ldots,m_N^{i^\star,b})$ be the messages guaranteed by condition (iii) of relation $\Rdecf$ for proof $\gamma_b$.
	Condition (i) for proof $\gamma_0$ (resp. $\gamma_1$) implies that the secret key $\SK_1'^0$ (resp. $\SK_2'^1$) is honestly computed and thus, the unique secret key property and the fact that it fulfills $\PKE.\PK_{i_1^0}=\PKE.\PK_{i^\star}$ (resp. $\PKE.\PK_{i_2^1}=\PKE.\PK_{i^\star}$) imply that for all $j\in[N]$,
	$\PKE.\Dec(\Ct_{j,i^\star},\PKE.\SK_1'^0)=\PKE.\Dec(\Ct_{j,i^\star},\PKE.\SK_2'^1)$.
	
	Furthermore, condition (ii) and (iii) for proof $\gamma_0$ (resp. $\gamma_1$) imply that for all $j\in[N]$, either $m_j^{i^\star,0}=\bot$
	or $m_j^{i^\star,0}=\PKE.\Dec(\Ct_{j,i^\star},\PKE.\SK_1')\in\M$ (resp. either $m_j^{i^\star,1}=\bot$ or $m_j^{i^\star,1}=\PKE.\Dec(\Ct_{j,i^\star},\PKE.\SK_2'^1)\in\M$).
   	
	Hence, for all $j\in[N]$, $m_j^{i^\star,0}=m_j^{i^\star,1}\in\M\cup\{\bot\}$.
	Now, condition (iv) for proof $\gamma_0$ (resp. $\gamma_1$) implies that
	either $y_0=F(m_1^{i_1^0,0},\ldots,m_N^{i_1^0,0})$ or $y_0=\bot$
	(resp. either $y_1=F(m_1^{i_2^1,1},\ldots,m_N^{i_2^1,1})$ or $y_1=\bot$) and, as by hypothesis $y_0,y_1\neq\bot$, it holds that $y_0=y_1$.

\item{$i^\star=i_2^0=i_1^1$.}
	This case is identical to the first one, except that we replace $i_1^0$ with $i_2^0$ and $i_2^1$ with $i_1^1$.
\item{$i^\star=i_1^0=i_1^1$.} 	
	This case is identical to the first one, except that we replace $i_2^1$ with $i_1^1$.
\item{$i^\star=i_2^0=i_2^1$.} 	
	This case is identical to the first one, except that we replace $i_1^0$ with $i_2^0$.
\end{enumerate}
\end{enumerate}
In all cases,  if $\VerifyTallyf(\PK, \allowbreak\BT_1,\allowbreak\ldots,\allowbreak\BT_N,\allowbreak y_0,\allowbreak\gamma_0)\allowbreak=\allowbreak\VerifyTallyf(\PK,\allowbreak\BT_1,\allowbreak\ldots,\allowbreak\BT_N,\allowbreak y_1,\allowbreak\gamma_1)\allowbreak=\allowbreak\OK$ then $y_0\allowbreak =\allowbreak y_1$. In conclusion, the claim is proved.

From the previous claim, it follows that there exists a {\em unique} value $y^\star$ such that,
		for all $(y,\gamma)$ such that $y\neq\bot$,
		if $\VerifyTallyf(\PK,\BT_1,\allowbreak \ldots,\BT_N,y,\gamma)=\OK$
		then $y=y^\star$ (1). Moreover, it is easy to see that,
		for all $(y,\gamma)$,
		if $\VerifyTallyf(\PK,\allowbreak\BT_1,\allowbreak \ldots,\allowbreak\BT_N,\allowbreak y,\allowbreak \gamma)\allowbreak=\allowbreak\OK$, there exist messages $m_1,\allowbreak \ldots,\allowbreak m_N\in\M\cup\{\bot\}$ such that $y \allowbreak=\allowbreak F(m_1,\allowbreak \ldots,\allowbreak m_N)$ (2).

		Now, we have two mutually exclusive cases.
		\begin{itemize}
			\item For all $(y,\gamma)$ such that $y\neq\bot$,
					$\VerifyTallyf(\PK,\BT_1,\allowbreak \ldots,\BT_N,y,\gamma)=\bot$.
				Then, letting $m_1,\ldots,m_N$ in the statement of the theorem be arbitrary messages in $\M\cup\{\bot\}$, the statement is verified with respect to $\PK$ and $\BT_1,\ldots,\BT_N$.
			\item There exists $(y',\gamma)$ such that $y'\neq\bot$ and $\VerifyTallyf(\PK,\BT_1,\allowbreak \ldots,\BT_N,y',\allowbreak\gamma)\allowbreak=\allowbreak\OK$.
				In this case, (2) implies that there exist $m_1',\ldots,m_N'\in\M\cup\{\bot\}$ such that $y'=F(m_1',\ldots,m_N')$ (3).
				Hence, (1) and (3) together imply that $y^\star=F(m_1',\ldots,m_N')$ (4).

				Therefore, for all $(y,\gamma)$ such that $y\neq\bot$, if $\VerifyTallyf(\PK,\BT_1,\BT_N,y,\gamma)=\OK$ then (by (1)) $y=y^\star=$ (by (4)) $=F(m_1',\ldots,m_N')$.

				Then, for $m_1\defeq m_1',\ldots,m_N\defeq m_N'$, the statement of condition (1) of weak verifiability is verified with respect to
				$\PK$ and $\BT_1,\ldots,\BT_N$.

		\end{itemize}
		In both cases, for $m_1\defeq m_1',\ldots,m_N\defeq m_N'$, the statement of condition (1) of weak verifiability is verified with respect to $\PK$ and $\BT_1,\ldots,\BT_N$.

		As  $\PK$ and $\BT_1,\ldots,\BT_N$ are arbitrary strings, the statement of condition (1) of weak verifiability is proven.

	It is also easy to check that condition (2) of weak verifiability is satisfied. This follows straightforwardly from the perfect soundness of $\NIWIDecf$. Thanks to $\NIWIDecf$, the authority always proves that the public key of the PKE scheme is honestly generated. Therefore, by the perfect correctness of the PKE scheme, an honestly computed ballot for message $m$ for the $j$-th voter is decrypted to $m$ (because an honestly computed ballot, by definition, consists of three ciphertexts that encrypt the same message, and thus the value committed to in $Z$ is not relevant).  Consequently, if the tally $y$ is different from $\bot$ (i.e., if the evaluation of the tally function is equal for all indices), then $y$ has to be compatible with $m$ at index $j$ (cf. Def.~\ref{def:restriction}).

In essence, condition (2) is satisfied because the degree of freedom of the authority in creating a dishonest public key only allows it to set up the commitment dishonestly. This does not affect how honest ballots are decrypted and ``counted''.
\end{proof}
Note that, for the proof of the theorem above, the security of the commitment scheme $\Com$ is not needed, i.e., the theorem holds for any PPT algorithm $\Com$, even insecure ones.


\subsection{Privacy of the Construction}\label{sec:privacyfull}

\begin{theorem}\label{thm:privfull}
	For all $N>0$, all sets $\M,\Sigma\subset\zu^\star$, and all tally functions $F:(\M\cup\{\bot\})^N\rightarrow\Sigma\cup\{\bot\}$, if $\PKE$ is a perfectly correct PKE scheme with unique secret key (cf. Def. \ref{def:pke}), $\Com$ is a computationally hiding commitment scheme (cf. Def. \ref{def:com}), and $\NIWIDecf$ and $\NIWIEncf$  are (one-message) NIWIs (cf. Def. \ref{def:niwi}), respectively, for the relations $\Rdecf$ and $\Rencf$, then $\EVOTEF^{N,\M,\Sigma,F,\PKE,\Com,\NIWIEncf,\NIWIDecf}$ is \IND-Secure (cf. Def. \ref{def:evotepriv}).
\end{theorem}
	\begin{proof}
		Consider the following experiment $H_{\adv}^Z(1^\lambda)$ between a challenger and $\adv$ (henceforth, we often omit the parameters).
		\paragraph{{\bf Experiment $H^Z$.}} $H^Z$ is equal to the experiment $\SecGame^{N,\M,\Sigma,F,\EVOTEF}_\adv$ except that the challenger sets the commitment $Z$ in the public key to be a commitment to $0$ instead of $1$. We define the output of the experiment to be a bit that is $1$ if and only if all winning conditions are satisfied.
Then, consider the following claim.
		\begin{claim}\label{clm:expz}
			The probability $P_0$ that $\adv$ wins the experiment
			$\SecGame^{N,\M,\Sigma,F,\EVOTEF}_\adv$
			is negligibly different from the probability $P_1$ that $\adv$ wins game $H^Z$.
\end{claim}
\begin{proof}
Suppose towards a contradiction that the difference between $P_0$ and $P_1$ is some non-negligible function $\epsilon(\lambda)$.
We construct an adversary $\B$ that breaks the computationally hiding property of $\Com$ with non-negligible probability.

$\B$ receives as input a commitment $\com$ that is either a commitment to $0$ or to $1$.
For $l\in[3]$, $\B$ runs $\PKE.\Setup(1^\lambda)$ to compute $(\PKE.\PK_l,\PKE.\SK_l)$ and sets the public key $\PK = (\PKE.\PK_1,\ldots,\PKE.\PK_3,Z = \com)$. $\B$ follows the challenger of  $\SecGame^{N,\M,\Sigma,F,\EVOTEF}_\adv$ to compute the remaining messages that are sent to the adversary. Finally, $\B$ gets the output $b'$ from $\adv$. $\B$ outputs $1$ if and only if all winning conditions are satisfied.


By hypothesis, if $\com$ is a commitment to $1$, the probability that $\B$ outputs $1$ equals the probability that $\adv$ wins in
			$\SecGame^{N,\M,\Sigma,F,\EVOTEF}_\adv$,
and if $\com$ is a commitment to $0$, the probability that $\B$ outputs $1$ equals the probability that $\adv$ wins in $H^Z$. Thus, the advantage of $\B$ in breaking the computational hiding property of $\Com$ is $\epsilon(\lambda)$, which contradicts the assumption that the commitment scheme is computationally hiding.

	\end{proof}


	Before continuing with the proof, we would like to remark a subtle point. In the previous claim, we implicitly assumed that the adversary $\B$ is able to check all of the winning conditions efficiently.
	This is possible if $\M$ is efficiently enumerable and its cardinality, as well as the number of voters $N$, are {\em constant} in the security parameter. This could seem like resorting to ``complexity leveraging'' arguments. In fact, one could ask if our proof would break down if $N$ and $\M$ depend on the security parameter. However, the whole proof can be generalized to the case of $N$ and $|\M|$ polynomial in the security parameter by using the following observation. Let $A$ be the event that $\adv$ submits challenges that satisfy the winning condition. Then, if the probability that $\adv$ wins the $\SecGame^{N,\M,\Sigma,F,\EVOTEF}_\adv$ is non-negligible, then the event $A$ must occur with non-negligible probability and, conditioned on it, $\A$ wins with non-negligible probability as well. Therefore, the rest of the proof would follow analyzing the probability that $\adv$ wins in the next hybrid experiments conditioned under the occurrence of the event that, in such experiments, $\adv$ submit challenges satisfying the winning condition. As we will see now, a similar ``conditioning'' argument will be anyhow necessary for the rest of the proof.

	\ignore{Let $E^0$ be the event that in $\SecGame^{N,\M,\Sigma,F,\EVOTEF}_\adv$, $\adv$ submits as challenge
	two tuples $M_0\defeq(m_{0,1},\ldots,m_{0,N})$ and
	$M_1\defeq(m_{1,1},\ldots,m_{1,N})$, and a set $S\subset[N]$ such that there exists $j\in S$ such that $m_{0,j}=m_{1,j}$ and, letting $B\defeq m_{0,j}\defeq (\Ct_1,\ldots,\CT_3)$ (and supposing that it can be parsed in a such way), it holds that $\VerifyBallot(\PK,B)=\OK$ but there exist $i_1,i_2\in[3],i_1\neq i_2$ such that $\PKE.\Dec(\Ct_{i_1},\SK_{i_1})\neq\PKE.\Dec(\Ct_{i_2},\SK_{i_2})$. That is, the event that the adversary submits a message representing a possible dishonest ballot that passes the verification ballot test but such that the three ciphertexts of which it is made up cannot be decrypted to the same message $\in\M\cup\{\bot\}$.

	By the perfect soundness of $\NIWIEncf$ and the definition of relation $\Rencf$, it is straightforward to see that the probability that event $E^0$ occurs is $0$.
}
	Let $E^1$ be the event that, in experiment $H^Z,$ $\adv$ submits as challenge
	two tuples $M_0 = (m_{0,1},\ldots,m_{0,N})$ and
	$M_1 = (m_{1,1},\ldots,m_{1,N})$ and a set $S\subset[N]$ that fulfill the following condition: there exists $j\in S$ such that $m_{0,j}=m_{1,j}$ and, letting $B = m_{0,j} = (\Ct_1,\ldots,\CT_3)$ (suppose that $m_{0,j}$ can be parsed that way), it holds that $\VerifyBallotf(\PK,j,B)=\OK$ but there exist $i_1,i_2\in[3],i_1\neq i_2$ such that $\PKE.\Dec(\Ct_{i_1},\SK_{i_1})\neq\PKE.\Dec(\Ct_{i_2},\SK_{i_2})$.

	\begin{claim}\label{clm:eventeone}
			The probability that $E^1$ occurs is negligible.
\end{claim}
\begin{proof}
Suppose towards a contradiction that the probability of occurrence of $E^1$ be some non-negligible function $\epsilon(\lambda)$.
We construct an adversary $\B$ that breaks the computationally hiding property of $\Com$ with non-negligible probability.

$\B$ receives as input a commitment $\com$ that is either a commitment to $0$ or to $1$.
For $l\in[3]$, $\B$ runs $\PKE.\Setup(1^\lambda)$ to compute $(\PKE.\PK_l,\PKE.\SK_l)$ and sets the public key $\PK = (\PKE.\PK_1,\ldots,\PKE.\PK_3,Z = \com)$.  $\B$ follows the challenger of  $\SecGame^{N,\M,\Sigma,F,\EVOTEF}_\adv$ to compute the remaining messages that are sent to the adversary. $\B$ receives two tuples $M_0 = (m_{0,1},\ldots,m_{0,N})$ and
	$M_1 = (m_{1,1},\ldots,m_{1,N})$  and a set $S\subset[N]$ from the adversary.

	
	For all $j\in S$, $\B$ checks whether the following conditions are all satisfied: $m_{0,j}=m_{1,j}$ and, after setting $B = m_{0,j}$, $B$ can be parsed as $(\Ct_1,\ldots,\CT_3)$ and it holds that $\VerifyBallotf(\PK,j,B)=\OK$ but there exist $i_1,i_2\in[3],i_1\neq i_2$ such that $\PKE.\Dec(\Ct_{i_1},\SK_{i_1})\neq\PKE.\Dec(\Ct_{i_2},\SK_{i_2})$.
If for some $j\in S$ the conditions are satisfied, $\B$ outputs $0$, otherwise it outputs $1$.

If $\com$ is a commitment to $1$, the perfect soundness of $\NIWIEncf$ and the definition of relation $\Rencf$ guarantee that the conditions above are never satisfied for any $j\in S$. Therefore,  if $\com$ is a commitment to $1$ $\B$ outputs $1$ with probability $1$.

On the other hand, if $\com$ is a commitment to $0$, the probability that the conditions are satisfied for some $j\in[S]$ equals the probability of $E^1$. Therfore, $\B$ outputs $0$ with probability $\epsilon$ and $1$ with probability $1-\epsilon$. In conclusion, the advantage of $\B$ in breaking the computationally hiding property of $\Com$ is $\epsilon(\lambda)$, which contradicts the assumption that the commitment scheme is computationally hiding.
	\end{proof}

From Claim~\ref{clm:expz} and Claim~\ref{clm:eventeone}, we now know that, for some negligible function $\negl(\cdot)$, the following equations hold:
\begin{equation}
	\left|\Prob{\SecGame=1}-\Prob{H^Z=1}\right|\le \negl(\lambda),
\end{equation}
\begin{equation} \label{eqn:one}
\Prob{E^1}\le \negl(\lambda),
\end{equation}
\begin{multline} \label{eqn:three}
\Prob{H^Z=1}=\Prob{H^Z=1|E^1}\Prob{E^1}+\Prob{H^Z=1|\bar E^1}\Prob{\bar E^1} \le \\ \negl + \Prob{H^Z=1|\bar E^1}(1-\negl).
\end{multline}

(Here and henceforth, we omit the parameters, but it is meant that the experiments are parameterized by $\lambda$ and $\negl(\cdot)$.)

Thus, to show that $\Prob{\Priv=1}$ equals $1/2$ plus a negligible quantity, it is sufficient to show that $\Prob{H^Z=1|\bar E^1}$ equals $1/2$ plus a negligible quantity. We prove the latter by means of a series of hybrid experiments. The reader could still refer to Table~\ref{table} for a pictorial explanation of the hybrid experiments. However, the experiments in the table, though conceptually very similar, correspond to the security reduction for the weakly verifiable eVote. Moreover, in the following we analyze the behavior of the adversary conditioned on the occurrence of the event $\bar E^1$.
		\paragraph{{\bf Hybrid $H_1$.}} Experiment $H_1$ is equal to the experiment $H^Z$ except that the challenger sets $b = 0$.
	
\paragraph{{\bf Hybrid $H_2^k,$} for $k=0,\ldots,N$.}
For all $k=0,\ldots,N$, experiment $H_2^k$ is identical to experiment $H_1$ except that, for all $j=1,\ldots,k$ such that $j\notin S$, the challenger computes $\Ct_{k,3}$ on input $m_{1,k}$.
Note that $H_2^0$ is identical to $H_1$.

\begin{claim}\label{clm:hybridtwofull}
	For all $k=1,\ldots,N$, $\left|\Prob{H_2^{k-1}=1|\bar E^1}-\Prob{H_2^k=1|\bar E^1}\right|$ is negligible.
\end{claim}
\begin{proof}
	Suppose toward a contradiction that the difference between such probabilities is non-negligible function $\epsilon(\lambda)$.
	We construct an adversary $\B$ that has advantage at most $\epsilon(\lambda)$ against the IND-CPA security of $\PKE$.

	$\B$ receives from the challenger of IND-CPA a public key $\pk$ and sets $\PK_3 = \pk$. For $l\in[2]$, $\B$ runs $\PKE.\Setup$ to compute $(\PKE.\PK_l,\PKE.\SK_l)$, computes $Z\from\Com(0)$ and runs $\A$ on input $\PK = (\PKE.\PK_1,\PKE.\PK_2,\PKE.\PK_3,Z)$.

	$\A$ outputs two tuples $(m_{0,1},\ldots,m_{0,N})$ and $(m_{1,1},\ldots,m_{1,N})$ and a set $S$. If $k\in S$, $\B$ sends $(0,0)$ as its pair of challenge messages to the IND-CPA challenger, which returns the challenge ciphertext $\ct^\star$ to $\B$.
	If $k\notin S$, $\B$ sends $(m_{0,k},m_{1,k})$ as its pair of challenge messages to the IND-CPA challenger, which returns  the challenge ciphertext $\ct^\star$ to $\B$.

	If $k\in S$, $\B$ sets $\BT_j$ as the challenger in the real experiment would do, else $\B$ computes $\BT_k = (\Ct_{k,1},\Ct_{k,2},\ct^\star)$ by computing $\Ct_{k,1}$ and $\Ct_{k,2}$ on input $m_{0,j}$.
	For all $j\in[N] (j\neq k)$, $\B$ computes the ballots $\BT_j$  exactly as the challenger in both experiments would do.
	$\B$ computes $y$ using $\EvalTallyf$ and uses the $2$ secret keys $\PKE.\SK_1,\PKE.\SK_2$ to compute a proof $\gamma$ exactly as the challenger in both experiments would do.
	$\B$ sends $\A$ the computed ballots along with $(y,\gamma)$ and returns the output of $\A$.
	
	It is easy to see that, if $\ct^\star$ is an encryption of $m_{0,k}$ and if $k\notin S$, then $\B$ simulates experiment $H_2^{k-1}$ and if
	$\ct^\star$ is an encryption of $m_{1,k}$ and $k\notin S$, then $\B$ simulates experiment $H_2^{k}$.
	If $k\in S$ the advantage of $\adv$ is $0$.

	Therefore, $\B$ has non-negligible probability of winning the IND-CPA game, which contradicts the assumption that the PKE scheme fulfills the IND-CPA property.
\end{proof}

\paragraph{{\bf Hybrid $H_3$}.} Experiment $H_3$ is identical to experiment $H_2^N$ except that the challenger computes the proof $\gamma$ on input a witness that contains indices $(1,3)$ and secret keys $(\SK_1,\SK_3)$ (precisely, the witness contains the randomness used to compute those secret keys, but henceforth, for simplicity, we omit this detail).
\begin{claim}\label{clm:hybridthreefull}
	$\left|\Prob{H_2^{N}=1|\bar E^1}-\Prob{H_3=1|\bar E^1}\right|$ is negligible.
\end{claim}
\begin{proof}
	The proof follows from the WI property of $\NIWIDecf$. We observe that both the randomness used to compute $(\SK_1,\SK_2)$ and the randomness used to compute $(\SK_1,\SK_3)$ constitute valid witnesses for the statement $(\BT_1, \allowbreak \ldots, \allowbreak \BT_N, \allowbreak \PKE.\PK_1, \allowbreak \ldots, \allowbreak \PKE.\PK_3, \allowbreak y)$. Additionally, we observe that, if event $\bar E^1$ occurs, any ballot in the set $S$ is in both experiments either replaced by $\bot$, if $\VerifyBallotf$ refuses it, or decrypted to the same value. Consequently, the tally is identical in both experiments.
	\end{proof}

	\paragraph{{\bf Hybrid $H_4^k,$} for $k=0,\ldots,N$.}
	For all $k=0,\ldots,N$, experiment $H_4^k$ is identical to experiment $H_3$ except that, for all $j=1,\ldots,k$ such that $j\notin S$, the challenger computes $\Ct_{k,2}$ on input $m_{1,k}$.
Note that $H_4^0$ is identical to $H_3$.

\begin{claim}
	For all $k=1,\ldots,N$, $\left|\Prob{H_4^{k-1}=1|\bar E^1}-\Prob{H_4^k=1|\bar E^1}\right|$ is negligible.
\end{claim}
\begin{proof}
	The proof is identical to the one for Claim~\ref{clm:hybridtwofull} except that the third index and the second index are swapped.
	\end{proof}

	\paragraph{{\bf Hybrid $H_5$}.} Experiment $H_5$ is identical to experiment $H_4^N$ except that the challenger computes the proof $\gamma$ on input a witness that contains indices $(2,3)$ and secret keys $(\SK_2,\SK_3)$.
\begin{claim}
	$\left|\Prob{H_4^{N}=1|\bar E^1}-\Prob{H_5=1|\bar E^1}\right|$ is negligible.
\end{claim}
\begin{proof}
	This follows straightforwardly from the WI property of $\NIWIDecf$. We observe that both the randomness used to compute $(\SK_1,\SK_3)$ and the randomness used to compute $(\SK_2,\SK_3)$ constitute valid witnesses for the statement  $(\BT_1, \allowbreak \ldots, \allowbreak \BT_N, \allowbreak \PKE.\PK_1, \allowbreak \ldots, \allowbreak \PKE.\PK_3, \allowbreak y)$. Additionally, we observe that, if event $\bar E^1$ occurs, any ballot in the set $S$ is in both experiments either replaced by $\bot$, if $\VerifyBallotf$ refuses it, or decrypted to the same value. Consequently, the tally is identical in both experiments.
	\end{proof}
	\paragraph{{\bf Hybrid $H_6^k,$} for $k=0,\ldots,N$.}
For all $k=0,\ldots,N$, experiment $H_6^k$ is identical to experiment $H_5$ except that, for all $j=1,\ldots,k$ such that $j\notin S$, the challenger computes $\Ct_{k,1}$ on input $m_{1,k}$.
Note that $H_6^0$ is identical to $H_5$.
\begin{claim}
	For all $k=1,\ldots,N$, $\left|\Prob{H_6^{k-1}=1|\bar E^1}-\Prob{H_6^k=1|\bar E^1}\right|$ is negligible.
\end{claim}
\begin{proof}
	The proof is identical to the one for Claim~\ref{clm:hybridtwofull} except that the third index and the first index are swapped.
	\end{proof}
	\paragraph{{\bf Hybrid $H_7$}.} Experiment $H_7$ is identical to experiment $H_6^N$ except that the challenger sets $b=1$ (so that the winning condition be computed differently) and computes the proof $\gamma$ on input a witness that contains indices $(1,2)$ and secret keys $(\SK_1,\SK_2)$.
	\begin{claim}\label{clm:lasthybrid}
	$\left|\Prob{H_6^{N}=1|\bar E^1}-\Prob{H_7=0|\bar E^1}\right|$ is negligible.
\end{claim}
\begin{proof}
	The proof follows straightforwardly from the WI property of $\NIWIDecf$. We observe that both the randomness used to compute $(\SK_1,\SK_2)$ and the randomness used to compute $(\SK_2,\SK_3)$ constitute valid witnesses for the statement  $(\BT_1, \allowbreak \ldots, \allowbreak \BT_N, \allowbreak \PKE.\PK_1, \allowbreak \ldots, \allowbreak \PKE.\PK_3, \allowbreak y)$. Additionally, we observe that, if event $\bar E^1$ occurs, any ballot in the set $S$ is in both experiments either replaced by $\bot$, if $\VerifyBallotf$ refuses it, or decrypted to the same value. Consequently, the tally is identical in both experiments.

	Note that according to the proof received, an adversary against NIWI can emulate experiment $H_6^N$ or $H_7$, and return the output of $\adv$.
      In the first case, the probability that $\adv$ outputs $0$ is exactly $\Prob{H_6^{N}=1|\bar E^1}$ because the winning condition is computed with respect to $b=0$, whereas in the second case it is $\Prob{H_7=0|\bar E^1}$ because the winning condition is computed with respect to $b=1$.
	\end{proof}

	Now, consider Equation~\ref{eqn:four} in Fig.~\ref{fig:equation}.
\begin{figure}
\begin{framed}
\tiny
	\begin{equation}\label{eqn:four}
	\begin{split}
	\Prob{H^Z=1|\bar E^1}=\\
	\Prob{H^Z=1|\bar E^1 \wedge b=0}\Prob{b=0}+\Prob{H^Z=1|\bar E^1 \wedge b=1}\Prob{b=1}=&\\
	=1/2\cdot\left(\Prob{H^Z=1|\bar E^1 \wedge b=0}+\Prob{H^Z=1|\bar E^1 \wedge b=1}\right)=&\\
	(\text{since $H_1$ is identically distributed to $H^Z$ with bit $b=0$ and $H_7$ to $H^Z$ with $b=1$})&\\
	=1/2\cdot\left(\Prob{H_1=1|\bar E^1}+\Prob{H_7=1|\bar E^1}\right)=&\\
	=1/2+1/2\cdot\left(\Prob{H_1=1|\bar E^1}-\Prob{H_7=0|\bar E^1}\right)=\\
	(\text{since $H_1$ (resp. $H_3,H_5$) is identically distributed to $H_2^0$ (resp. $H_4^0, H_6^0$)})&\\
	=1/2+1/2\cdot(
		\sum_{k=0}^{N-1}(\Prob{H_2^k=1|\bar E^1}-\Prob{H_2^{k+1}=1|\bar E^1})+
		(\Prob{H_2^N=1|\bar E^1}-\Prob{H_4^0=1|\bar E^1})+&\\
		\sum_{k=0}^{N-1}(\Prob{H_4^k=1|\bar E^1}-\Prob{H_4^{k+1}=1|\bar E^1})
		(\Prob{H_4^N=1|\bar E^1}-\Prob{H_6^0=1|\bar E^1})+&\\
		\sum_{k=0}^{N-1}(\Prob{H_6^k=1|\bar E^1}-\Prob{H_6^{k+1}=1|\bar E^1})
		(\Prob{H_6^N=1|\bar E^1}-\Prob{H_7=0|\bar E^1}))\le&\\
	1\le/2+1/2\cdot|(
		\sum_{k=0}^{N-1}(\Prob{H_2^k=1|\bar E^1}-\Prob{H_2^{k+1}=1|\bar E^1})+
		(\Prob{H_2^N=1|\bar E^1}-\Prob{H_4^0=1|\bar E^1})+&\\
		\sum_{k=0}^{N-1}(\Prob{H_4^k=1|\bar E^1}-\Prob{H_4^{k+1}=1|\bar E^1})
		(\Prob{H_4^N=1|\bar E^1}-\Prob{H_6^0=1|\bar E^1})+&\\
		\sum_{k=0}^{N-1}(\Prob{H_6^k=1|\bar E^1}-\Prob{H_6^{k+1}=1|\bar E^1})
		(\Prob{H_6^N=1|\bar E^1}-\Prob{H_7=0|\bar E^1}))|\le&\\
		(\text{by the triangle inequality})&\\
		\le1/2+1/2\cdot(
		\sum_{k=0}^{N-1}|\Prob{H_2^k=1|\bar E^1}-\Prob{H_2^{k+1}=1|\bar E^1}|+
		|(\Prob{H_2^N=1|\bar E^1}-\Prob{H_4^0=1|\bar E^1})|+&\\
		\sum_{k=0}^{N-1}|\Prob{H_4^k=1|\bar E^1}-\Prob{H_4^{k+1}=1|\bar E^1}|
		|\Prob{H_4^N=1|\bar E^1}-\Prob{H_6^0=1|\bar E^1}|+&\\
		\sum_{k=0}^{N-1}|\Prob{H_6^k=1|\bar E^1}-\Prob{H_6^{k+1}=1|\bar E^1}|
		|\Prob{H_6^N=1|\bar E^1}-\Prob{H_7=0|\bar E^1}|)\le&\\
	(\text{by Claims \ref{clm:hybridtwofull}\ -\ \ref{clm:lasthybrid}})&\\
	\le3k\cdot\negl,
\text{where $\negl$ is the sum of the negligible functions guaranteed by Claims \ref{clm:hybridtwofull}\ -\ \ref{clm:lasthybrid}}.
\end{split}
\end{equation}
\end{framed}
\caption{Equation~\ref{eqn:four}}
\label{fig:equation}
\end{figure}
Finally, Claim \ref{clm:expz} and equations \ref{eqn:one},\ref{eqn:three} and \ref{eqn:four} imply that $\Prob{\SecGame=1}\le \nu$ for some negligible function $\nu$ and the theorem is proven.
	\end{proof}

\begin{corollary} \label{cor:full}
	\ifnum\fullversion=1
	If the Decision Linear assumption (see Section \ref{app:dlin}) holds, then there exists a (fully) verifiable eVote.
	\else
	If the Decision Linear assumption (see Appendix \ref{app:dlin}) holds, then there exists a (fully) verifiable eVote.
	\fi
	\end{corollary}
	\begin{proof}
		Boneh \etal\ \cite{C:BonBoySha04} show the existence of a PKE with perfect correctness and unique secret key that fulfills the IND-CPA property under the Decision Linear assumption. Groth \etal\ \cite{C:GroOstSah06} show the existence of (one-message) NIWI (with perfect soundness) for all languages in $\NP$ and of statistically binding commitments. Both constructions are secure under the Decision Linear assumption.	Then, because Theorem~\ref{thm:fullver} and Theorem~\ref{thm:privfull} are proven, the corollary follows.
	\end{proof}




\fi
\end{document}